\documentclass[%
 reprint,
 amsmath,amssymb,
 aps,
]{revtex4-2}

\usepackage{graphicx}
\usepackage{dcolumn}
\usepackage{bm}
\usepackage{gensymb}

\def\be{\begin{equation}}
\def\ee{\end{equation}}
\def\ba{\begin{array}}
\def\ea{\end{array}}
\def\bea{\begin{eqnarray}}
\def\eea{\end{eqnarray}}
\def\bi{\begin{itemize}}
\def\ei{\end{itemize}}
\def\bra#1{\langle #1|}
\def\ket#1{|#1\rangle}
\def\expe#1{\langle #1 \rangle}
\def\half{{\textstyle{1\over2}}}

\begin{document}

\preprint{APS/123-QED}

\title{Glassy quantum nuclear pasta in neutron star crusts}

\author{William G. Newton}
\email{William.Newton@tamuc.edu}
\affiliation{Department of Physics and Astronomy, Texas A\&M University-Commerce, Commerce, TX 75429-3011, USA}

\author{Mark Alexander Kaltenborn}
\affiliation{Department of Physics, The George Washington University, Washington DC 20052, USA}
\affiliation{The Computational Physics and Methods Group, Los Alamos National Laboratory, Los Alamos, NM 87545 USA}

\author{Jirina Rikovska Stone}
\affiliation{Department of Physics and Astronomy, University of Tennessee, Knoxville, TN 37996, USA}
\affiliation{Physics Division, Oak Ridge National Laboratory, Oak Ridge, TN 37831, USA}
\affiliation{Department of Physics, University of Oxford, Oxford, UK}

\author{Sarah Cantu}
\affiliation{Department of Physics and Astronomy, Texas A\&M University-Commerce, Commerce, TX 75429-3011, USA}

\author{Shuxi Wang}
\affiliation{Department of Physics and Astronomy, Texas A\&M University-Commerce, Commerce, TX 75429-3011, USA}

\author{Amber Stinson}
\affiliation{Department of Physics and Astronomy, Texas A\&M University-Commerce, Commerce, TX 75429-3011, USA}

\date{\today}

\begin{abstract}
We investigate the nuclear pasta phases in neutron star crusts by conducting a large number of three-dimensional Hartree-Fock+BCS calculations at densities leading to the crust-core transition. We survey the shape parameter space of pasta at constant pressure. Spaghetti, waffles, lasagna, bi-continuous phases and cylindrical holes occupy local minima in the resulting Gibbs energy surfaces. The bi-continuous phase, in which both the neutron gas and nuclear matter extend continuously in all dimensions and therefore protons are delocalized, appears over a large range of depths. 

Our results support the idea that nuclear pasta is a glassy system. Multiple pasta configurations coexist in a given layer of the crust. At a characteristic temperature, of order $10^8$-$10^9$K, different phases become frozen into domains whose sizes we estimate to be 1-50 times the lattice spacing and over which the local density and electron fraction can vary. Above this temperature, there is very little long-range order and matter is an amorphous solid. Electron scattering off domain boundaries may contribute to the disorder resistivity of the pasta phases. Annealing of the domains may occur during cooling; repopulating of local minima during crustal heating might lead to temperature dependent transport properties in the deep layers of the crust. 

We identify 4 distinct regions: (1) nuclear pasta first appears as a local minima, but spherical nuclei are the ground state; (2) nuclear pasta become the absolute minimum, but spherical nuclei are still a local minimum (3) only nuclear pasta appears in local minima, and protons are still localized in at least one dimension (4) only pasta appears, and protons are delocalized. The whole pasta region can occupy up to 70\% of the crust by mass and 40\% by thickness, and the layer in which protons are delocalized could occupy 45\% of the crust mass and 25\% of its thickness.

\end{abstract}

\maketitle


\section{\label{Intro}Introduction}

As one moves deeper into the solid neutron star crust, passing through the outer crust stabilized by degenerate electrons and the inner crust stabilized by neutrons that have leaked out of the neutron-rich nuclei forming the crystal lattice, the nuclei occupy an increasing volume of the matter. Soft-matter systems on Earth give clues about what to expect as the separation between nuclei become comparable with their size \cite{Caplan:2017aa}. When two fluid phases exist with one increasingly concentrated with respect to another, matter is self-organized into a number of different geometries. For example, aggregates of amphiphilic molecules called micelles arrange themselves into spherical, cylindrical and planar phases with increasing concentration in water \cite{Jones_2002}. This phenomenon of self-orgnization can be understood to arise from frustration: the microscopic components of the system are subject to a competition between interactions operating over similar length scales. The total energy cannot be minimized with respect to all these microscopic interactions simultaneously. There is not single potential well for the system to fall into; instead a rich energy landscape of local minima separated by energy barriers of varying heights emerges. Small changes in the initial conditions of the system could lead to the microscopic constituents arranging themselves in different ways. Such materials are expected to have complex low-energy dynamics, giving rise to correspondingly complex behaviors of thermal conductivities, electrical resistivities and elastic properties.

Our system of interest comprises a Coulomb lattice of nuclei immersed in a fluid of neutrons. The microscopic interactions in play are the nuclear force and the Coulomb interaction. Throughout most of the inner crust, the short range nuclear interaction binds nuclei while the long rang Coulomb interaction drives the formation and stability of the lattice, determining the spacing between nuclei. As we move deeper towards the crust-core transition, however, the separation between nuclei becomes comparable with the length scale of the nuclear interaction $\sim 10$fm and frustration is expected to ensue.

Consequences of this convergence of length scales was first studied by examining the compromise between surface and bulk Coulomb energies of the nuclei in a liquid drop model and led to the conclusion that a sequence of exotic nuclear geometries became preferable at $\approx 0.5 \rho_{\rm cc}$ where $\rho_{\rm cc}$ is the crust core transition density \cite{Lamb:1978lr,Ravenhall:1983fp,Hashimoto:1984lq,Oyamatsu:1984eu}. Cylindrical, planar, cylindrical hole and spherical hole configurations was found to be energetically favored with increasing density; the community has embraced a terminology based on the resemblance of the shapes to different forms of pasta \footnote{A question is whether to continue the proliferation of pasta nomenclature. We come done on the side of yes!}, with the above sequence often referred to as spaghetti, lasagna, anti-spaghetti (or bucatini \cite{Fattoyev:2017aa}) and anti-gnocchi (with spherical nuclei cast as gnocchi). More recently, phases intermediate to this canonical sequence have been explored, with perforated planar configurations appearing between cylindrical and planar phases (nuclear waffles \cite{Schneider:2014hh}), planar phases with helical connectors between sheets (parking garage structures \cite{Berry:2016wf}) and bi-continuous-P and gyroid phases mediating the transition between between planar and cylindrical holes \cite{Nakazato:2009aa,Nakazato:2011aa,Schuetrumpf:2015gf} (we are awaiting the development of pasta forms that mimic many of these phases). Many of these find counterparts in the world in soft condensed matter \cite{Pethick:1998aa,Berry:2016wf}. The behavior of soft condensed matter systems suggests we should expect complex, potentially non-isotropic elastic and transport properties in nuclear pasta.


The state-of-the-art simulating nuclear pasta over the past two decades encompasses three dimensional quantum mean-field simulations using relativistic and non-relativistic energy density functionals (EDFs) \cite{Magierski:2002fk,Gogelein:2007rt,Gogelein:2008fr,Grygorov:2010ys,Newton:2009eu,Pais:2012lq,Schuetrumpf:2013zr,Schuetrumpf:2014mz,Schuetrumpf:2015ly,Schuetrumpf:2015gf,Sagert:2016aa,Schuetrumpf:2016ve,Schuetrumpf:2019aa,Schuetrumpf:2020aa}, and classical and quantum molecular dynamics simulations which access larger computational volumes but incorporate more schematic nuclear forces \cite{Maruyama:1998om,Sonoda:2008wu,Horowitz:2004cj,Horowitz:2005ca,Horowitz:2008ys,Schneider:2013tk,Schneider:2014hh,Caplan:2015qr,Schneider:2016qd,Berry:2016wf,Caplan:2018aa,Schneider:2018aa,Lin:2020aa,Nandi:2016aa,Nandi:2018aa,Nandi:2018ab,Dorso:2012fb,Gimenez-Molinelli:2014df,Dorso:2018aa,Lopez:2021aa}. Microscopic quantum mean-field calculations demonstrated the complexity of the energy landscape of nuclear pasta which was shown to be enhanced by quantum shell effects \cite{Magierski:2001yq}. This strongly suggests the possibility that pasta is highly disordered and amorphous with multiple different nuclear shapes coexisting at a given depth in the crust. Larger scale molecular dynamics simulations reveal other sources of disorder: for example, from topological defects in pasta \cite{Schneider:2016qd}, in which planar phases develop defects in the form of bridges between adjacent sheets.

As the crust cools, microscopic domains of different pasta phases may form at the same crust depth \cite{Schneider:2018aa} which may persist on long timescales before annealing \cite{Caplan:2020aa}. Individual domains could have highly anisotropic elastic and transport properties \cite{Yakovlev:2018aa,Kobyakov:2018aa,Pethick:2020aa}, for example arising because of the difference between electron scattering parallel to spaghetti and lasagna structures and perpendicular to them. However, averaging over domains may render them more isotropic at the macroscopic level and reduce the resistive effect of pasta \cite{Yakovlev:2018aa}. However, the existence of domains could give rise to another source of resistivity: electron scattering off domain boundaries. Thermal fluctuations may destroy the long-range order of pasta \cite{Watanabe:2000aa,Caplan:2020aa} and may set the length scale of domains; those length scales will determine if the rate of electron scattering rate off domain boundaries is important for the overall resistivity of the crust. 

Pasta could account for 50\% of the crust by mass \cite{Lorenz:1993aa,Balliet:2020aa}, so there are observational consequences to the microscopic organization of pasta. The increased resistivity of disordered pasta could lead to potentially observable effects on the cooling curves of X-ray binaries \cite{Horowitz:2015rt} and the evolution of pulsar magnetic fields \cite{Pons:2013ly}. 

Given that there are many low lying minima separated by energy barriers, perhaps the best terrestrial analogue is a glass: solids which, when heated, pass through an amorphous phase before melting. One possible scenario is as follows: as the neutron star cools below a characteristic temperature set by the energy barriers between local minima (either early in its life or after a period of accretion-induced crustal heating), amorphous nuclear pasta undergoes a transition in which it becomes frozen into coexisting domains of a certain length scale. Annealing may then take place on a timescale that is uncertain but could be long compared to, for example, the cooling timescale of the crust. The energy spectrum of the pasta phases, and the typical temperature and length scales in play in this scenario, are the subject of this paper.

We aim to map out the energy surfaces of nuclear pasta at a variety of densities by performing a large number of three-dimensional Skyrme-Hartree-Fock+BCS (3DHF+BCS) simulations at zero temperature and at proton fractions around $\beta$-equilibrium. Although our simulations are restricted to smaller computational volumes than molecular dynamics simulations, we can access zero temperature and lower proton fractions. Most calculations are performed at a given density; however, a given depth in the crust is defined by its pressure. In order to examine coexisting domains, energy surfaces at constant pressure must be calculated; the Gibbs free energy is the relevant quantity to compare different phases. We will interpolate between calculations over a range of densities to find the configurations at a constant pressure and thus map out the Gibbs energy surfaces.


We aim to answer the following questions:

\begin{enumerate}
    \item How many phases of pasta could coexist at a given depth in the crust?
    \item How does the structure of the energy surfaces evolve with depth?
    \item What is the characteristic temperature below which pasta could be locked into microscopic domains?
    \item What are the characteristic length scales of the domains?
\end{enumerate}

We perform around 500,000 CPU hours of 3DHF+BCS calculations, using a quadrupole constraint to control the nuclear geometry. The only bias in our results is the imposition of parity conservation in all three dimensions to restrict the computation to one octant of the unit cell and make such a wide survey of parameter space possible. This means that certain topologically distinct phases (such as the gyroid) will be omitted. It is important to note that the constraint itself does not bias us; as we shall see, the constraint phase space we explore admits all triaxially symmetric pasta shapes, and by systematically exploring that space we allow for the appearance of all possible shapes. 

In section~II we describe the numerical method. In section~III we describe our results, starting with a detailed description of our analysis of one particular layer, then displaying the rest of our results layer-by-layer. In section IV we analyze what our results imply for the deep layers of the neutron star crust, and in section V we give out conclusions.

\section{Numerical method}

There are many detailed accounts of the Skyrme-Hartree-Fock method \cite{Bender:2003aa}, and we have outlined the method used in our previous paper \cite{Newton:2009eu}. Here we highlight details of the implementation that are important for this study.

\subsection{Skyrme Hartree-Fock}

Approximating the ground state many-body wavefunction as a Slater determinant $\Phi$, and minimizing the Skyrme energy density functional $\mathcal{E}_{\rm Skyrme}[\Phi] = \langle \Phi| \hat{H}_{\rm Skyrme} | \Phi \rangle$ with respect to the single particle wavefunctions obtains the Skyrme Hartree-Fock equations. We write them with the inclusion of a quadrupole constraining potential, explained below, as

\be \label{2:E21a} [ - \nabla {\hbar^2 \over 2 m_q^*}
\nabla + u_q({\bf r}) + \lambda_{\rm c} \cdot Q ] \phi_{i,q}({\bf r}) = \epsilon_{i,q} \phi_{i,q} ({\bf r}). \ee

\noindent where $q =$ p,n label the isospin states, $i$ the single particle states. $u_q$ are the one-body Hartree-Fock potentials:

\begin{align*} 
u_q &= t_0 (1+ \half x_0) \rho - t_0(\half + x_0) \rho_q \notag
\\
       & \;\;\; + {1 \over 12} t_3 \rho^{\alpha} \bigg[(2 +
       \alpha)(1 + \half x_3)\rho - 2(\half + x_3) \rho_q \notag \\
       &\;\;\;\;\;\;\;\;\;\;\;\;\;\;\;\;\;\;\; - \alpha(\half + x_3) {\rho^2_{\rm p} + \rho^2_{\rm n} \over \rho}
       \bigg] \notag \\
       & \;\;\; + {1 \over 4}[ t_1 (1 + \half x_1) + t_2 (1+ \half
       x_2)] \tau \notag \\
       &\;\;\;\;\;\;\;\;\;\;\;\;\;\;\;\;\;\;\; - {1 \over 4}[ t_1(\half + x_1) - t_2 (\half +
       x_2)] \tau_q \notag \\
       & \;\;\; - {1 \over 8}[3 t_1 ( 1 + \half x_1) - t_2(1 +
       \half x_2)] \nabla^2 \rho \notag \\
       &\;\;\;\;\;\;\;\;\;\;\;\;\;\;\;\;\;\;\; + {1 \over 8} [3 t_1 (\half + x_1) + t_2 (\half +
       x_2)] \nabla^2 \rho_q,
\end{align*}

\noindent and $m_q^*$ is the effective mass. 

\begin{align} {\hbar^2 \over 2 m_q^*} = {\hbar^2 \over 2
m_q} &+ {1 \over 4}[ t_1 (1 + \half x_1) + t_2 (1 + \half
       x_2)] \rho \notag \\ &- {1 \over 4}[ t_1(\half + x_1) - t_2 (\half +
       x_2)] \rho_q \; .
\end{align}

\noindent Here $t_i$,$x_i$ and $\alpha$ are parameters of the Skyrme interaction, $\rho = \rho_{\rm p} + \rho_{\rm n}$ are the nucleon densities and $\tau = \tau_{\rm p} + \tau_{\rm n}$ are the kinetic energy densities. 

A key feature of our simulations is the quadrupole potential term we have added to the single particle Hamiltonian $\lambda_{\rm c} \cdot Q$ to control the geometry of the nucleon density distribution so we can systematically survey the shape space of the pasta configurations. The quadrupole operator has elements in coordinate space

\begin{equation} \label{Eqn: Qmoment1}
    Q_{ab} = 3x_ax_b - r^2 \delta_{ab}
\end{equation}

\noindent with $\{x_a\}=x,y,z$. 

Restricting ourselves to triaxial shapes, the quadrupole operator becomes diagonal $Q_a=3x_a^2-r^2$. The strength of the constraining force has components $\lambda_{c,a}$. This is an artificial potential whose strength is reduced to zero as we approach convergence, so that it doesn't give an artificial contribution to the total energy \cite{Cusson:1985aa}.

\subsection{The computational grid}

We solve the Hartree-Fock equations in coordinate space. The computational domain is taken to be a cube defined by Cartesian coordinates $x_a$ with the origin at the
centre of the cell. Each co-ordinate runs over $-l_a
\le x_a \le l_a$ so that the length of the cell in
each direction is 2$l_a$. The space is discretized to form a grid of
collocation points $x_{a,i}$ with even spacings in each
direction $\Delta x_a$, defined by

\be \label{Eqn: grid2} x_a = (i + \frac{1}{2})
\Delta x_a \;\;\;\;\;\;\;\; i = -N_a, -N_a +
1, ... N_a-1 .\ee

\noindent where $N_a$ is the number of collocation points we use in direction $a$. In this work we use cubic cells so $N_x$=$N_y$=$N_z$=$N$. To make a large set of calculations feasible, we take parity in all three directions to be a good quantum number. It is thus sufficient to calculate just one octant of the computational cell. This means we limit ourselves to surveying tri-axial shapes only. Simple periodic boundary conditions mean the wavefunctions obey

\be \label{bloch3} \phi_{i,q}({\bf r} + {\bf T}) = \phi_{i,q}({\bf r}),
\ee
\noindent where ${\bf T}$ is the translation vector from the position ${\bf r}$ to
the equivalent positions in the adjacent cells. The boundary conditions are enforced by representing the derivatives and solving for the Coulomb potential in Fourier space. The Coulomb solver is implemented using the FFTW software package~\cite{Frigo:2012aa}. Integrals on the grid are performed using the trapezoidal rule which is exact for functions represented by Fourier series \cite{Baye:1986aa}.

The restrictions we impose above on our computational space are necessary for the large survey of densities, proton fractions and nuclear shapes to be tractable. They impose limitations noted below.

\begin{itemize}
    \item We must set the spin-orbit contribution to the potentials to be zero. This has the added benefit of the wavefunctions being entirely real functions and therefore reducing the computational cost. Although the spin-orbit interaction is important in determining the details of the energy spectrum in finite nuclei, it has been shown to make a much smaller contribution under crust conditions at high density, especially for the unbound neutron gas \cite{Negele:1973aa}.
    \item Enforcing simple periodic boundary conditions rather than the full Bloch boundary conditions makes the simulations susceptible to spurious shell effects \cite{Schuetrumpf:2016ve}. These arise in large part from the artificial discretization of the unbound neutrons' single particle spectrum caused by restricting the calculations to a finite volume. This makes comparing results at different cell sizes unreliable as in low proton fraction matter the variation of the spurious contribution to the shell energy is of similar magnitude to the the physical contribution, and consequently we cannot find the energetically preferred unit cell size. We will therefore restrict ourselves to probing the shape of nuclear configurations at constant cell sizes.
\end{itemize}

\subsection{Single particle states}

Our solution to the Hartree-Fock equations begins with an initial guess for the wavefunctions. We have found that starting the neutron wavefunctions as plane waves and the proton wavefunctions as harmonic oscillator wavefunctions leads to most efficient convergence; other combinations of initial wavefunctions were tested, and it was verified that they lead reliably to the same ground state configurations so long as we impose the quadrupole constraint to control the shape of the configuration. The number of possible single particle states we can represent on our grid is $(N-1)^3$, and we evolve all of them; those that start out unoccupied often evolve to be occupied in the converged final state.

We impose BCS pairing, so that single particle states are occupied according to the distribution function

\be \label{4:Pair1} w_{k,q}^{\rm pair} = \half \bigg( 1 - {\epsilon_{k,q} - \epsilon_{F,q} \over \sqrt{(\epsilon_{k,q} - \epsilon_{{\rm F},q})^2 + f_{k,q}^2 \Delta_q^2}} \bigg) . \ee

\noindent where $f_{k,q}$ is a function that acts to cutoff coupling to continuum states and confine the active pairing space to the vicinity of the Fermi surface \cite{Bender:2000aa}. The pairing gap $\Delta_q$ is taken to be a constant, set as $\Delta_q=11.2$MeV$/\sqrt{A}$ \cite{Bonche:1985aa}. Other than the fact that pairing is a physical feature of our system, it significantly improves convergence of our iterations.

\subsection{Iterative solution}

We solve the HF equations iteratively, forming densities and potentials from the current wavefunctions and solving the HF equations to obtain new wavefunctions repeatedly until the wavefunctions converge. We combine two different algorithms to achieve the convergence to the ground state with maximum efficiency. 

We start off using the imaginary time step iteration~\cite{Davies:1980aa}, an adaptation of the time dependent HF iteration $\phi^{(n+1)}_{i,q} = e^{-i \hat{h}^{(n+1/2)}_{\rm HF} \Delta t / \hbar} \phi^{(n)}_{i,q}$ which evolves the wavefunctions by a time interval $\Delta t$. Here, $\hat{h}^{(n+1/2)}_{\rm HF}$ is a numerical approximation to the Hamiltonian at the half time step $(n + 1/2)\Delta t$. The imaginary time step is obtained by replacing $\Delta t$ with $-i \Delta t$ and $\hat{h}^{(n+1/2)}_{\rm HF}$ with $\hat{h}^{(n)}_{\rm HF}$. Defining the parameter $\lambda = \Delta t / \hbar$, and expanding the exponential in a power series, the $(n+1)^{\rm th}$ wavefunction is formed from the $n^{\rm th}$ by: 

\be \label{e13} \phi^{(n+1)}_{i,q} = e^{-\lambda \hat{h}^{(n)}_{\rm HF}}
\phi^{(n)}_{i,q} = \sum_{k=1}^{k_{\rm cut}} {1 \over k!} ( - \lambda
\hat{h}^{(n)}_{\rm HF})^k \phi^{(n)}_{i,q} \ee

\noindent where $\lambda$ controls the magnitude of the
imaginary time step, that is the size of the iterative step. We
can adjust the number of terms contained in the exponential
expansion through $k_{\rm cut}$.

The imaginary time step iteration is very robust - even if we
start with a set of initial wavefunctions that are far from those of the ground state solution, it
will remain stable. It converges quickly initially,
but as one approaches convergence, it slows down exponentially. Thus, when we get closer to the ground state we switch to the damped gradient iteration \cite{Reinhard:1982aa,Cusson:1985aa}. Here, the $(n+1)^{\rm th}$ wavefunction is formed from the $n^{\rm th}$ by: 

\be \label{e14} \phi^{(n+1)}_{i,q} = \phi^{(n)}_{i,q} - x_0
\hat{\text{D}}(e_0) (\hat{h}^{(n)}_{\rm HF} - \epsilon_{i,q}^{(n)}) \phi_{i,q} ,\ee

\noindent where the damping operator 

\be \label{e15} \hat{\text{D}} = \bigg[ 1 + {\hat{t}_x \over e_0}
\bigg]^{-1} \bigg[ 1 + {\hat{t}_y \over e_0} \bigg]^{-1} \bigg[ 1
+ {\hat{t}_z \over e_0} \bigg]^{-1} ,\ee

\noindent acts to damp out large kinetic
energy components of the wavefunctions with kinetic energies above $e_0$ that slow down convergence. Here $\hat{t}_a$ are the one-dimensional kinetic energy operators. The damped gradient iteration requires initial
wavefunctions that are relatively close to the actual ground state wavefunctions otherwise it becomes unstable. It converges roughly linearly and so is more efficient than the imaginary time step at late times.

\subsection{The Quadrupole Constraint}

In order to systematically survey the spectrum of nuclear geometries and their corresponding energies, we need control the geometry of the ground state to which an iteration converges without excluding any of known pasta geometries. To do this we implement a quadrupole constraint. Given the reflection symmetry, the next order deformation, consistent with our boundary conditions, is hexadecapole. It is expected to give energy variations at least an order of magnitude smaller than that of the quadrupole deformation.

Taking the quadrupole operator to be diagonal, the quadrupole moments of the nucleon density distribution are the matrix elements

\be \label{Eqn: Qmoment2} q_{a} = \expe{\hat{Q}_a} = \sum_{i=1}^N \bra{\phi_i}
Q_{a} \ket{\phi_i}. \ee

The three non-zero quadrupole moments must also fulfil $q_x + q_y + q_z =
0$, so just two of them are independent. The nuclear shape can be parameterized in Cartesian or spherical polar coordinates

\begin{align} \label{Eqn: CartesianAlphas} R &= R_0 (1 + \alpha_x
\xi^2 + \alpha_y \eta^2 + \alpha_z \zeta^2) \notag \\
&= R_0 ( 1 + \alpha_{20}
Y_{20} (\theta, \phi) + \alpha_{2+2} Y_{2+2} (\theta, \phi) \notag \\
 &\;\;\;\;\;\;\;\;\;\;\;\;\;\;\;\;\;\;\;\;\;\;\;\;\;\;\;\;\;\;\;\;\;\;\;\; +
\alpha_{2-2} Y_{2-2} (\theta, \phi) ),\end{align}


\begin{figure}[t]
\includegraphics[scale=0.5]{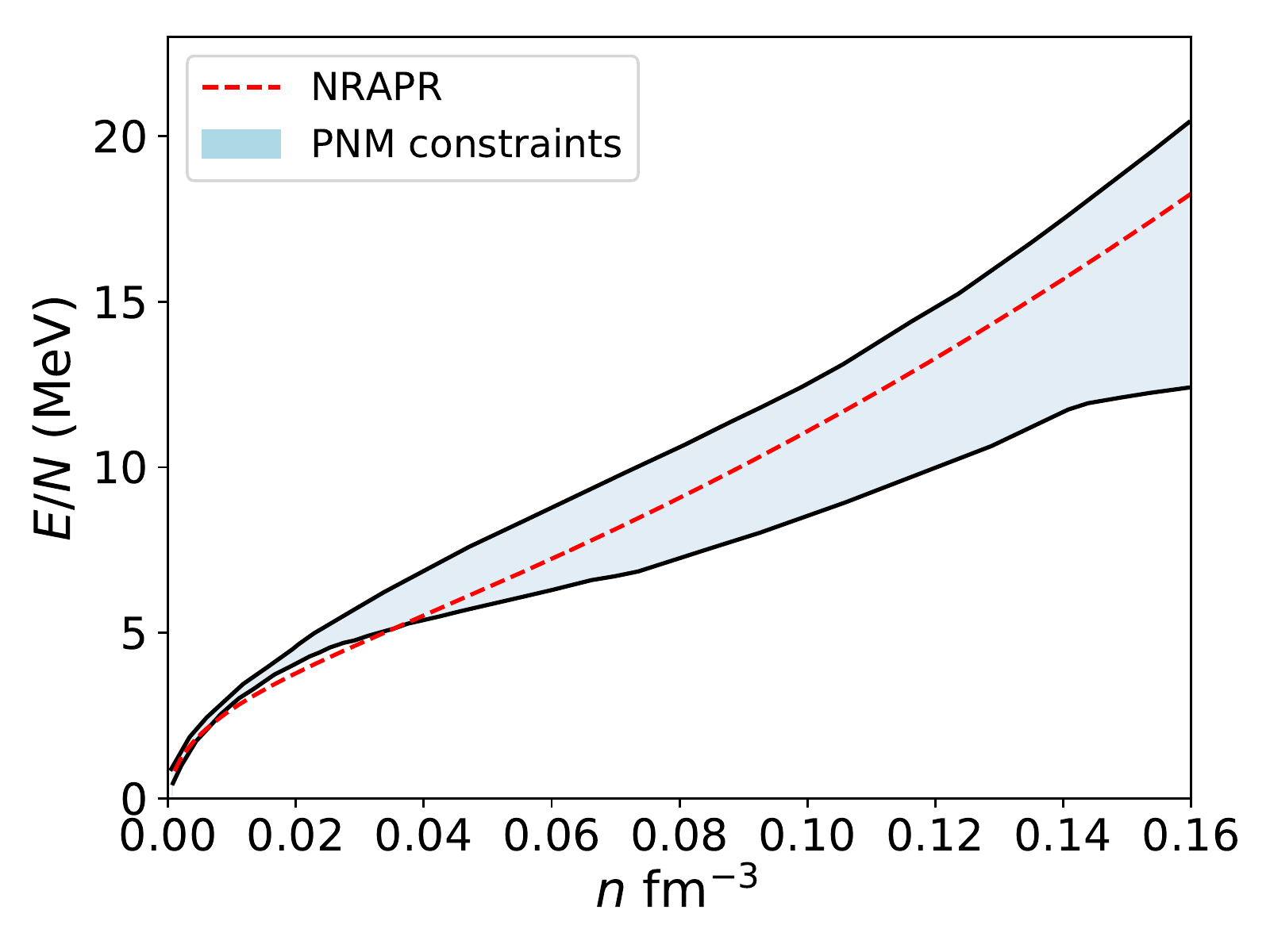}
\caption{The pure neutron matter EOS of the NRAPR Skyrme parameterization used in this work (red dashed line) compared to the region predicted by \emph{ab-initio} calculations of PNM \cite{Gandolfi:2015nr,Tews:2016ty}. Given the importance of the PNM EOS to crust properties, it is important to consider the consistency of the interaction used to model the crust and our best theoretical knowledge of PNM.} \label{fig:1}
\end{figure}


\noindent where $q_a = R_0 \alpha_a$ for $\{ a \}=x,y,z$, $R_0$
is the root mean square nuclear radius, $\xi = x/R_0$, $\eta = y/R_0$ and $\zeta = z/R_0$. The spherical polar moments are related to their Cartesian counterparts via 

\be \label{Eqn: alphax,y} \alpha_{2\pm2} = \sqrt{2\pi \over 15} (
\alpha_x - \alpha_y) \equiv \alpha_2 ,\ee

\be \label{Eqn: alphaz} \alpha_{20} = \sqrt{8\pi \over 90} (
2\alpha_z - \alpha_x - \alpha_y) \equiv \alpha_0 ,\ee

\noindent where $\alpha_0$ is the relative stretch along the z axis of the
nuclear cluster with respect to the x and y axes and $\alpha_2$ is the relative difference 
in length between the
x and y axes. We can define the parameters $\beta, \gamma$ \cite{Greiner:1997}:

\be \label{Eqn: betagamma} \alpha_0 = \beta \cos{\gamma}
\;\;\;\;\;\;\;\; \alpha_2 = {1 \over \sqrt{2}} \beta \sin{\gamma} ,\ee

\noindent analogous to polar co-ordinates
in ($\alpha_0$, $\alpha_2$) space: $\beta$ represents the magnitude of the deformation
of the configuration, and $\gamma$ the direction of the deformation from prolate $\gamma$ = 0$^o$ to oblate $\gamma = 60^o$.

We specify our desired quadrupole moments through the polar co-ordinates $\alpha, \beta$ given in equation~(\ref{Eqn: betagamma}). These are then turned into the moments $q_a$ through equations~(\ref{Eqn: alphax,y}) and~(\ref{Eqn: alphaz}), the requirement that $\alpha_x + \alpha_y + \alpha_z = 0$ and the definition $q_a = R_0 \alpha_a$.

The force strength needs to be updated iteration by iteration as it drives the quadrupole moments towards the desired values~\cite{Cusson:1985aa}. Denoting
the quadrupole moments we wish our nuclear configuration to converge towards by $q_{a,0}$, the procedure is as follows: An intermediate iteration is carried out $\ket{\tilde{\phi}_i^{(n)}} = \mathcal{O} I
[(\hat{h}_{HF} + x_0 \lambda^{(n)}_{\rm c} \cdot \hat{Q})
\ket{\phi_i^{(n)}}]$, where $I$ represents the operation of either the imaginary time step iteration or the damped gradient iteration, and $x_0$ is the same parameter that controls the damped gradient iteration step. Intermediate quantities $\ket{\tilde{\phi}_i^{(n)}}, \;\;
\tilde{\rho}^{(n)}, \;\; \tilde{q}_a^{(n)}, \;\;
\tilde{q}_a^{2 \; (n)}$ are calculated. The components of the constraining force strength are updated at each step according to

\be \label{Eqn: constraint1} \lambda_{{\rm c},a}^{(n+1)} = \lambda_{{\rm c},a}^{(n)} + {c_0 (\tilde{q}_a^{(n)} - q_a^{(n)}) \over 2 x_0
(q_a^{2 \; (n)} - (q_a^{(n)})^2/N_q) + d_0 }, \ee

\noindent where $N_q$ is the number of particles of species $q$.  We define
\be \label{Eqn: constraint2} \delta \lambda_{{\rm c},a} = {c_0 (q_a^{(n)} - q_{a,0})
\over 2 x_0 (q_a^{2 \; (n)} - (q_a^{(n)})^2/N_q)
+ d_0 }. \ee 

\noindent Finally, the ($n+1$)th wavefunctions are formed as $\ket{\phi_i^{(n+1)}} = \mathcal{O}
[(\ket{\tilde{\phi}_i^{(n)}} - x_0(\lambda_{{\rm c}}^{(n+1)} - \lambda_{{\rm c}}^{(n)} +
\delta \lambda_{{\rm c}}) \cdot \hat{Q}) \ket{\tilde{\phi}_i^{(n)}}]$
from which we finally obtain the updated quantities. Here, $c_0$ and $d_0$ are parameters adjusted
for optimum convergence of the constraint iteration. This procedure must be performed for the three components of the quadrupole moment $a \in \{x,y,z\}$. Note that as we converge to the ground state with the desired quadrupole moments, the strength of the constraining force tends to zero.

We apply the constraint only to the neutrons; the proton density distribution follows the neutron density distribution to a good degree of accuracy.

\begin{figure}[t]
\includegraphics[scale=0.55]{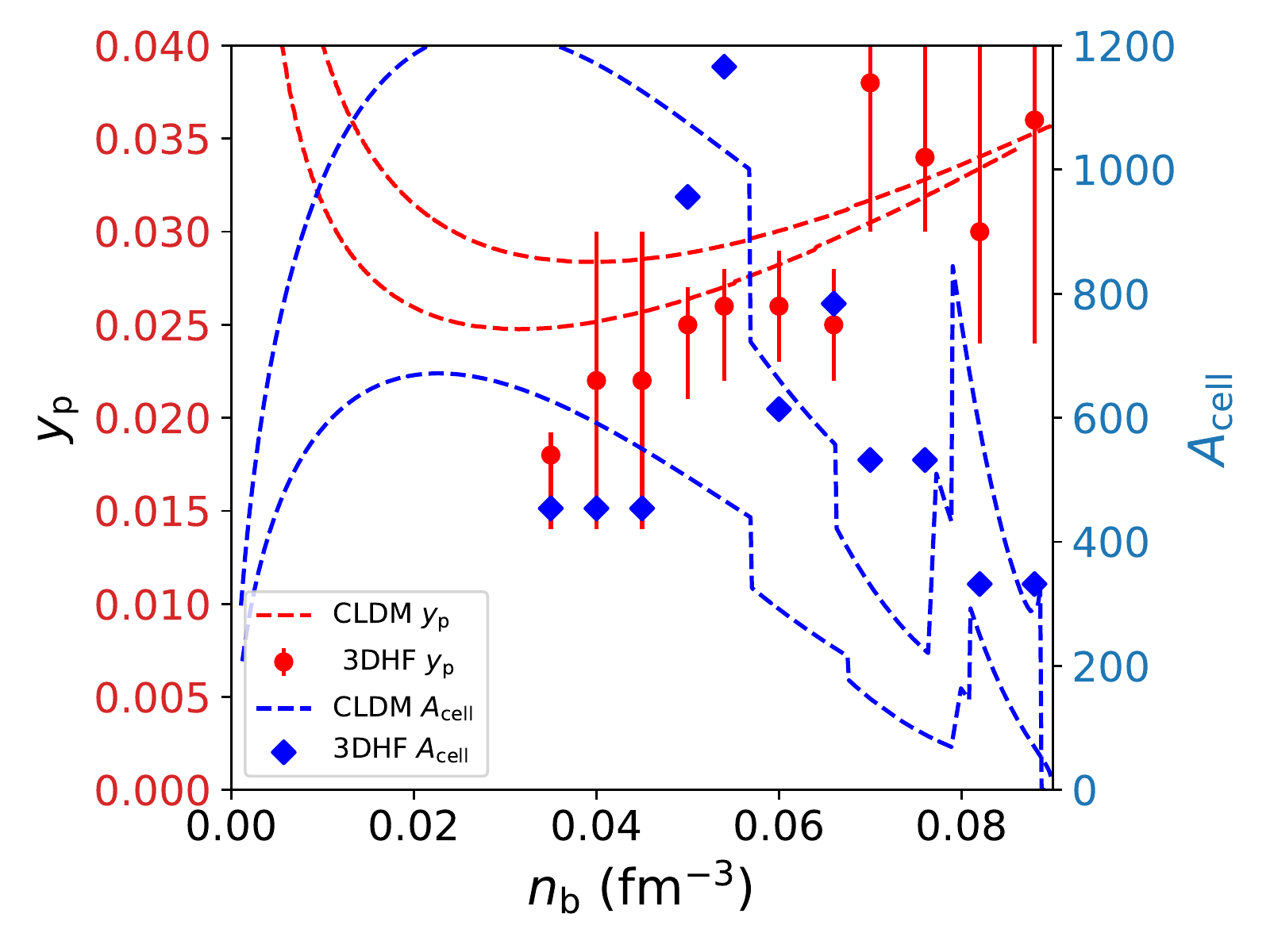}
\caption{The red and blue dashed lines are the bounds of the total nucleon number in the cell $A_{\rm cell}$ and the average proton fraction $y_{\rm p}$ from a compressible liquid drop model using the NRAPR interaction \citep{Steiner:2005aa} and varying the surface energy parameters over a reasonable range \citep{Newton:2013sp}. The blue diamonds are the values of $A_{\rm cell}$ we choose to perform our 3DHF calculations at. The red points indicate the beta-equilibrium proton fractions we at each density from our 3DHF calculations, and the ``error bars'' through them indicate the range of proton fractions we performed calculations at.} \label{fig:2}
\end{figure}

\begin{figure*}[!t]
\includegraphics[scale=0.35]{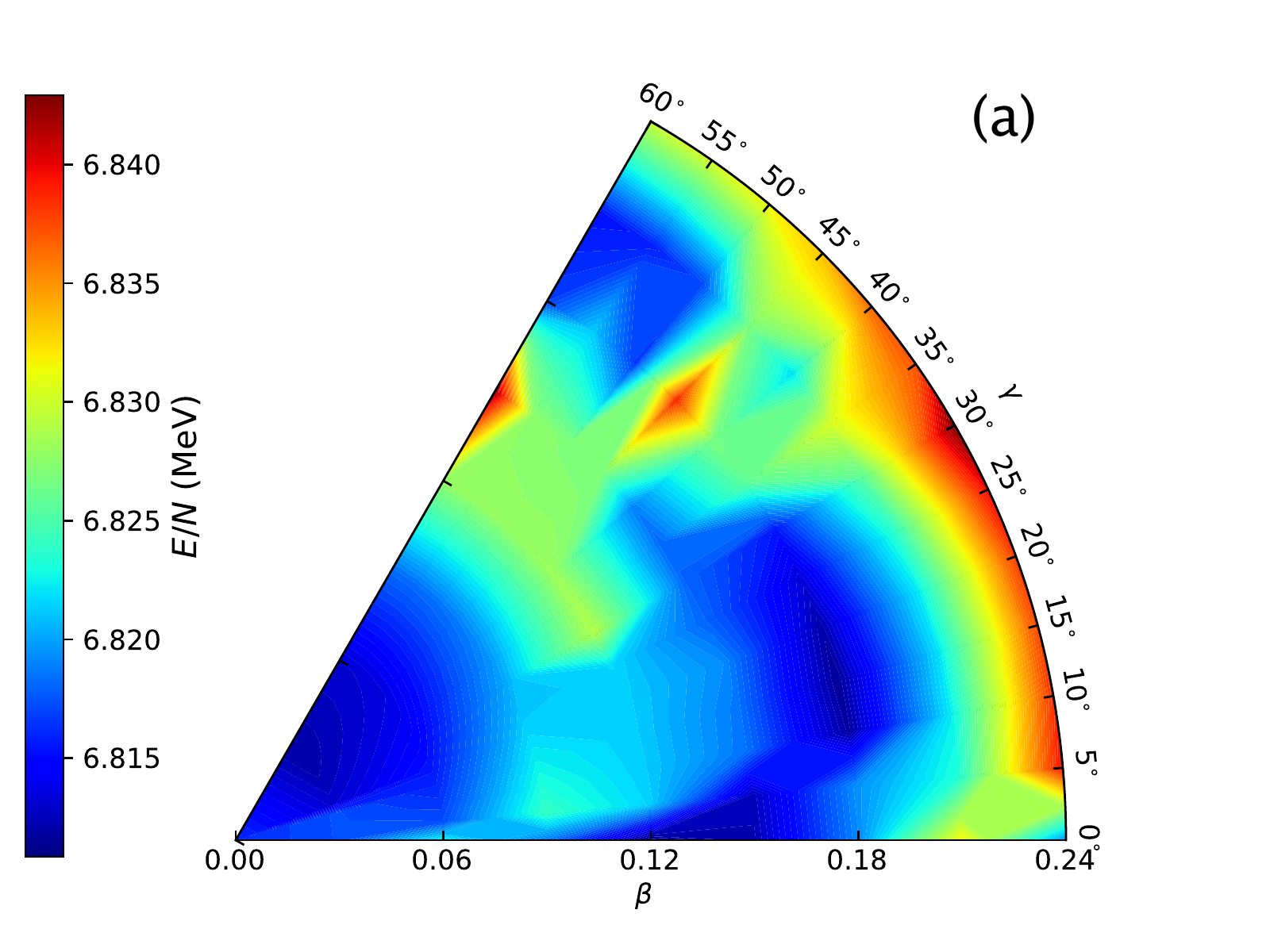}\includegraphics[scale=0.35]{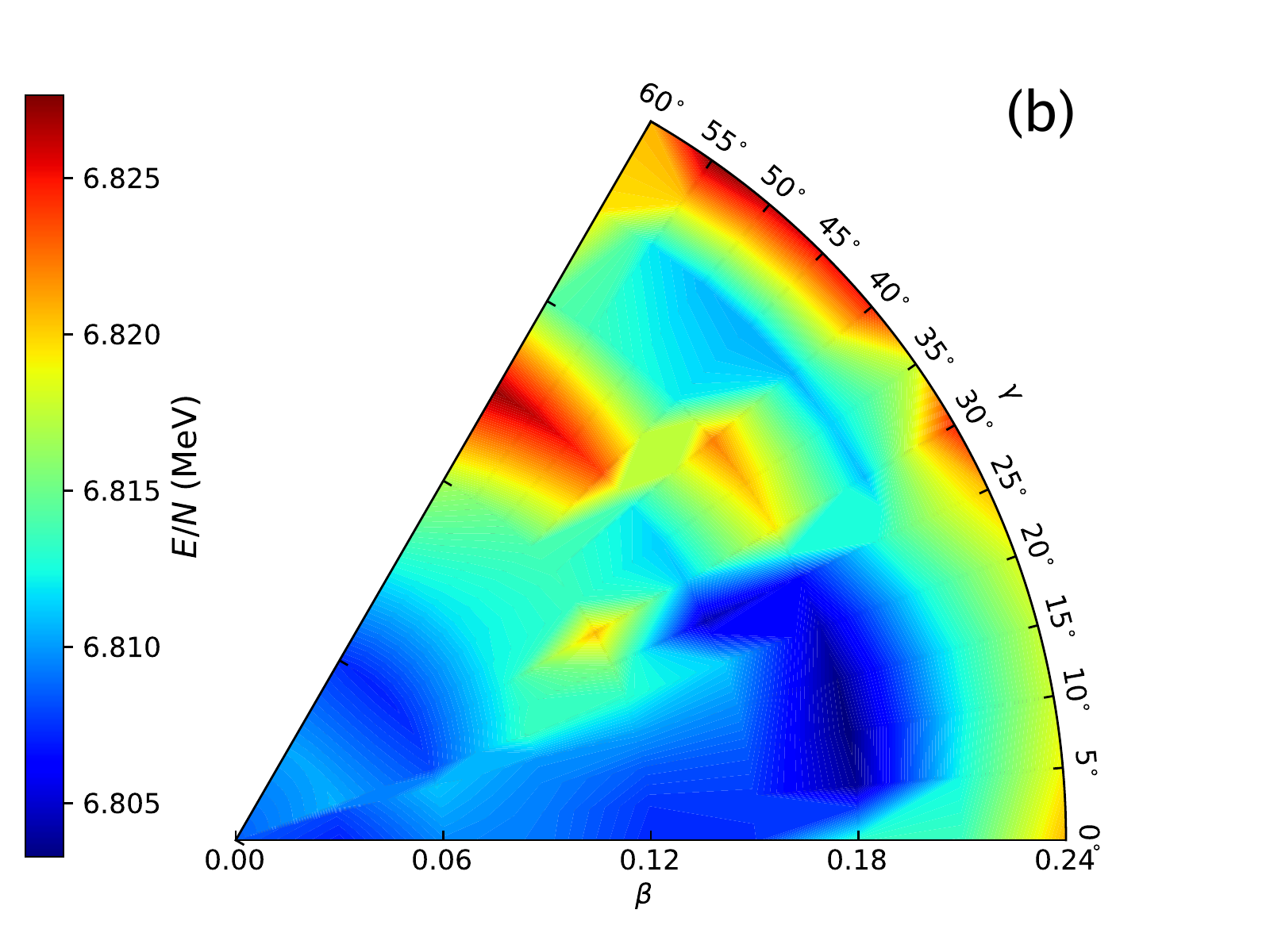}\includegraphics[scale=0.35]{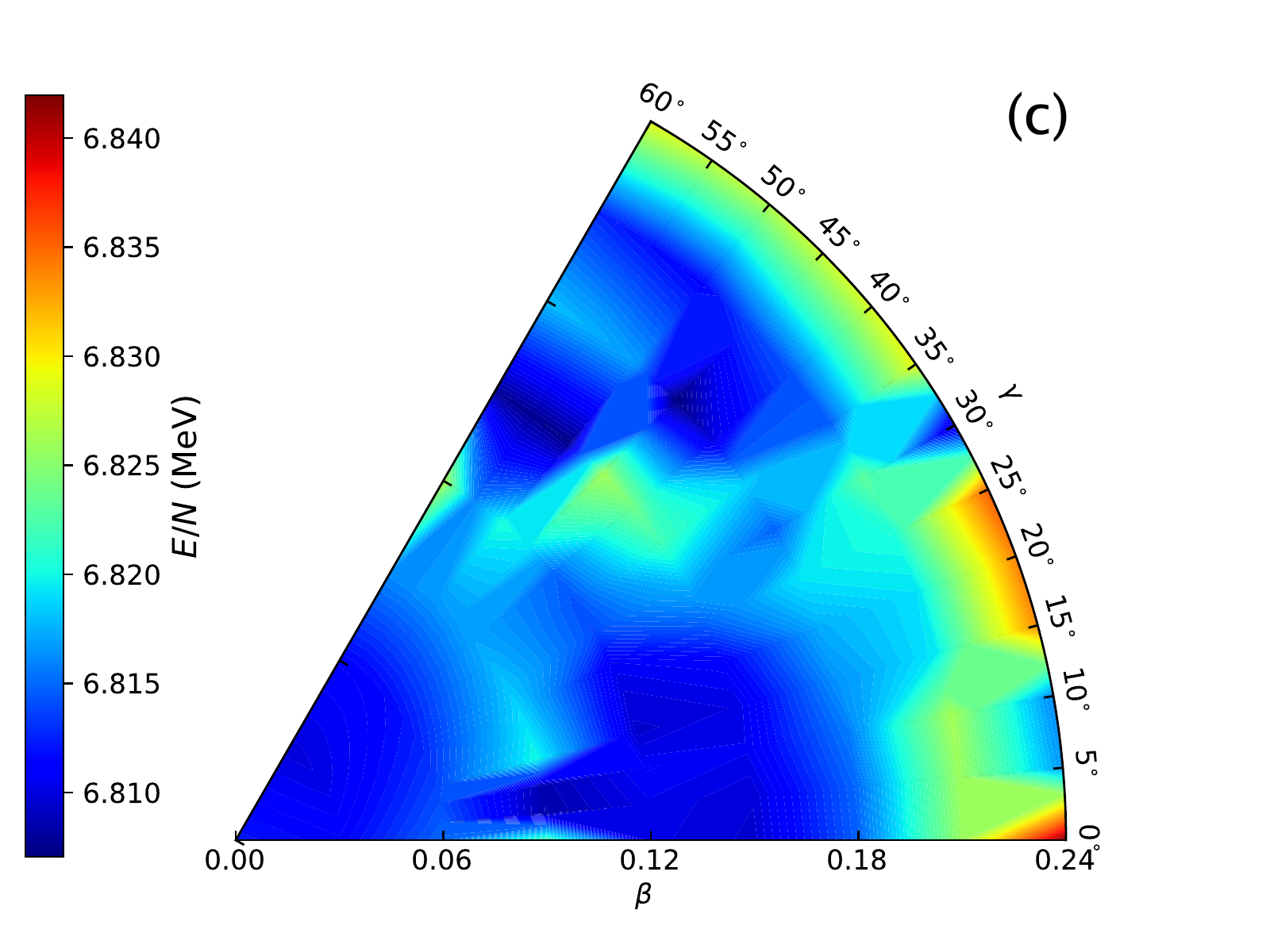}
\caption{Energy-deformation surfaces at an average density of $n_{\rm b}$ = 0.06 fm$^{-3}$ for cells containing $A_{\rm cell}$=614 nucleons and a proton fraction of 0.023 (a), 0.026 (b) and 0.029 (c). Many local minima (dark blue troughs) appear in the landscapes.} \label{fig:3}
\end{figure*}

\subsection{Numerical parameters and testing}

The simulations were conducted with the following parameter values. The imaginary time step iteration had a step size of $\lambda=5\times10^{-4}$ and a cutoff in the exponential at 5th order. The damped gradient iteration has a step size of $x_0=0.4$ and kinetic energy cutoff of 40 MeV. The quadrupole iteration has parameters $c_0$=0.03, $d_0$=0.9.

Convergence is achieved when the sum of the variance over all the wavefunctions $\sum_i w_{i,q} [  \bra{\phi_{i,q}} h_{\rm HF}^2 \ket{\phi_{i,q}} - (\bra{\phi_{i,q}} h_{\rm HF} \ket{\phi_{i,q}})^2]$ drops below 1 keV$^2$, which generally corresponds to convergence in the total energy of order 1 part in 10$^8$ or better.

The code was comprehensively tested and validated as outlined in \cite{Newton:2009eu}. It was established that the optimal grid spacing (the compromise between accuracy and computation time) is 1.2-1.3 fm, which we use here. We established that the periodic boundary conditions have a 1 part in 10$^4$ effect on the total energy of the cell, and that the nuclear shapes we obtain are not artifacts of the finite cell size (the same nuclear shapes are obtained when we double the cell size in in each direction).

\subsection{Nuclear Interaction Used}

The pressure, and hence stability, of the inner crust is provided by the fluid of dripped neutrons in which the lattice of nuclei and nuclear pasta is immersed. \emph{ab-initio} pure neutron matter (PNM) calculations are therefore an important guide for neutron star EOSs \cite{Hebeler:2013zl,Gandolfi:2014xy}.  In our investigation, it is important to choose a nuclear model that predicts a pure neutron matter EoS consistent with these calculations. We choose the NRAPR parameterization of the Skyrme interaction \citep{Steiner:2005aa}, which is fit to the APR neutron matter EOS. It gives a slope of the symmetry energy of $L=60$ MeV, which is of intermediate stiffness.  Figure~\ref{fig:1}a shows the NRAPR EOS and the band from recent \emph{ab-initio} calculations of the PNM matter EOS \cite{Gandolfi:2015nr,Tews:2016ty}. The extent of the pasta phases is sensitive to the EOS, particularly the symmetry energy parameters of nuclear matter; a follow-up study will examine the dependence of our results on the EOS. Preliminary results suggest the results are qualitatively similar.


\subsection{Choosing the cell size and proton fraction}

It is currently computationally prohibitive to conduct a full minimization over cell size. The presence of spurious shell effects make such a minimization unreliable, and in future a consistent minimization should be done using methods that minimize spurious shell effects, such as the use of twist-averaged boundary conditions \cite{Schuetrumpf:2016ve}. 

Instead, we choose to conduct the calculations at cell sizes and proton fractions guided by the compressible liquid-drop model (CLDM) \cite{Newton:2013sp}. We have checked the proton fractions corresponding to beta equilibrium for both the 3DHF model and CLDM, and they agree well. In Figure~2 we show the predictions of the proton fraction $y_{\rm p}$ and total nucleon number in the unit cell $A_{\rm cell}$ from the CLDM varying the surface energy of the CLDM over a wide range, together with total nucleon numbers $A_{\rm cell}$ and range of proton fractions we choose to perform calculations at in this work. The total nucleon number characterizes the cell size at a given baryon density $n_{\rm b}$, since the computational volume we use is calculated as $V=A/n_{\rm b}$. Our choice of $A_{\rm cell}$ as a function of density follows the rough trend predicted by the CLDM.

At each density, we perform calculations at two to three of values of $A_{\rm cell}$ to assess the $A_{\rm cell}$ dependence of our results. We find no qualitative dependence of our results on $A_{\rm cell}$, so will present our results for a single representative cell size. We also calculate a number of different proton fractions to make sure we can locate $\beta$-equilibrium within our quantum simulation: the variations with $y_{\rm p}$ will be explicitly presented. The range of proton fractions covered are indicated by the bars on the proton fraction points in Figure~2.

\section{Results}

To give a detailed example of our methods, we will first present an analysis of the layer of pasta at a baryon number density of around $n_{\rm b}$=0.06 fm$^{-3}$. We will then move to lower and higher densities to examine the layers where pasta first emerges and finally transitions to uniform matter, before summarizing results across the whole pasta region. 

\subsection{A case study: $n_{\rm b}\approx$ 0.06 fm$^{-3}$.}

We begin our investigation by calculating the minimum energy configurations at a constant cell size corresponding to a total nucleon number of $A_{\rm cell}$=614, which is in the range predicted by the CLDM calculations. We perform calculations over a range of deformation parameters: the full range of $\gamma$ from prolate configurations $\gamma=0\degree$ to oblate configurations $\gamma=60\degree$ with a step of $\Delta \gamma=5\degree$, and for a range of magnitudes of deformation $\beta$ that cover all unique local minima, which at this density ranges up to $\beta=0.24$. We use a step size of $\Delta \beta=0.03$. We perform these calculations at a range of different values of the proton fraction in order to locate the beta-equilibrium value. The results are shown for the three values of $y_{\rm p}$ around beta-equilibrium: $y_{\rm p}$=0.023, 0.026 and 0.029 (corresponding to proton numbers of $Z$=14,16 and 18).

\begin{figure*}[!t]
\centerline{\includegraphics[scale=0.35]{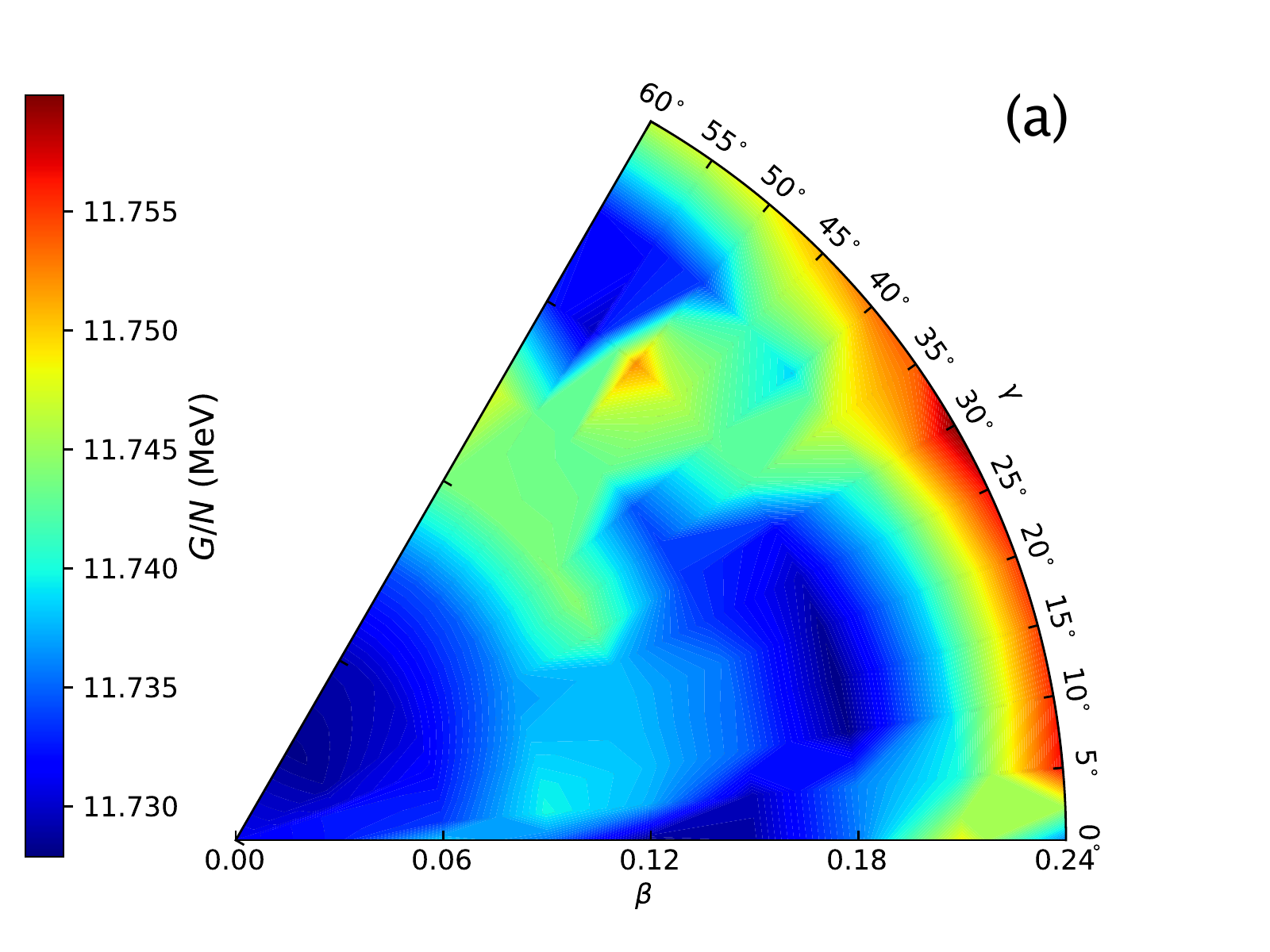}\includegraphics[scale=0.35]{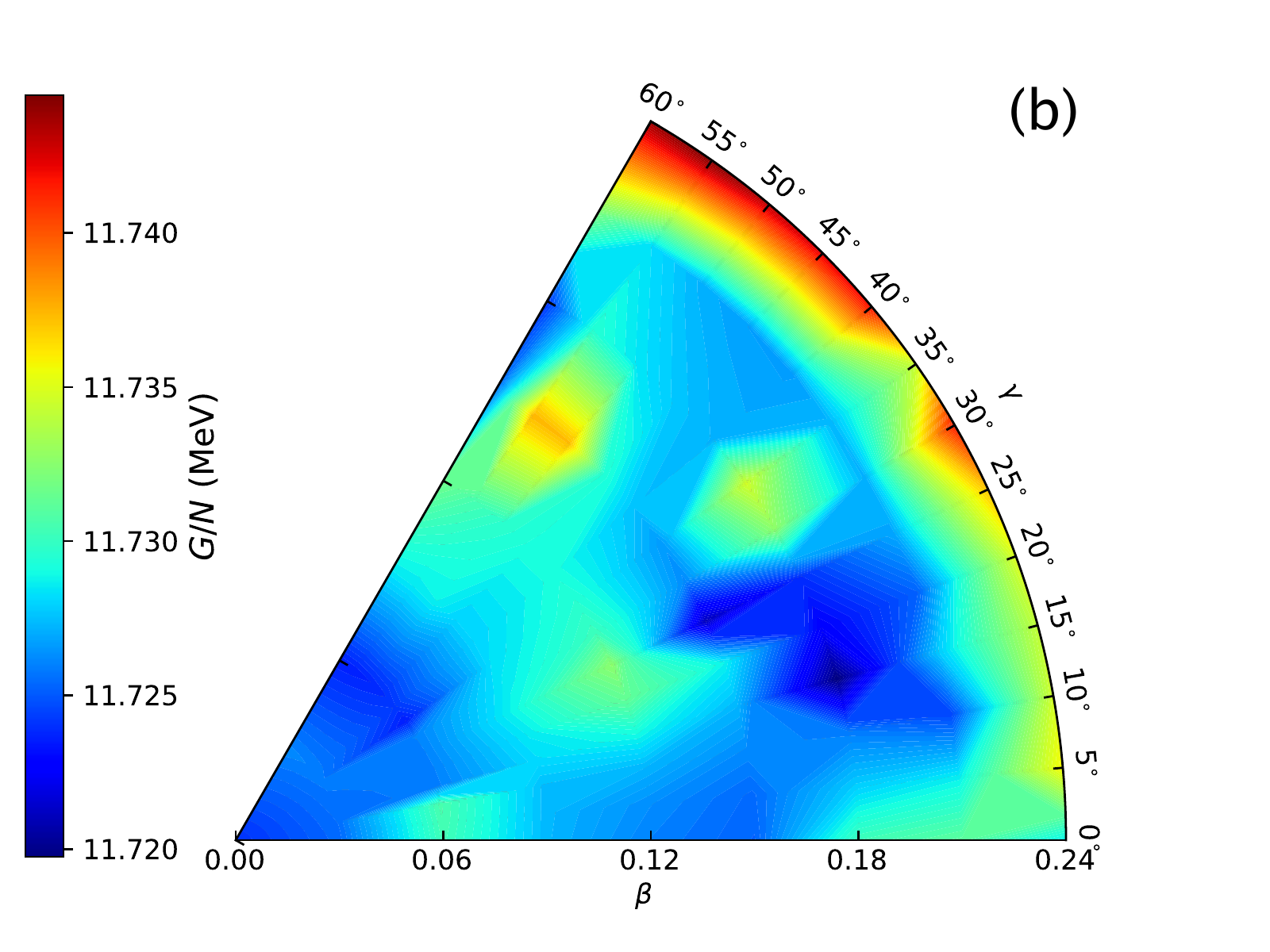}\includegraphics[scale=0.35]{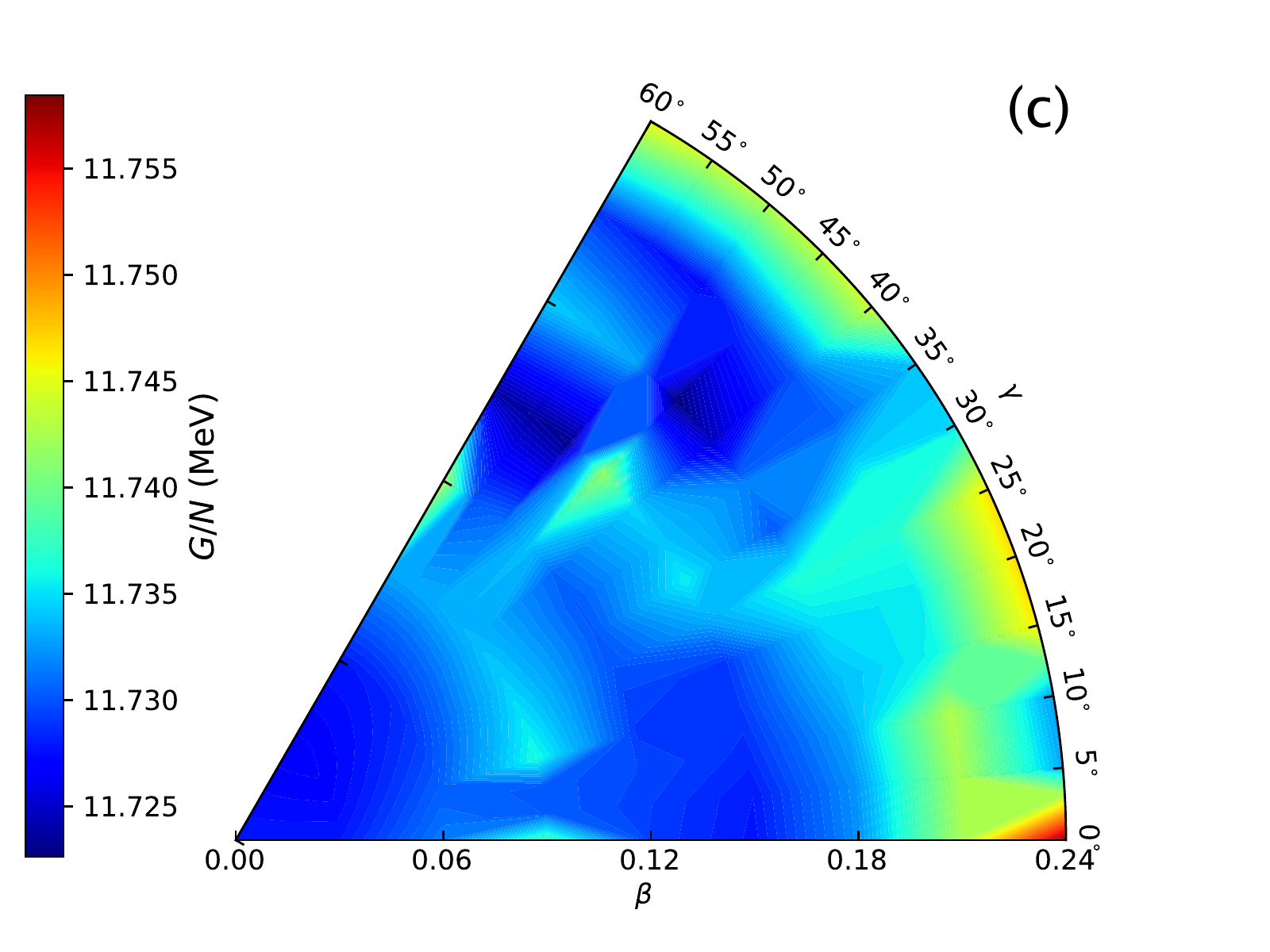}}
\centerline{\includegraphics[scale=0.35]{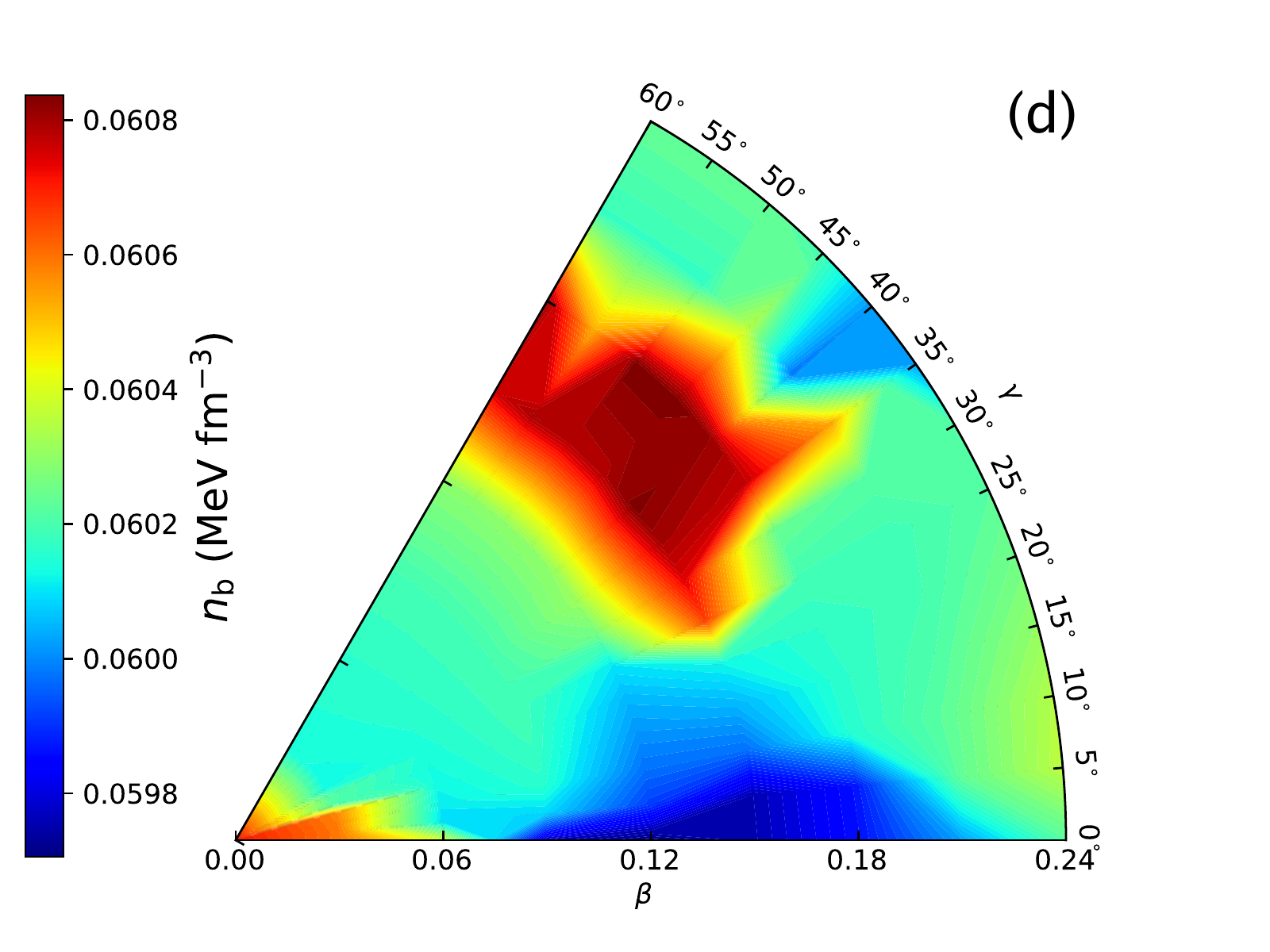}\includegraphics[scale=0.35]{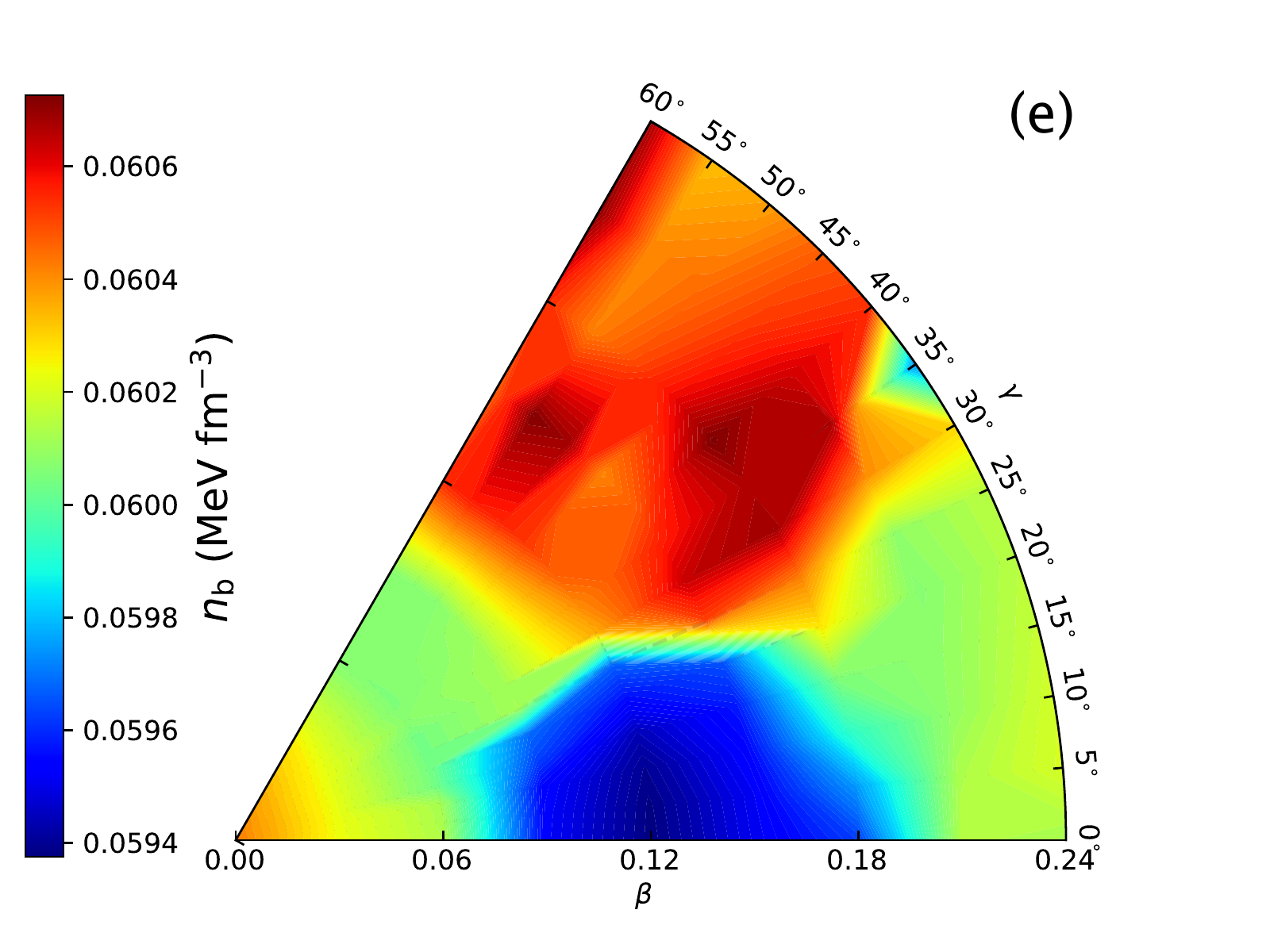}\includegraphics[scale=0.35]{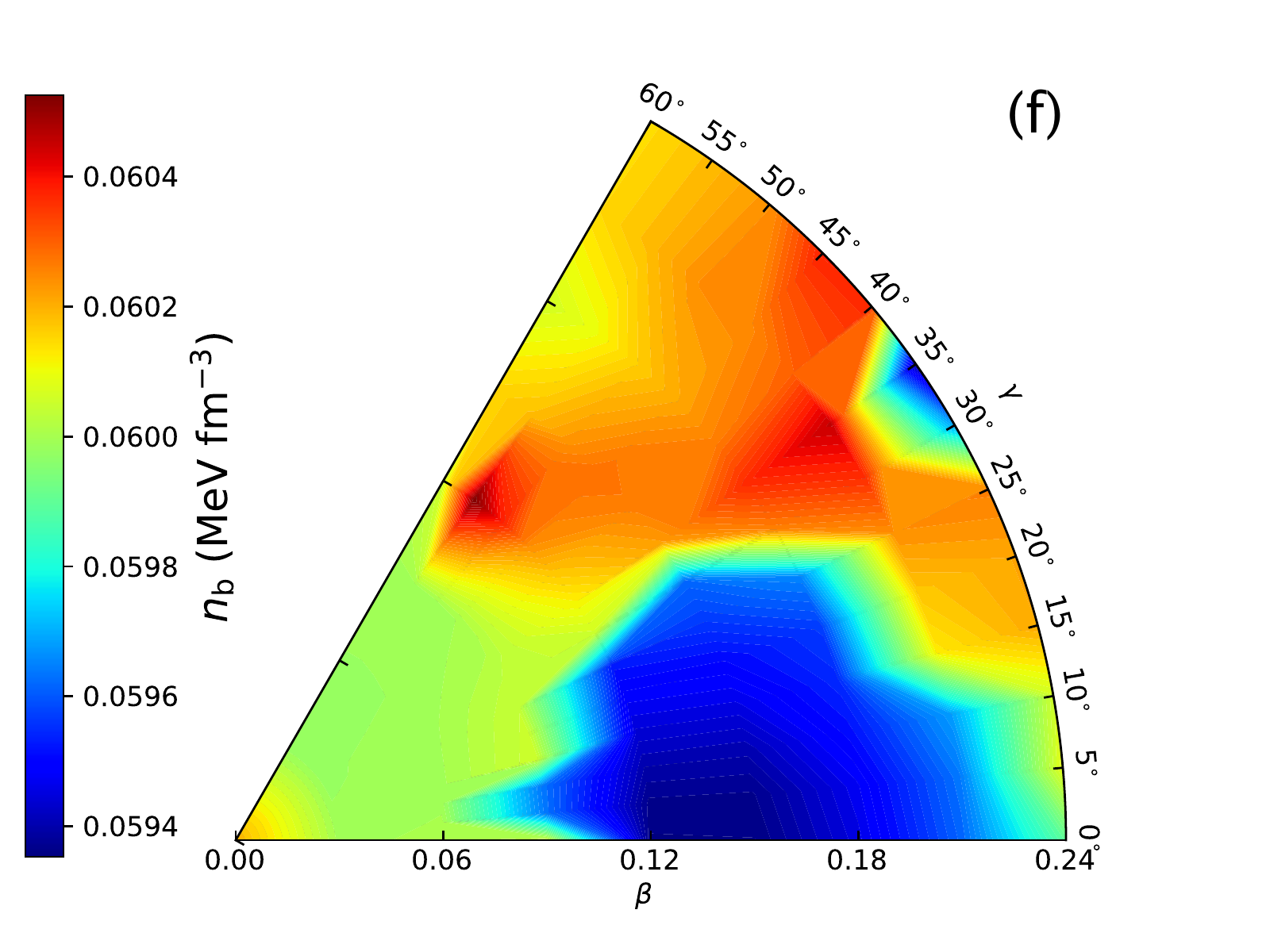}}
\caption{Gibbs energy-deformation surfaces (a-c) and average baryon density surfaces (d-f) at a constant pressure of $0.291$ MeV fm$^{-3}$ for cells containing $A_{\rm cell}$=614 nucleons. Results are shown for proton fraction of 0.023 (a,d), 0.026 (b,e) and 0.029 (c,f). The local minima differ in baryon density by of order 5\%.} \label{fig:4}
\end{figure*}

In Figure~3 we show the resulting energy surfaces as a function of deformation $(\beta,\gamma)$. The energy per baryon is plotted versus the magnitude $\beta$ and direction $\gamma$ of the deformation. The most important feature is that there are \emph{multiple} local minima in all three energy surfaces, visible as the darker blue regions. There are minima located in broadly the same regions of deformation space for each proton fraction: at small deformations in the intermediate to oblate sense ($\beta=0.03-0.06, \gamma\approx30-60\degree$), at stronger prolate deformations ($\beta=0.12-0.18, \gamma=0-25\degree$) and high deformations in the oblate direction ($\beta=0.15-0.21, \gamma=50-60\degree$). The lowest energies occur for the proton fraction of $y_{\rm p}=0.26$. However, the energy separation of local minima is small and there is no obviously pronounced ground state. This initial calculation confirms a number of other microscopic studies that have shown that matter deep in the crust is frustrated, and might have an amorphous, heterogeneous structure characterized by many local energy minima \cite{Magierski:2002fk}. 


It is worth thinking about how matter with this structure will behave as the crust of the neutron star cools. Different local regions of a given layer inside the star will fall into different local minima, and when the temperature falls below some critical value associated with the energy barriers between minima they will be trapped in those minima for some quantum tunneling timescale (which is likely to be related to the very uncertain pycnonuclear fusion timescale of heavy nuclei just above the pasta layers). On longer timescales, the crust might be able to anneal and eventually all the pasta at a given density could be converted into the ground state configuration. Depending on the temperature scales set by the energy barriers between local minima, crustal heating later in the neutron star's life might repopulate the local minima. Thus the pasta layers might plausibly transition between a single pasta configuration and multiple coexisting pasta configurations at different stages in the star's life.

Here we explore the possibility that different phases corresponding to the local minima coexist in microscopic domains, at a given depth in the crust. Such domains would exist in equilibrium at constant pressure; however, the calculations we perform are at constant density rather than constant pressure. In order to determine the local minima that will coexist, we need to calculate the Gibbs free energy at constant pressure as a function of $\beta$ and $\gamma$.

\begin{figure*}[!t]
\centerline{\includegraphics[scale=0.35]{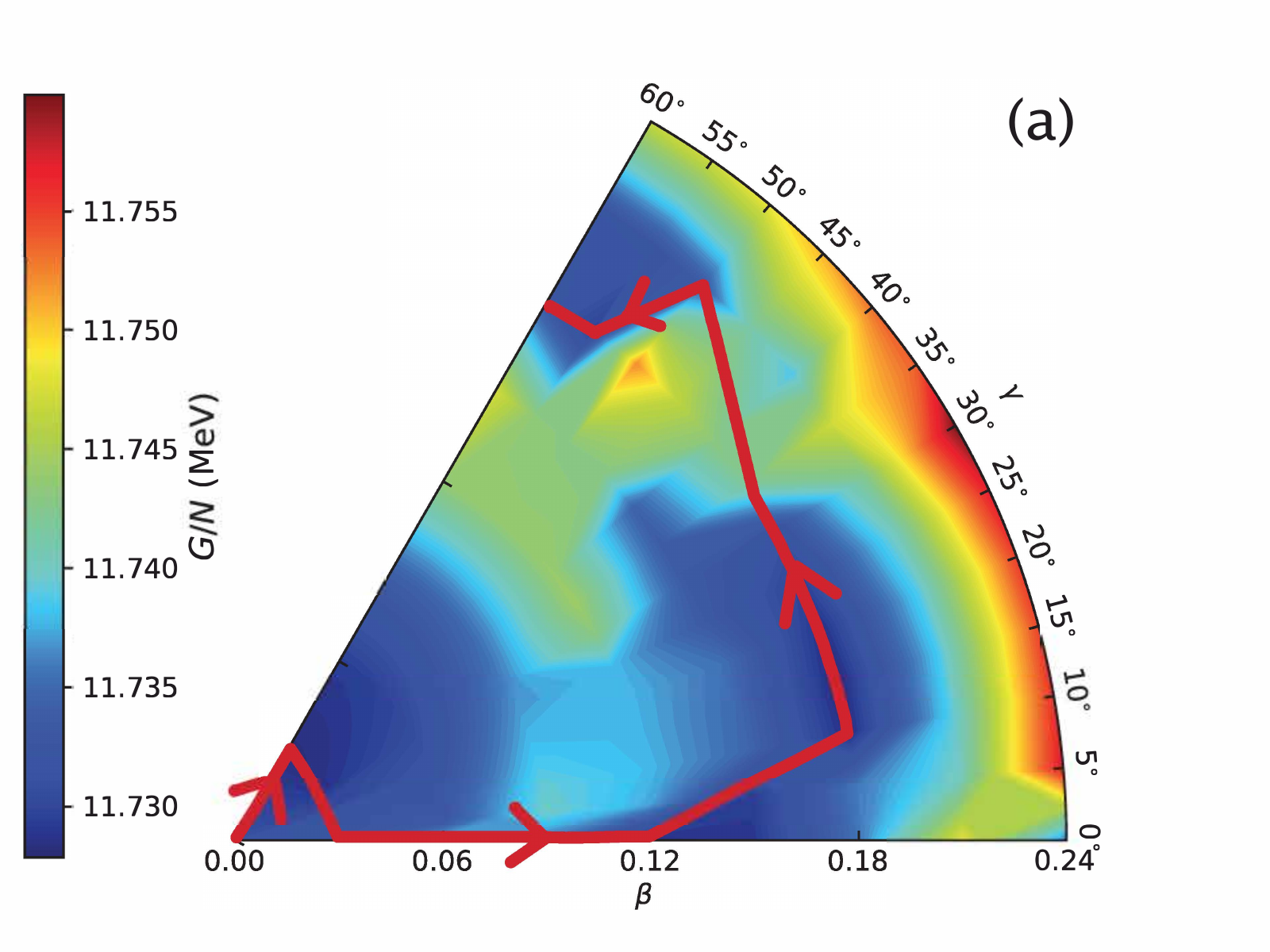}\includegraphics[scale=0.35]{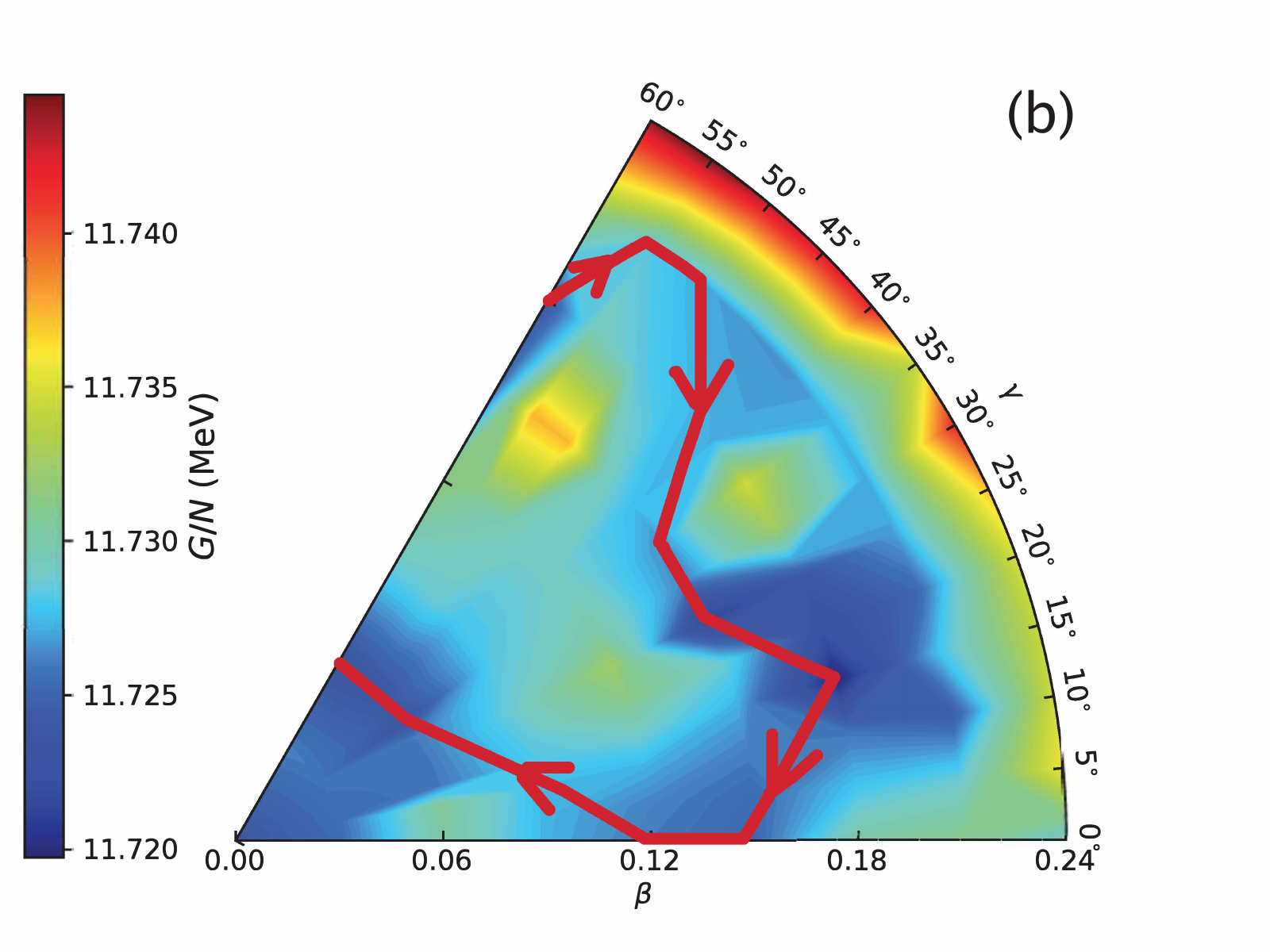}\includegraphics[scale=0.35]{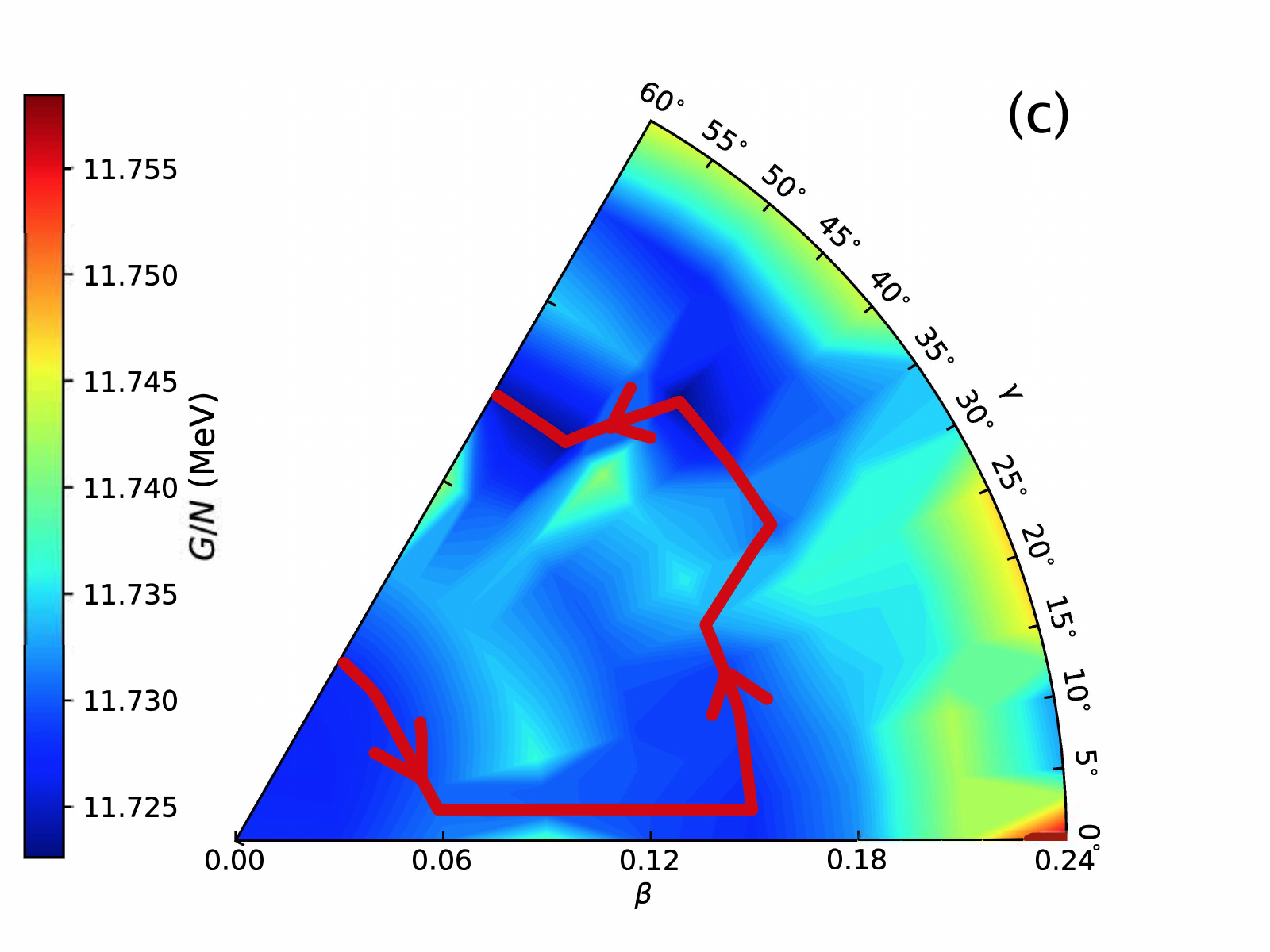}}
\centerline{\includegraphics[scale=0.65]{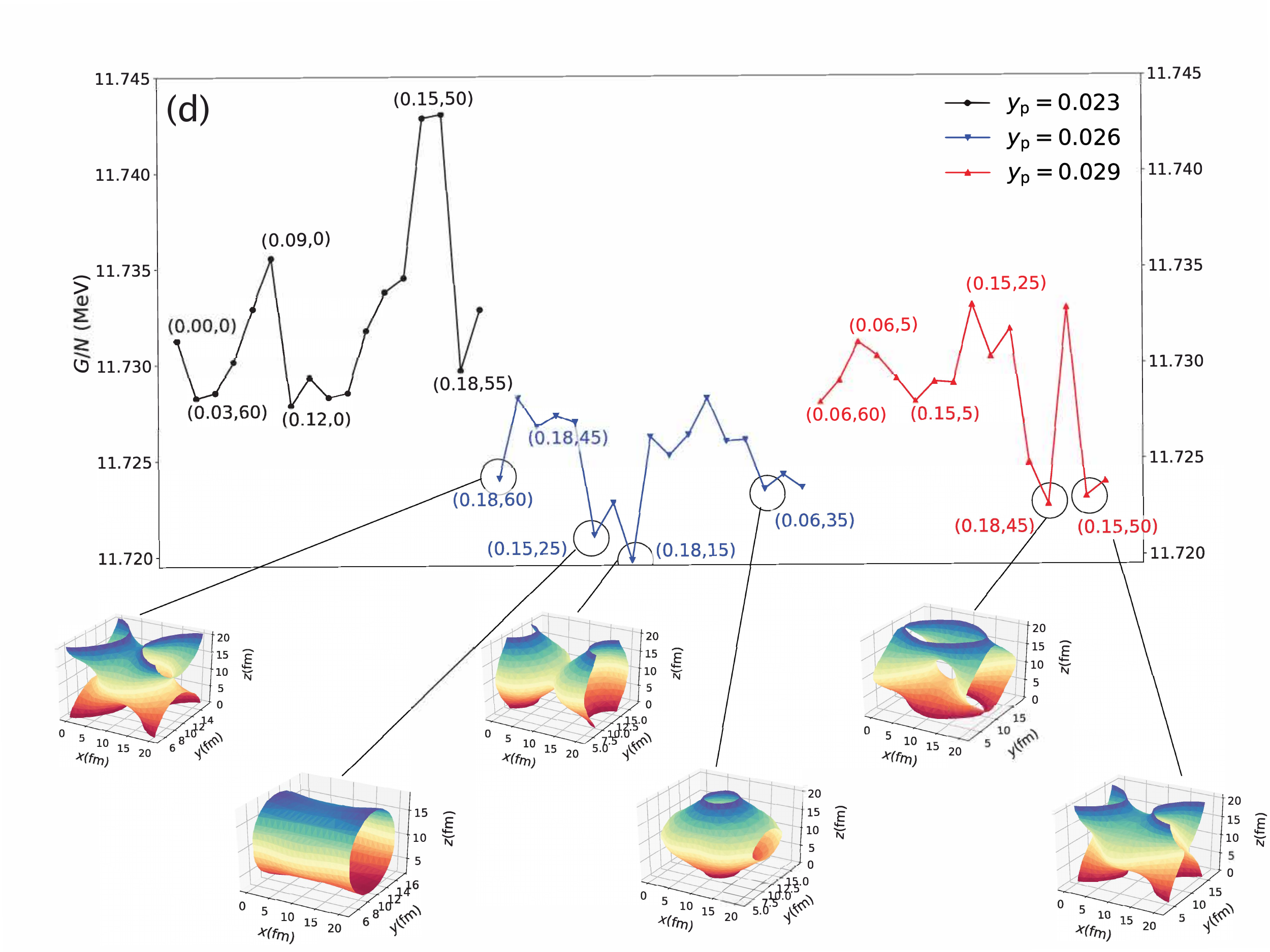}}
\caption{The top row a-c shows the Gibbs free energy surfaces from Figure~\ref{fig:4} at a constant pressure of $0.291$ MeV fm$^{-3}$ (densities around 0.06 fm$^{-3}$) with red arrows indicating paths between local minima that pass over the smallest energy barriers. Underneath is the corresponding one-dimensional plot of the Gibbs energy along those paths (d). Selected $(\beta,\gamma)$ coordinates are shown along the one-dimensional plots. 3D surfaces at constant neutron density in the cell show the shape of the nuclear cluster in the cell at the local minima indicated. The other apparent local minima can transition exothermically to one of the highlighted minima by beta decay or electron capture.} \label{fig:5}
\end{figure*}

In order to do this, we perform calculations of the energy surfaces over a range of densities in the range $n_{\rm b}$ = 0.058fm$^{-3}$ - 0.062fm$^{-3}$. We pick a reference pressure, which we choose to be that of the zero-deformation configuration $(\beta,\gamma)=(0,0\degree)$, at $n_{\rm b}$ = 0.06fm$^{-3}$: $P_{\rm ref} = 0.291$ MeV fm$^{-3}$. We then use interpolation to find the density at which the pressure is equal to the reference pressure for all other deformation values, and the energy at that density. We then calculate the specific Gibbs free energy $G = E + P/n_{\rm b}$. The interpolations assume there is no discontinuous change in energy with density at a particular point in deformation space, which would be associated with a change of shape. This can't be guaranteed, but so long as the interpolation window is small and we use a sufficiently large number of points in deformation space, we can reasonably assume it will occur infrequently enough that it will not affect the global features of the energy surface. Such discontinuities would show up as artifacts in the energy surfaces, and we see no such features.

\begin{figure*}[!t]
\centerline{\includegraphics[scale=0.65]{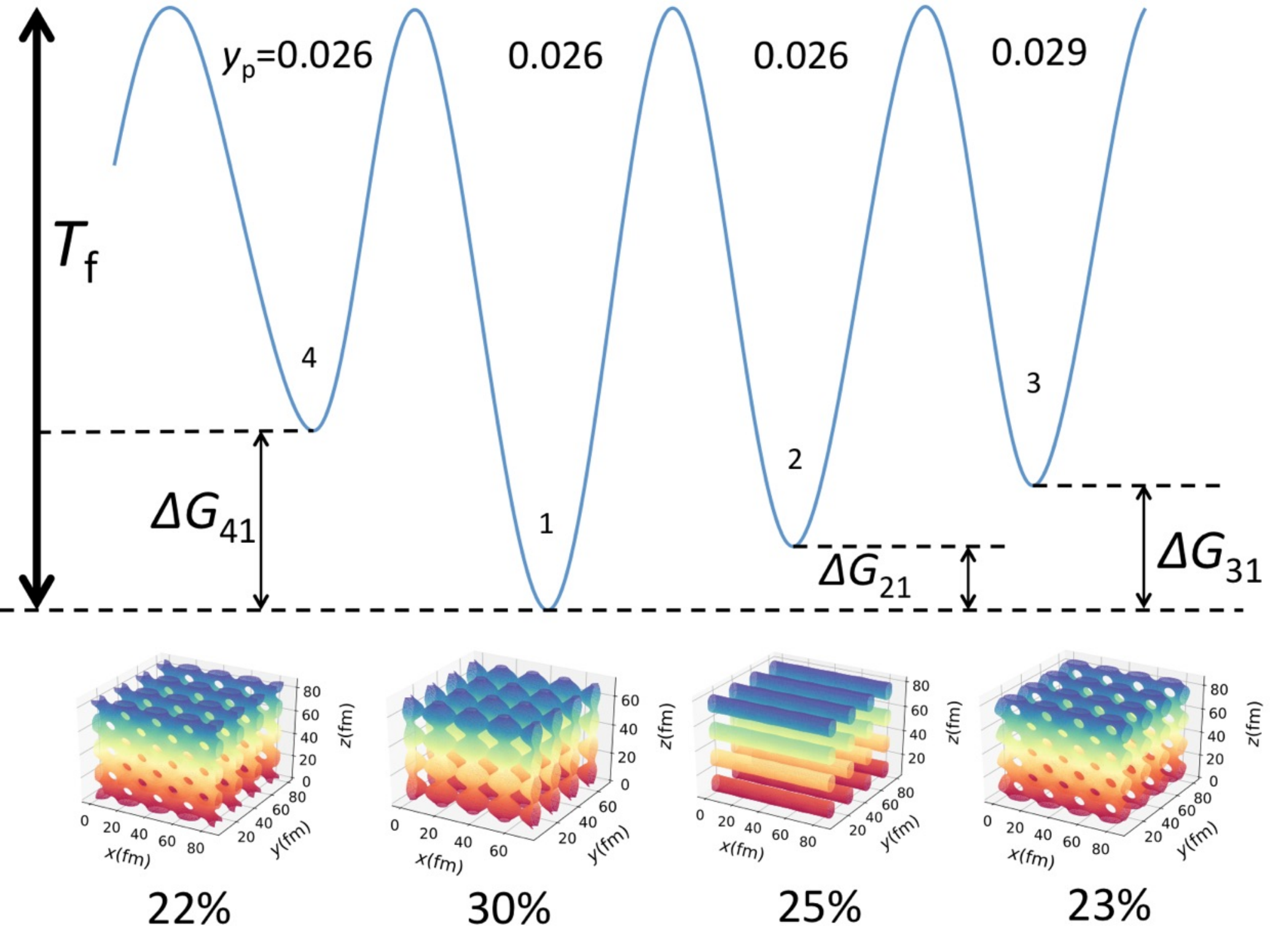}}
\caption{Illustration of our schematic model for pasta domains in local Gibbs energy minima at densities around 0.06fm$^{-3}$ where calculations present four physically distinct local minima. In order to visualize the phases better we have plotted a region of four by four unit cells, but it is important to remember we only simulate one single unit cell. The proton fractions corresponding to the minima we found are shown above each minima.The fictive temperature $T_{\rm f}$ is given by the height of the energy barriers, and the energy differences between he local minima are indicated as $\Delta G$ The relative abundances of the phases at a temperature equal to the fictive temperature is shown below the visualizations of the phases. In the simplified model we take a single barrier height to be the fictive temperature. This is determined as the average of the various barriers in our calculation. These abundances will be frozen in as the temperature drops further unless processes such as quantum tunneling anneal the matter.} \label{fig:6}
\end{figure*}

In Figure~4a-c we show the Gibbs free energy surfaces at constant pressure in deformation space at the three values of the proton fraction, and in Figure~4d-f we show the corresponding density variations over each surface. In all three cases, local minima appear at broadly the same locations as we found in the internal energy surface at constant density $n_{\rm b}$. The range of average baryon density is smaller than the range over which we performed the interpolations, indicating the interpolations are robust. Domains in different regions of deformation space will have different average baryon densities, so if they coexist at a given density in the crust, there will be fluctuations in density of order $10^{-3}$ fm$^{-3}$, or $\Delta n_{\rm b}/n_{\rm b} \sim$ 1\% of the total density. These fluctuations will have length scales corresponding to the size of the domains, which we will estimate shortly. The three regions where minima occur are large oblate deformations (higher density), small intermediate deformations (intermediate density) and intermediate prolate deformations (lower density).

Local minima in the Gibbs energy surfaces are separated by energy barriers. In order to more clearly see the relative height of the barriers and the difference in energy between local minima, we now plot the energy along one-dimensional paths across the deformation landscapes. We choose plots that link local minima by continuous deformation and require traversing the smallest energy barriers possible. The top of Fig.~5 shows the Gibbs energy surfaces from Figure~4 with the paths marked. The central plot on Fig.~5 shows the Gibbs energy along those paths. In order to orient the reader, the directions marked by the arrows on the surface plots show the direction moving left to right in the main plot of the one-dimensional trajectories. We also show selected values of the deformation coordinates $(\beta,\gamma)$ along the one dimensional trajectories.

The plot of the energy along the trajectories shows that indeed the Gibbs energy is on average lowest for the proton fraction of $y_{\rm p}$=0.026. However, two minima at a proton fraction $y_{\rm p}$=0.029 have smaller Gibbs energies than some local minima at $y_{\rm p}$=0.026. There may be a local variation of average proton fraction (and hence electron density) from domain to domain.

We have circled the 6 lowest local minima. All other local minima can access these six local minima by continuous deformation and adjusting their proton fraction through beta decay and electron capture without having to pass over an energy barrier. For example, matter at the local minimum at $y_{\rm p}$=0.29, $(\beta,\gamma)\approx(0.06,60\degree)$ can reach the minimum at $y_{\rm p}$=0.26, $(\beta,\gamma)\approx(0.06,60\degree)$ through electron capture.

The pasta configurations corresponding to each of the 6 minima are revealed by plotting surfaces of constant neutron density in our unit cell. These are shown in the six plots under the one dimensional energy plot. The neutron density at which to plot these surfaces is chosen to be the average neutron density $n_{\rm b} (1-y_{\rm p}$). The following pasta configurations are found:

\begin{itemize}
\item $y_{\rm p}$ = 0.026, $(\beta,\gamma)=(0.18,60\degree)$ - the nuclear waffle phase (the ``hole'' of the waffle is centered on the edge of each cell).

\item$y_{\rm p}$ = 0.026, $(\beta,\gamma)=(0.15,25\degree)$ - the nuclear spaghetti phase.

\item$y_{\rm p}$ = 0.026, $(\beta,\gamma)=(0.18,15\degree)$ - another form of nuclear waffle; large spherical nuclei with bridges connecting to adjacent cells in two different directions. The nucleus is centered on the y-boundary of the computational volume.

\item$y_{\rm p}$ = 0.026, $(\beta,\gamma)=(0.06,15\degree)$ - Similar to the previous configuration, a deformed nucleus with bridges to adjacent cells. The nucleus is now centered in the computational volume.

\item$y_{\rm p}$ = 0.029, $(\beta,\gamma)=(0.18,45\degree)$ - the nuclear waffle phase.

\item$y_{\rm p}$ = 0.029, $(\beta,\gamma)=(0.18,45\degree)$ - nuclear waffle phase (like the waffle configuration at $y_{\rm p}$ = 0.026, the ``hole'' of the waffle is centered on the edge of each cell).
\end{itemize}

The detailed structure of the minima is dependent on the nuclear interaction used and the total cell size. However, a precise extraction of the relative energy differences of minima and the barrier heights is not merited, as it is extremely unlikely that those details will ever be accessible through observational data. Hence, instead of using the exact values of the energies of all six minima and the barriers between them, we create a slightly simpler model based on these results. The model is depicted schematically in Fig.~\ref{fig:6}. We first reduce the number of minima from six to four, representing the four distinct configurations present. The two forms of nuclear waffle are distinct enough that we include them as separate configurations, though they may not give rise to any observable differences). The $y_{\rm p}$ = 0.026, $(\beta,\gamma)=(0.06,15\degree)$ and $(0.18,15\degree)$ minima are essentially the same, so we take the lowest of those two minima and ignore the other one (which, being the highest lying of the minima, will be least populated). Also, we treat the two waffle phases at $y_{\rm p}$ = 0.029 as a single minimum. Although these are similar to the waffle phase at $y_{\rm p}$ = 0.026, the different proton fractions makes them physically distinct. Finally, we make the simplification that there is just one single characteristic barrier height between phases, $G_{\rm barrier}$, taken to be the average of all the barriers between minima.

The four minima are represented schematically in Fig.~6. Under each minima we plot the neutron surfaces again for stacks of 4x4 unit cells of matter, in order to better see the structure of matter at larger scales (but note that the computation is done only in one unit cell). We have four distinct phases: the ground state is the waffle phase consisting of large nuclei connected to adjacent cells in two directions. The next lowest lying minimum is the spaghetti phase. Then the next two are nuclear waffle phases - intermediate phases between spaghetti and lasagna - at two different proton fractions.

In order to characterize the properties of these possibly amorphous phases of matter, we borrow a concept from the study of terrestrial amorphous materials. An effective or ``fictive'' temperature, $T_{\rm f}$ \cite{Mauro:2009} is defined as the temperature from which, if the material was instantaneously quenched to zero temperature, or any other temperature $T<T_{\rm f}$, its state would be that of the material at a temperature $T_{\rm f}$ (no reconfiguration of its microscopic degrees of freedom would occur). In our context, the fictive temperature is the temperature equivalent to the energy barrier height between minima. As the temperature falls below $T_{\rm f}$, thermal fluctuations can no longer rearrange matter, and so \emph{in the absence of quantum tunneling between the barriers} the matter is frozen into the state at $T_{\rm f}$. Of course, quantum tunneling will occur, but the timescales over which that would occur and rearrange potentially large regions of nuclear pasta are not well known (the closest timescales from the literature would be pycno-nuclear fusion timescales, but as we shall see whole domains of nuclei need to be rearranged for matter to substantially change its structure at the mesoscopic level.

In this work, as illustration of the concept and derive some order-of-magnitude implications, we choose the fictive temperature to be the height of the highest energy barrier relative to the lowest lying minimum, indicated as $T_{\rm f}$ in Figure~\ref{fig:6}. 

The ratio of the abundance of the pasta in the $i$th minimum to the $j$th minimum is given by

\be
{N_{\rm i} (T_{\rm f}) \over N_{\rm j}(T_{\rm f})} = e^{\Delta G_{\rm ij}/kT_{\rm f}},
\ee

\noindent where $\Delta G_{\rm ij} = G_{\rm i}-G_{\rm j}$.
Given that the occupation probability for minimum $i$ is given by $p_i = N_i/\sum_j N_j$ and $\sum_i p_i =1$, then the occupation probability of a particular minimum $j$ can be written

\be \label{Eq:18}
p_{\rm i}(T_{\rm f}) = {e^{\Delta G_{\rm i0}/kT_{\rm f}} \over 1 + \sum_{j \neq 0} e^{\Delta G_{\rm j0}/kT_{\rm f}} }.
\ee

\noindent Based on~\ref{Eq:18}, the abundances of pasta in each of the four phases in our simplified model in Figure~6 are shown as the percentages under the visualizations of the pasta.

Let us assume, to obtain a lower limit on the fictive temperature below which the pasta becomes frozen-in to the local minima, that the nucleons behave as free quasiparticles, Then differences in the energy \emph{per particle} between local minima and the energy barriers that separate them set the temperature scale required to transition from one pasta structure to another through thermal fluctuations. Realistically, this temperature scale will be modified upwards by a factor taking into account the extent to which the nucleons in the pasta and free neutron gas behave collectively during the rearrangement from one phase to another. A more accurate picture might suppose that a number of nucleons in the unit cell of order the proton number act collectively in the transition between phases, since the main driver of the shape formation is the electrostatic lattice energy. This number may also be modified by the number of unbound neutrons entrained by the cluster \cite{Chamel:2005vd}. Typically, this accounts for $\sim 10\%$ of the nucleons and so our lower limit could underestimate the fictive temperature by a factor of $\sim$10. To account for this factor, we can multiply the temperature by a factor $A_{\rm collective}$ which accounts for the number of nucleons in the unit cell that act collectively, and then $k T_{\rm f}$ = $A_{\rm collective} G_{\rm barrier}$. Note that the number of nucleons per unit cell that behave collectively upon shape rearrangement does not affect the equilibrium distribution, since both the energy difference between minima and the fictive temperature scale by $A_{\rm collective}$.

For pasta geometries continuous in one or more dimensions like spaghetti and lasagna, the unit cell does not have a physical meaning in that direction, but, for example, a transition between spaghetti and waffle phases requires the creation of connecting arms perpendicular to the spaghetti axis which are periodic according the unit cell size. Therefore the unit cell is still the relevant unit of matter when we think about rearranging nuclear pasta. 

In our simplified model, the fictive temperature is taken to be the average height of the barriers relative to the lowest minimum along the one-dimensional trajectories at $y_{\rm p}=0.026$ and $y_{\rm p}=0.029$ in Fig.~5: $k T_{\rm f}$ = 7.7 keV $\to$ $T_{\rm f} = 8.9\times 10^7$ K ($T_8$=0.89 where $T_8=T_{\rm f}/10^8$ K). Minima 1 and 2 are separated by energies $\Delta G_{21}=$ 1.6 keV/particle, minima 1 and 3 by $\Delta G_{31}=$ 2.0 keV/particle and minima 1 and 4 by $\Delta G_{31}=$ 2.4 keV/particle. 

At temperatures below $8.9\times 10^7$ K the composition of the domains will be frozen with respect to thermal fluctuations. The relative abundance of the pasta phases corresponding to the four minima is 0.30:0.25:0.23:0.22 respectively (see Figure~6). Thus about 70\% of the composition of this layer is at a proton fraction 0.026 and 30\% is at a proton fraction 0.029. We can thus expect fluctuations in average proton (and, correspondingly, electron) fraction at the microscopic level at the level of around 10\%.  Also around 75\% of matter is in waffle-like configurations, and 25\% in the spaghetti configuration.

\begin{figure*}[!t]
\centerline{\includegraphics[scale=0.35]{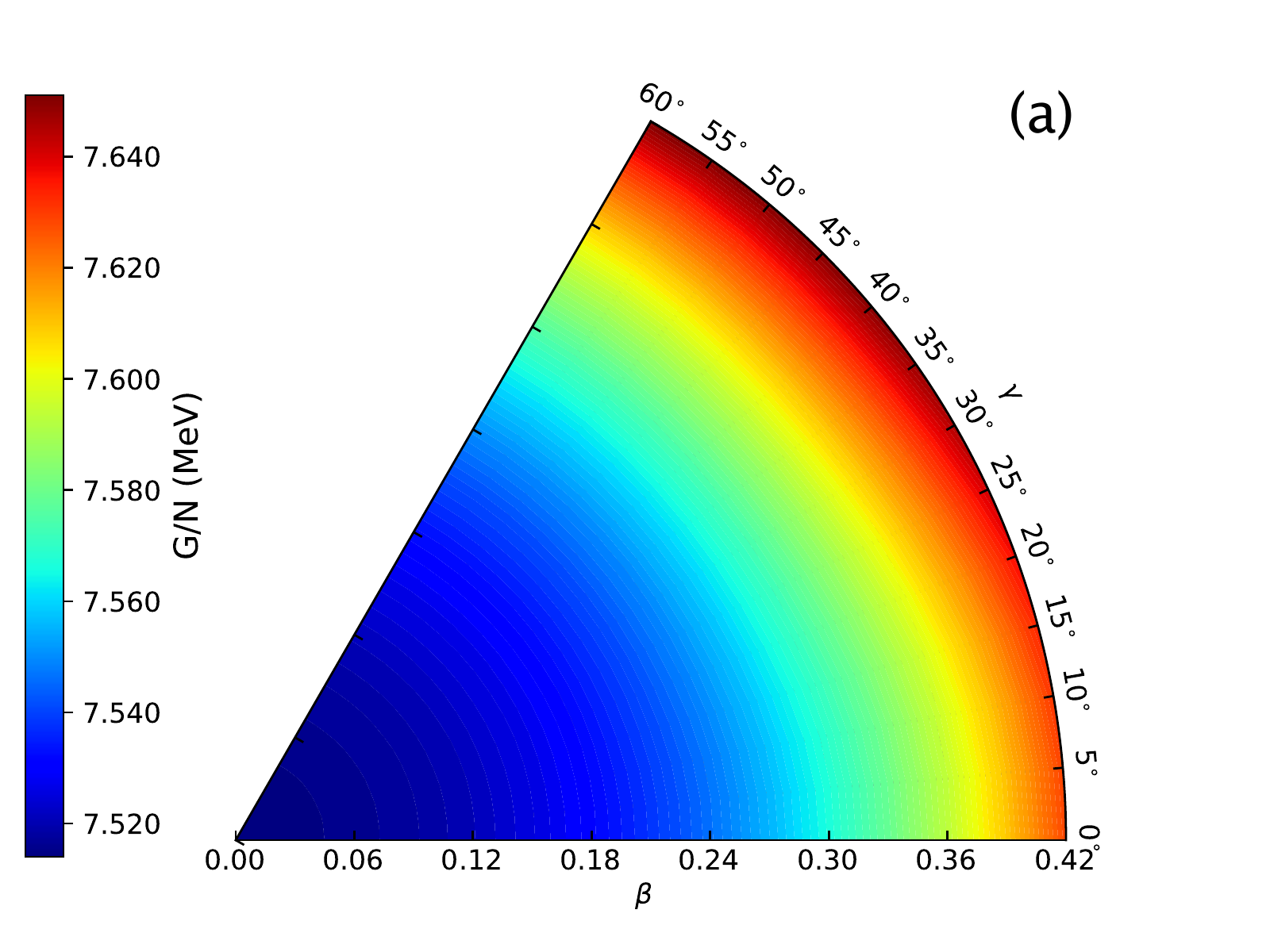}\includegraphics[scale=0.35]{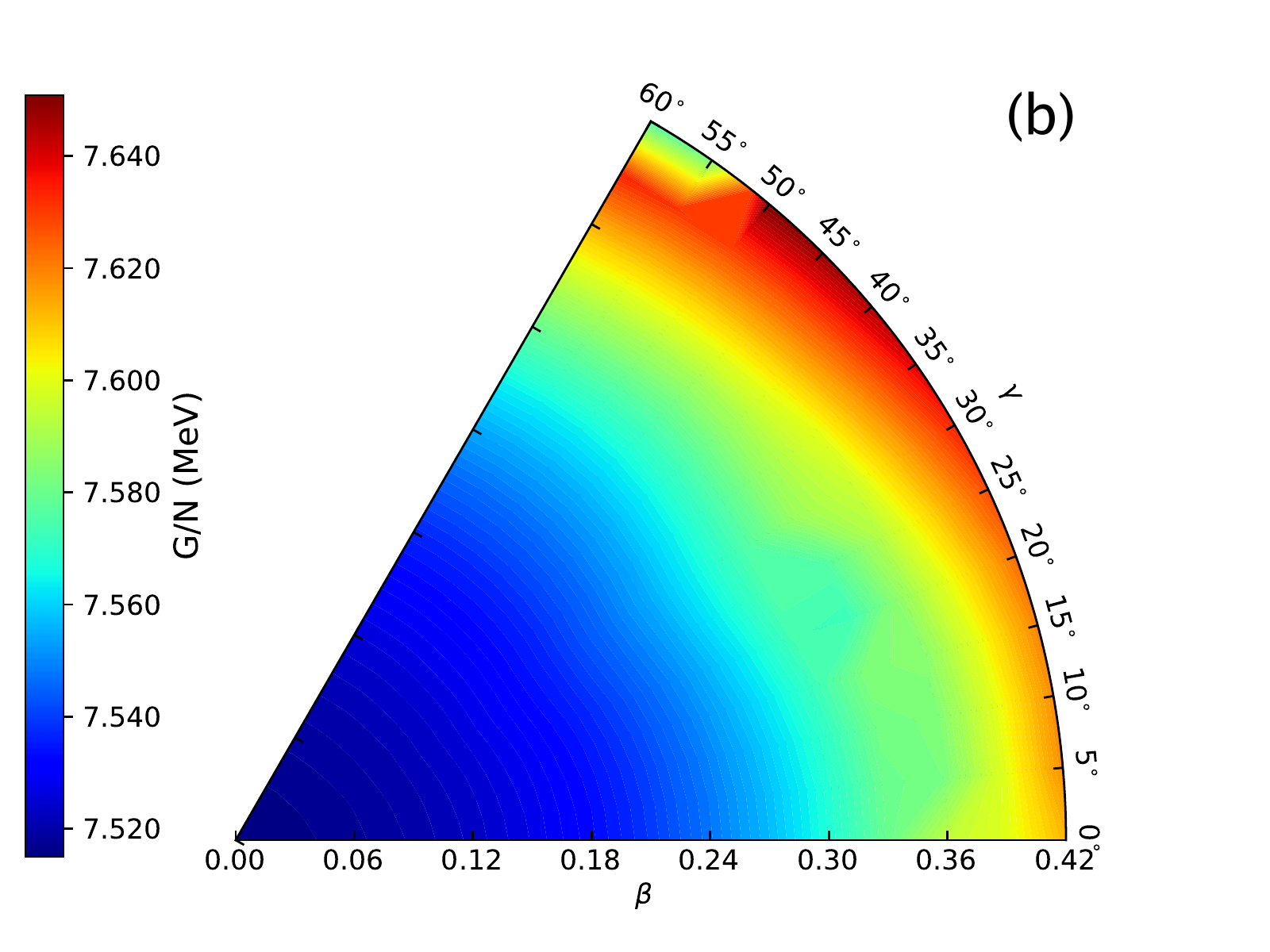}\includegraphics[scale=0.35]{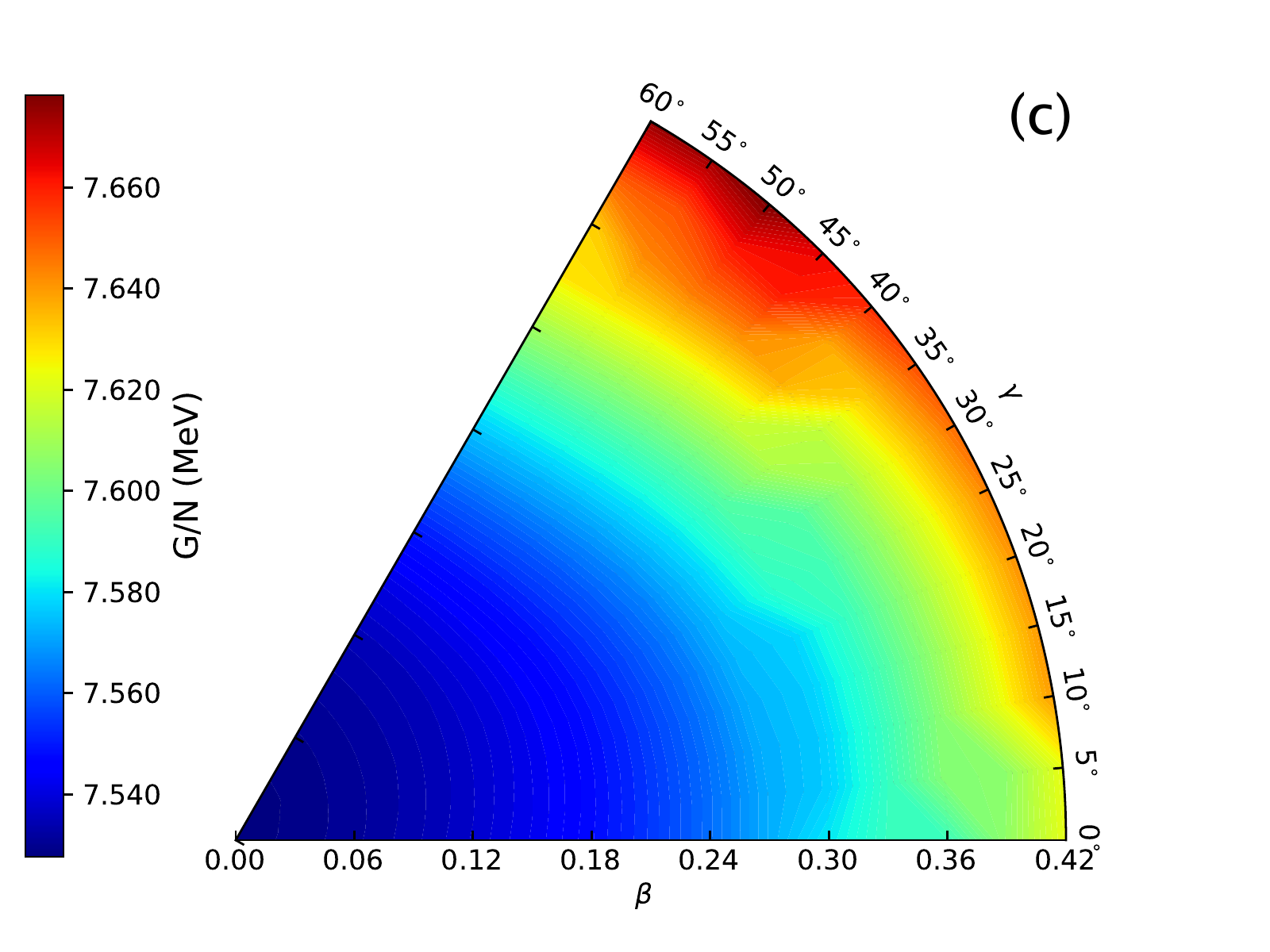}}
\centerline{\includegraphics[scale=0.65]{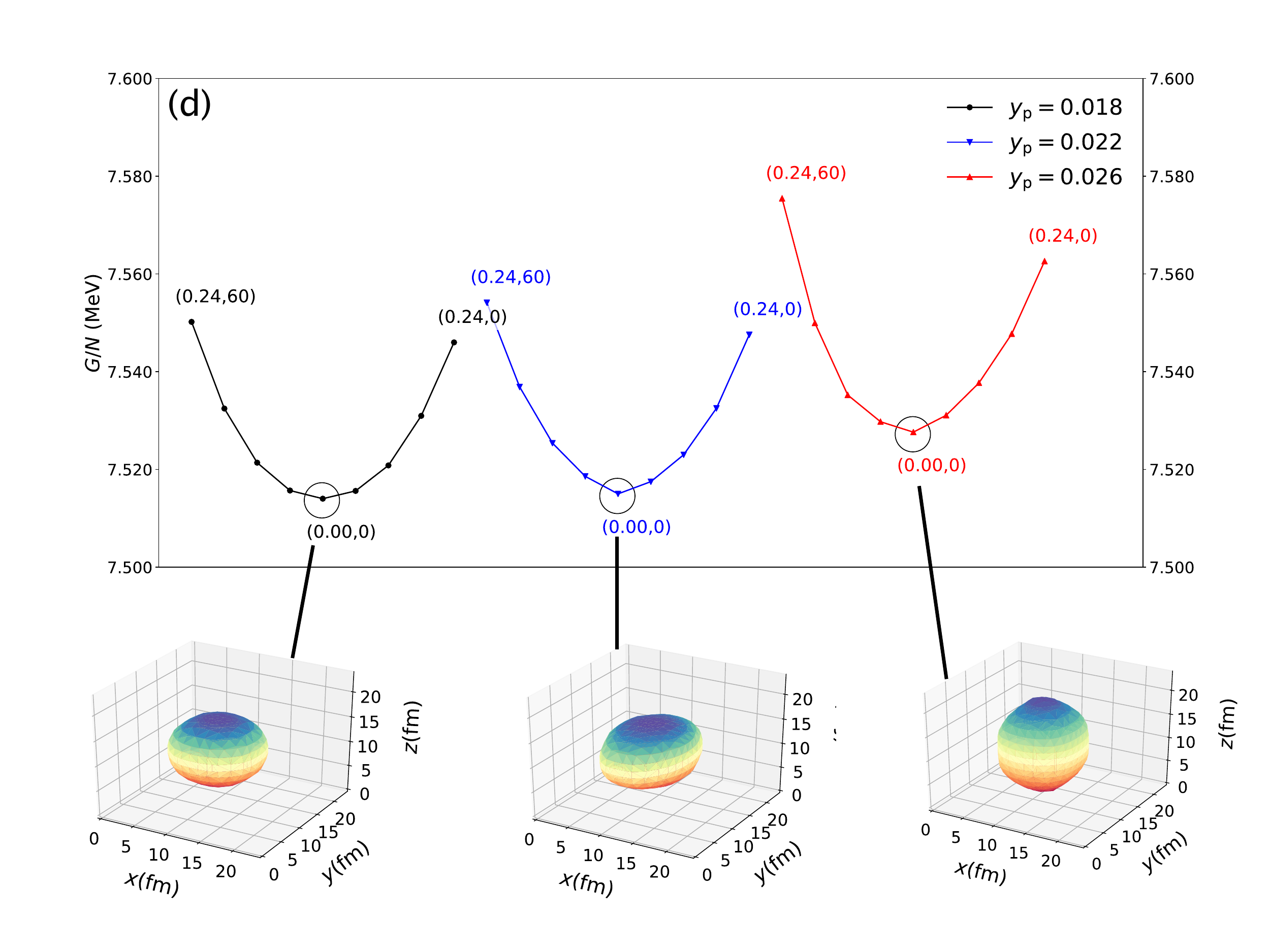}}
\caption{Top row: Gibbs free energy surfaces at a pressure of 0.094 MeV fm$^{-3}$ corresponding to a baryon density of $\approx 0.035$fm$^{-3}$ at proton fractions of 0.018 (a), 0.022 (b) and 0.026 (c), for cells containing $A_{\rm cell}$=454 nucleons. Below is the Gibbs free energy variation along one dimensional paths passing through the energy minimum (d). Selected $(\beta,\gamma)$ coordinates are shown along the one-dimensional plots. Visualizations of the minimum energy nuclear shapes are shown, obtained by plotting a surface of constant neutron density corresponding to the average neutron density in the cell. In all cases the minimum is a spherical nucleus.} \label{fig:7}
\end{figure*}

\begin{figure*}[!t]
\centerline{\includegraphics[scale=0.3]{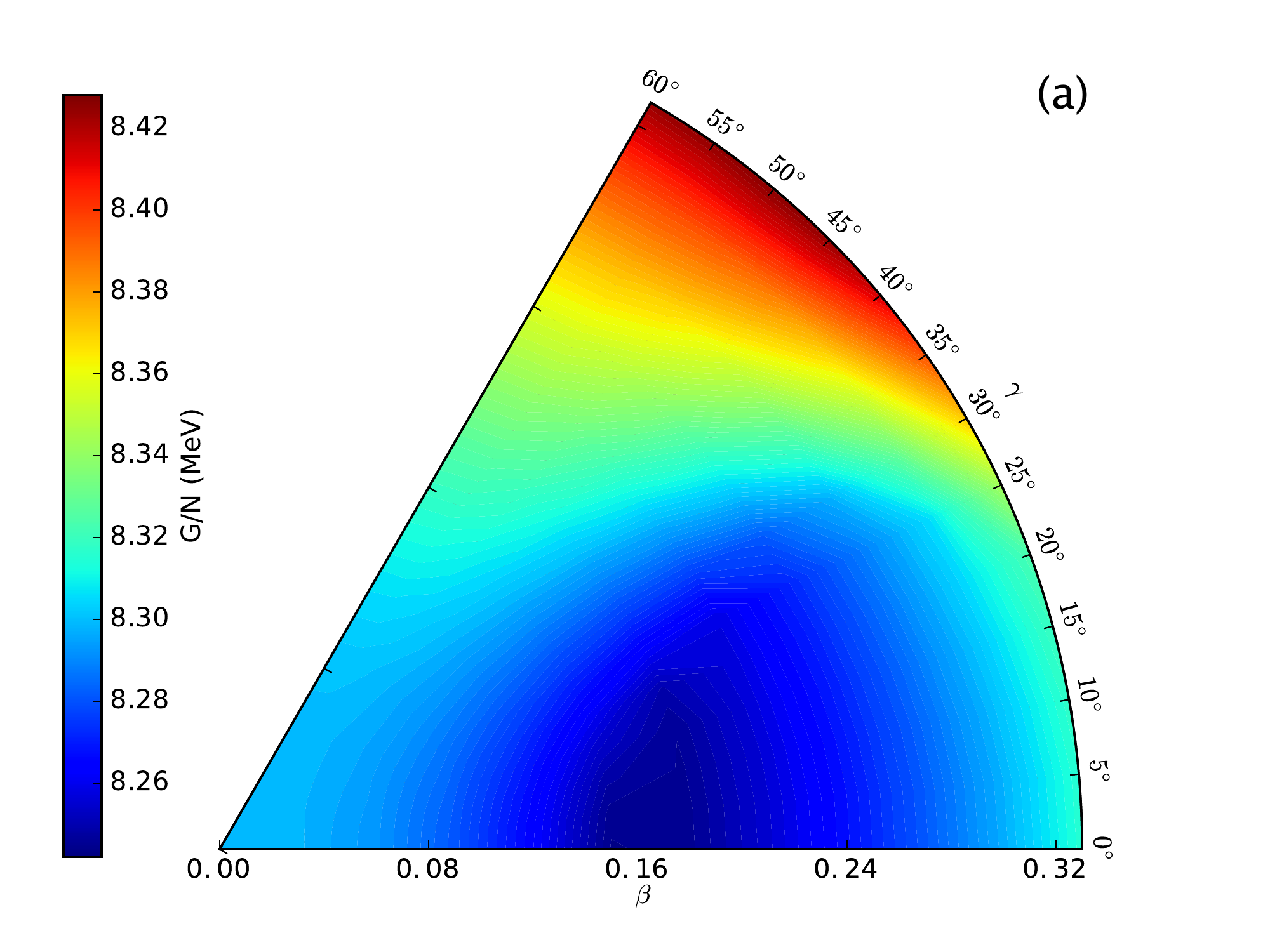}\includegraphics[scale=0.3]{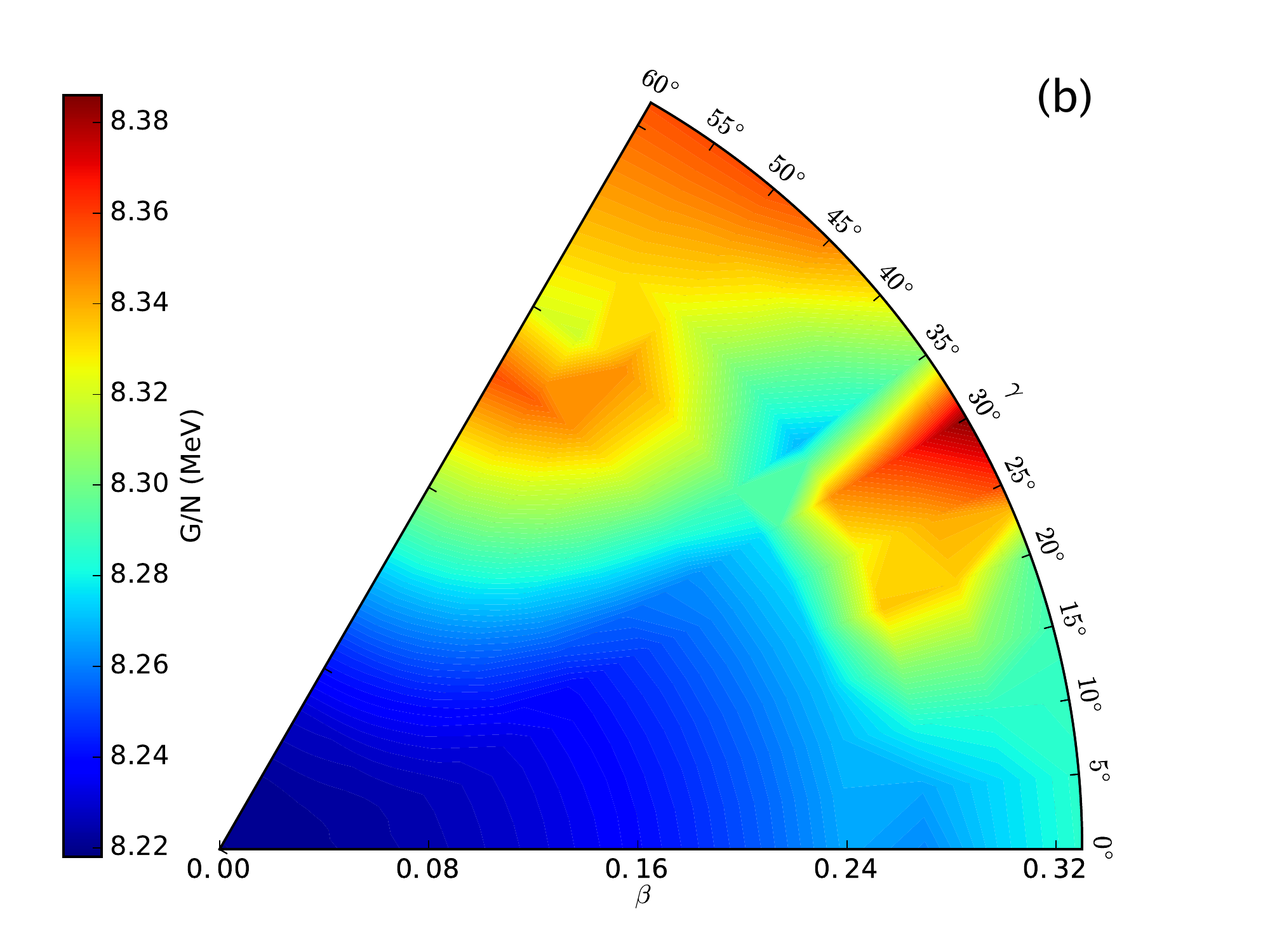}\includegraphics[scale=0.3]{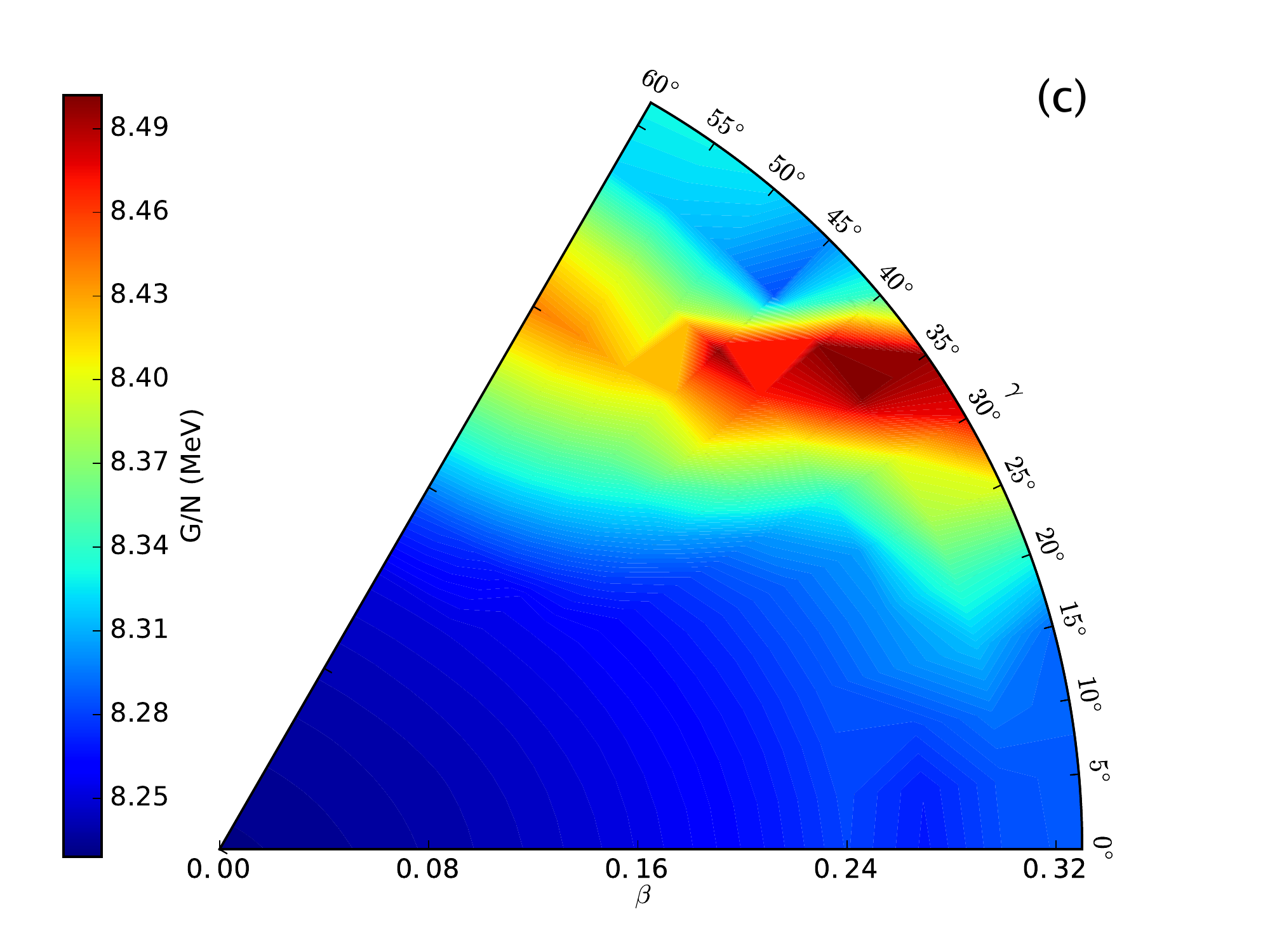}}
\centerline{\includegraphics[scale=0.45]{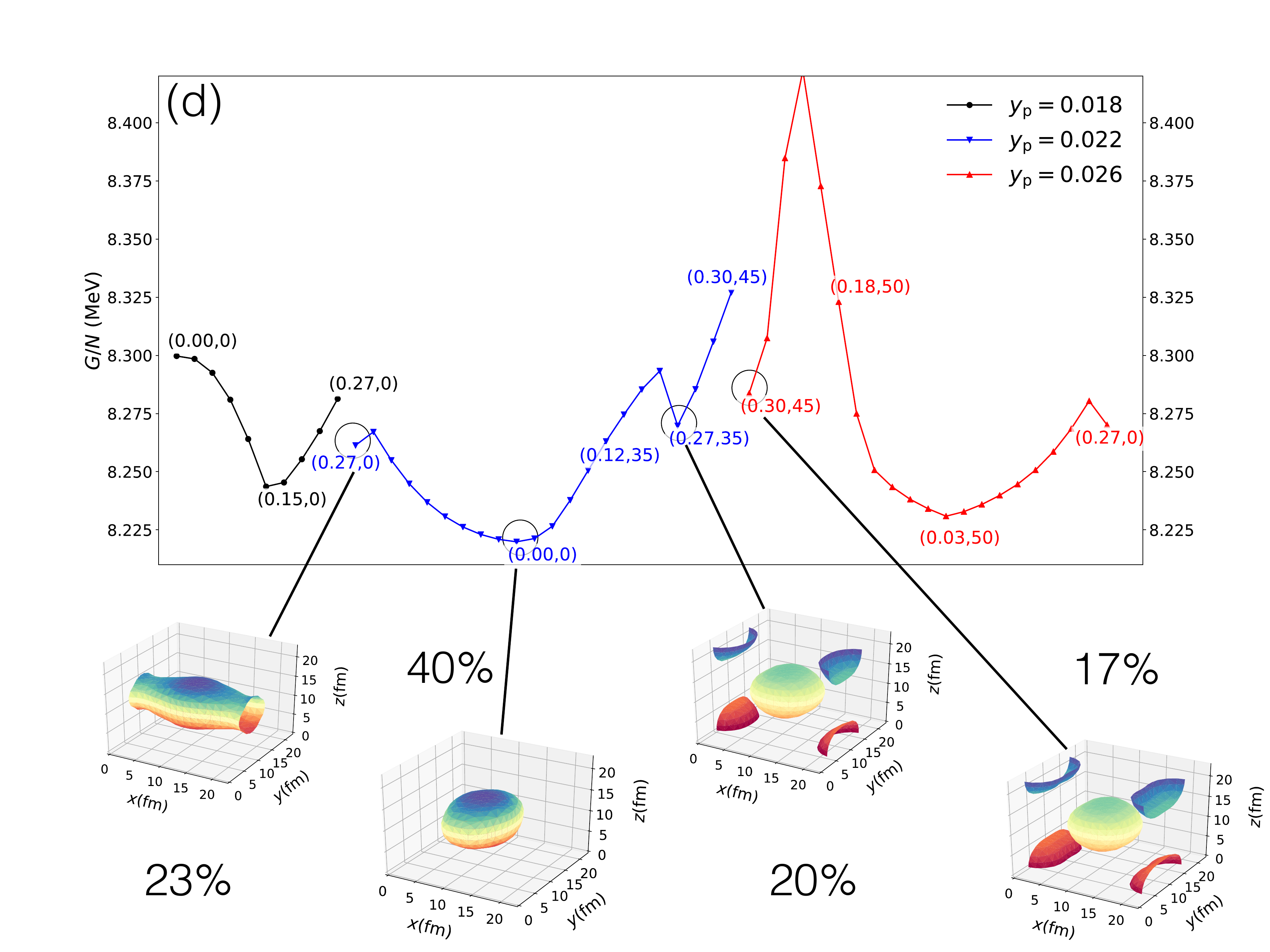}}
\caption{Top row: Gibbs free energy surfaces at a pressure of 0.120 MeV fm$^{-3}$, corresponding to a baryon density of $\approx 0.04$fm$^{-3}$, at proton fractions of 0.018 (a), 0.022 (b) and 0.026 (c), for cells containing $A_{\rm cell}$=454 nucleons. Below is the Gibbs free energy variation along one dimensional paths passing through the energy minimum (d). Selected $(\beta,\gamma)$ coordinates are shown along the one-dimensional plots. Visualizations of the minimum energy nuclear shapes are shown, obtained by plotting a surface of constant neutron density corresponding to the average neutron density in the cell. The minimum energy nuclear shape is roughly spherical, but deformed nuclear and cylindrical nuclear shapes appear as local minima. The relative abundances of the phases at a temperature equal to the fictive temperature is shown next to the visualizations of the phases.} \label{fig:8}
\end{figure*}

\begin{figure*}[!t]
\centerline{\includegraphics[scale=0.3]{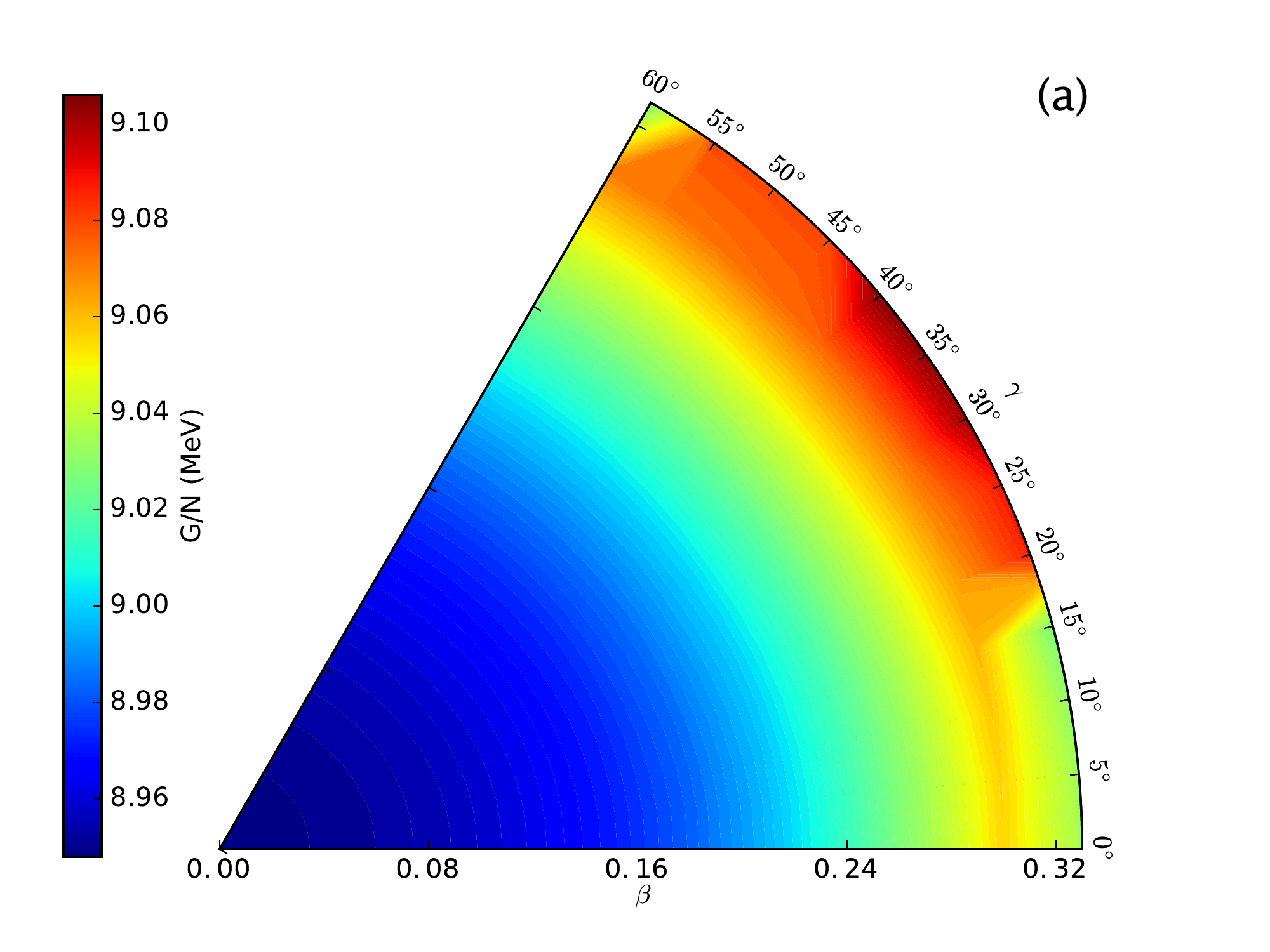}\includegraphics[scale=0.3]{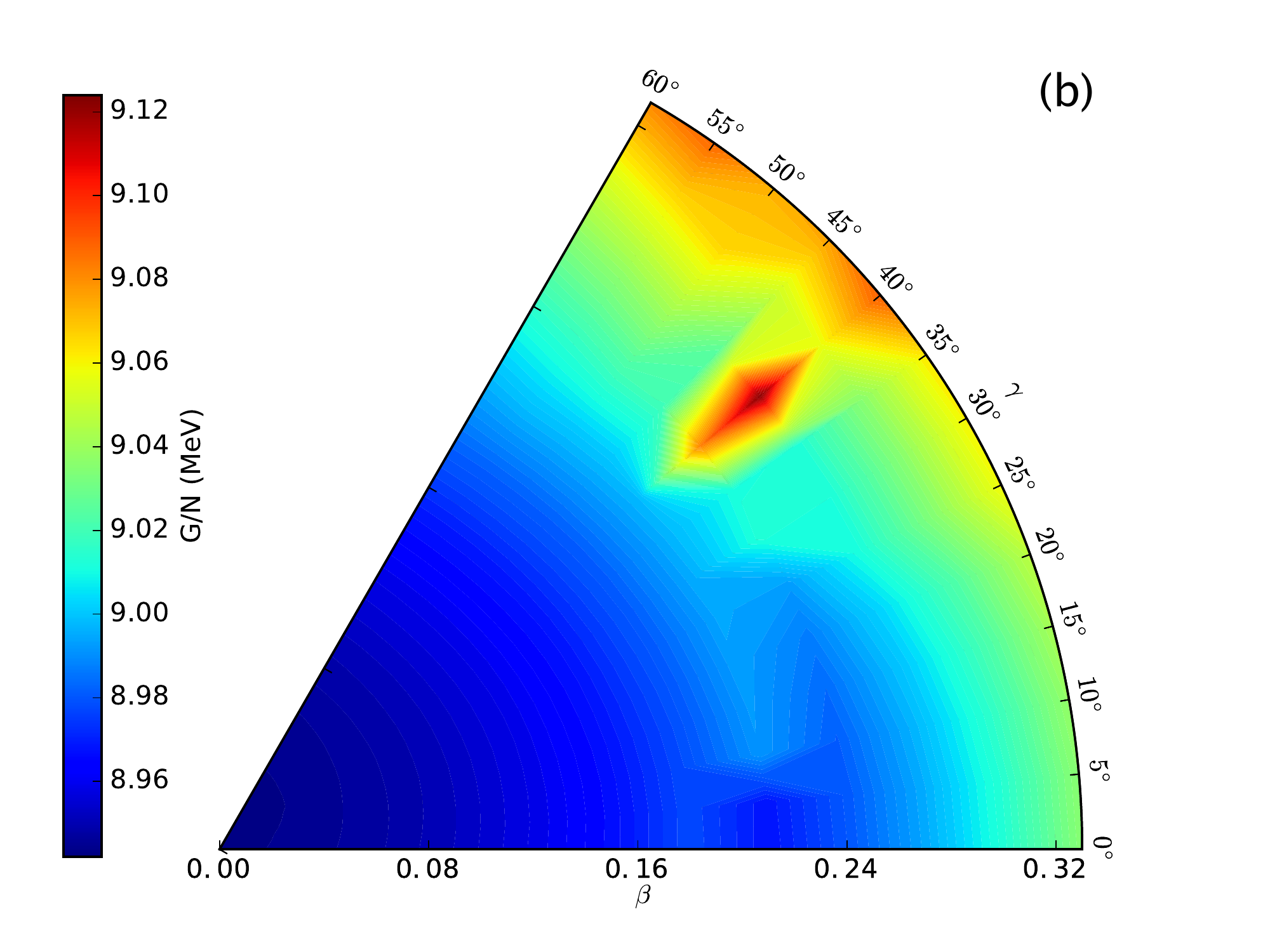}\includegraphics[scale=0.3]{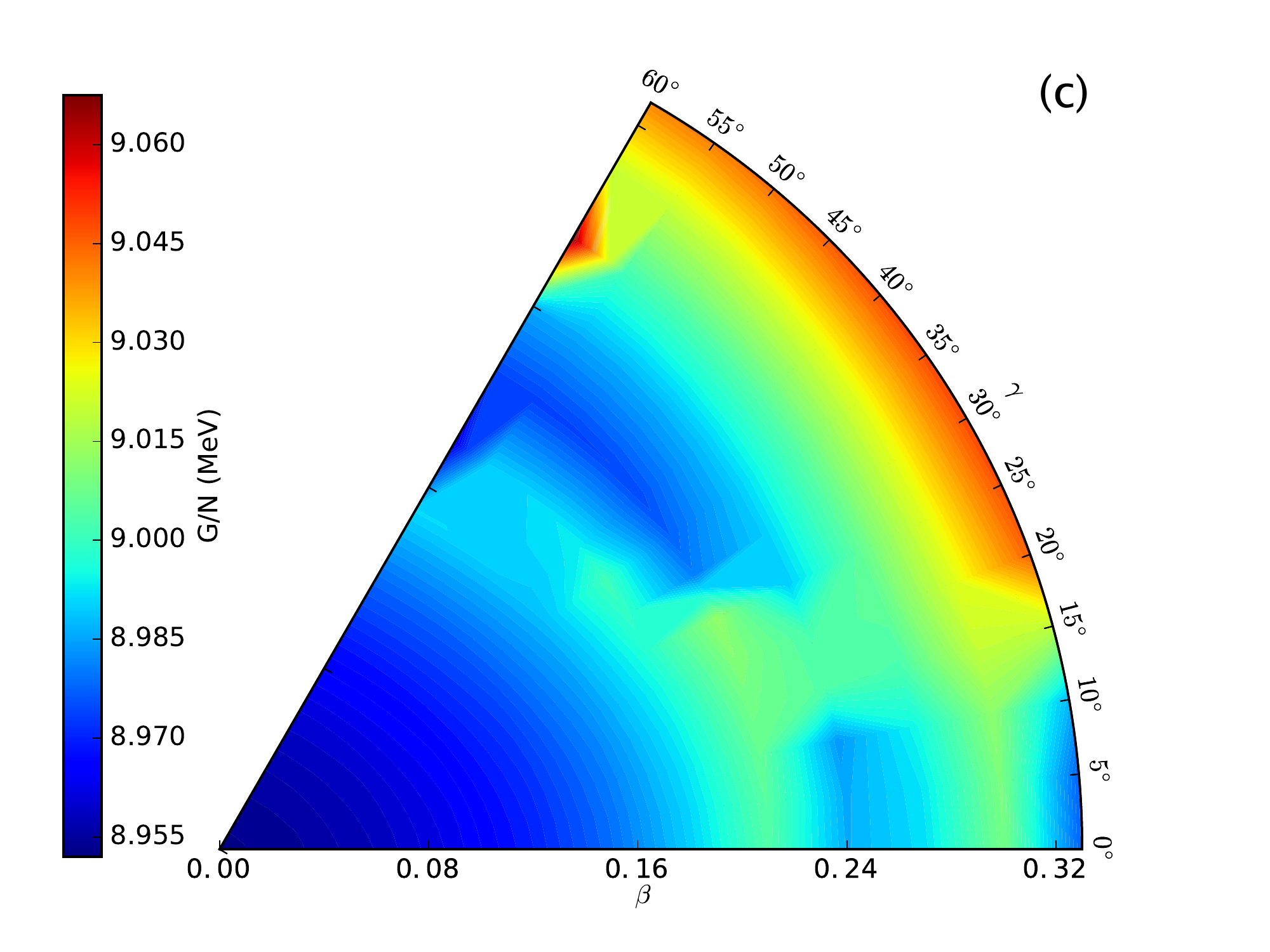}}
\centerline{\includegraphics[scale=0.47]{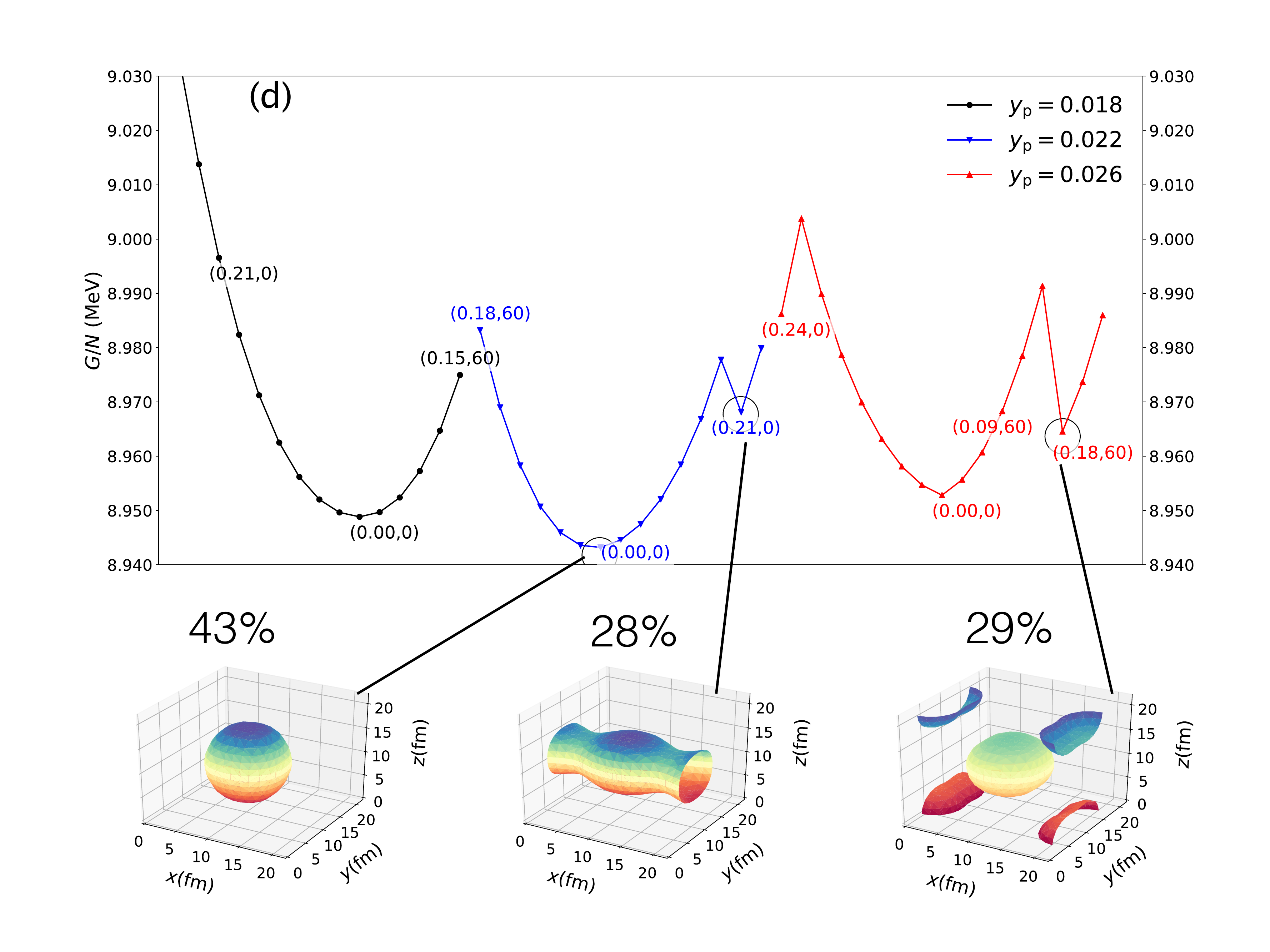}}
\caption{Top row: Gibbs free energy surfaces at a pressure of 0.150 MeV fm$^{-3}$, corresponding to a baryon density of $\approx 0.045$fm$^{-3}$, at proton fractions of 0.018 (a), 0.022 (b) and 0.026 (c), for cells containing $A_{\rm cell}$=454 nucleons. Below is the Gibbs free energy variation along one dimensional paths passing through the energy minimum (d). Selected $(\beta,\gamma)$ coordinates are shown along the one-dimensional plots. Visualizations of the minimum energy nuclear shapes are shown, obtained by plotting a surface of constant neutron density corresponding to the average neutron density in the cell. The minimum energy nuclear shape is roughly spherical, but deformed nuclear and cylindrical nuclear shapes appear as local minima. The relative abundances of the phases at a temperature equal to the fictive temperature is shown below the visualizations of the phases.} \label{fig:9}
\end{figure*}

\begin{figure*}[!t]
\centerline{\includegraphics[scale=0.3]{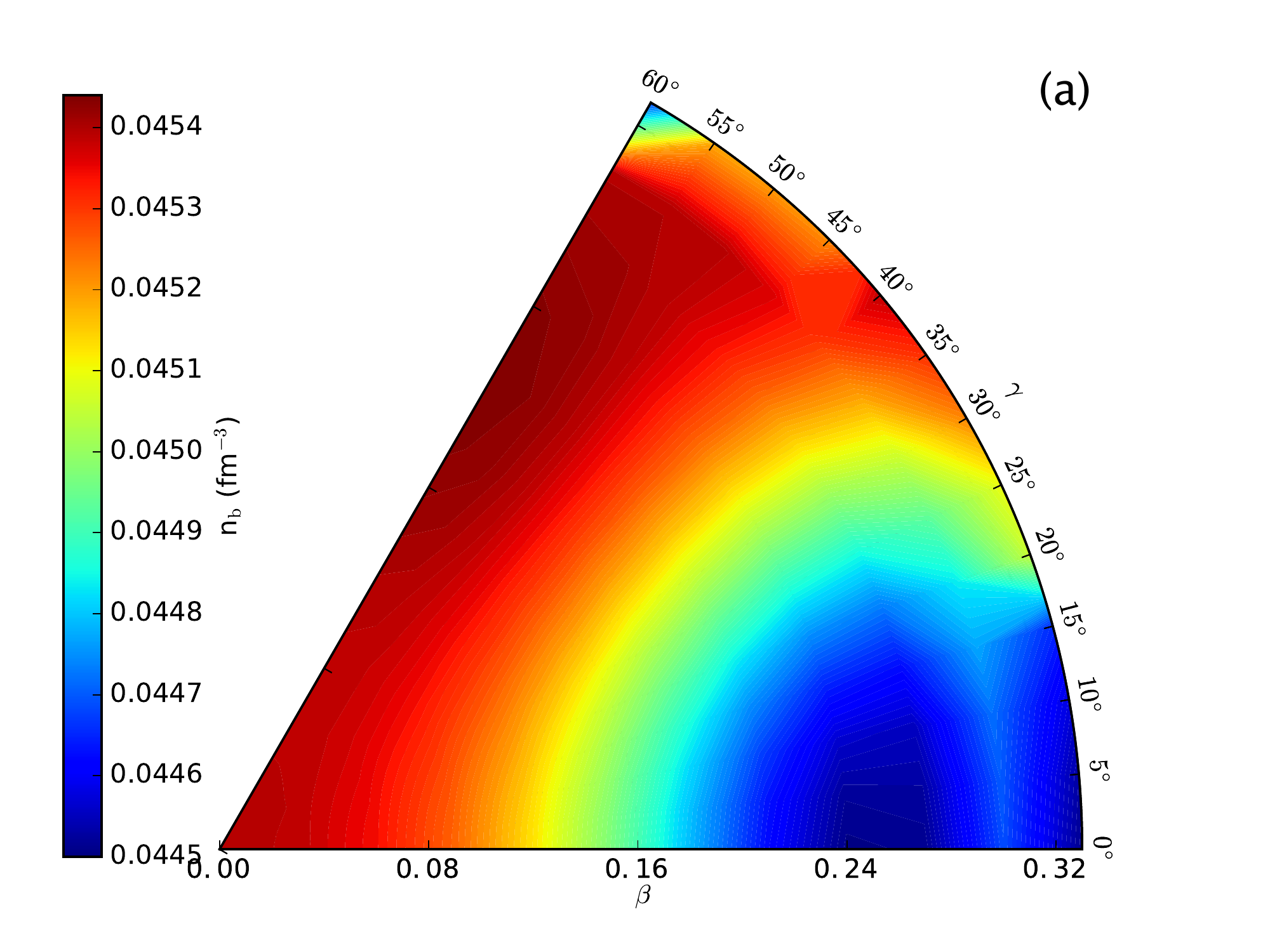}\includegraphics[scale=0.3]{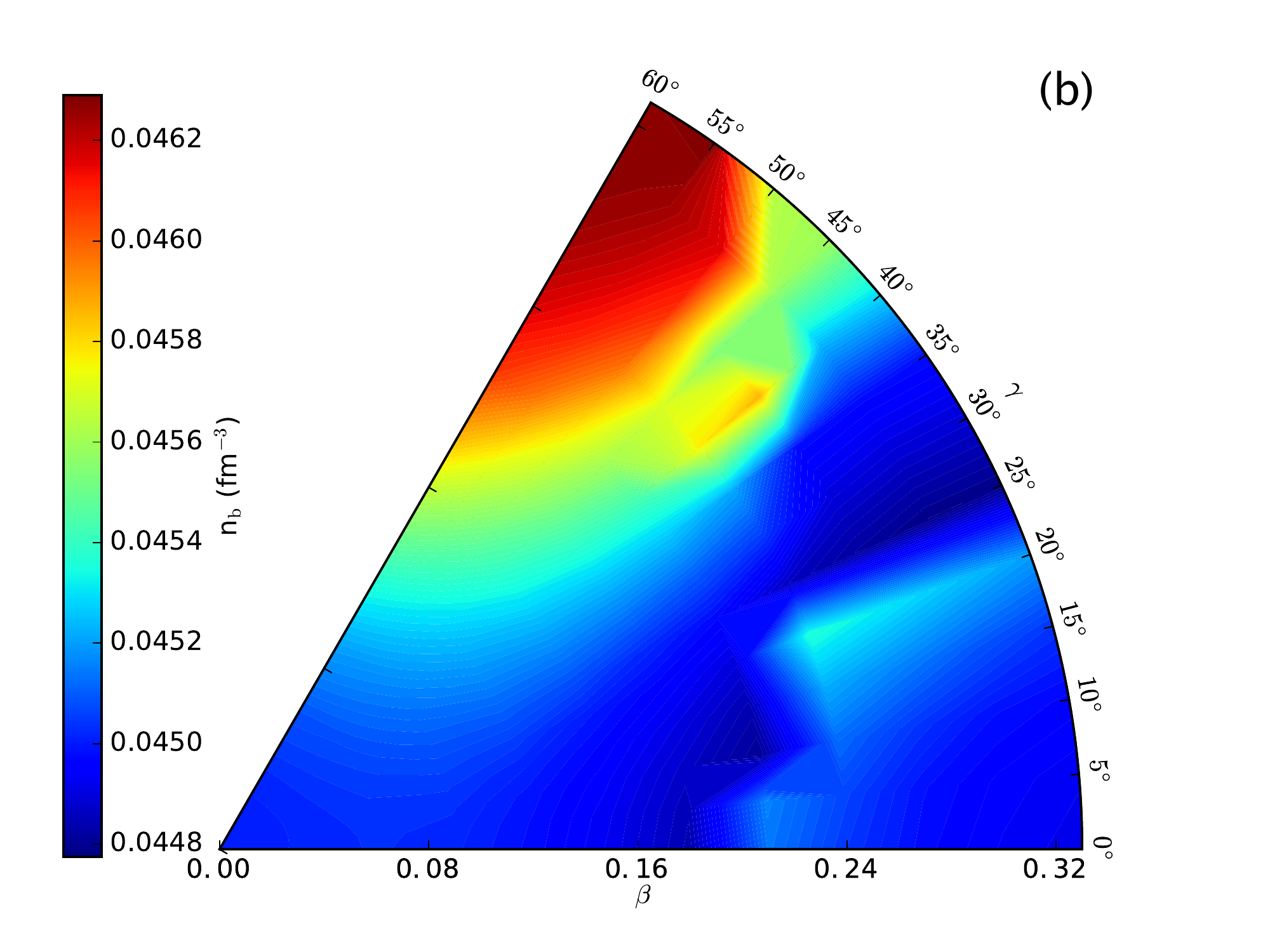}\includegraphics[scale=0.3]{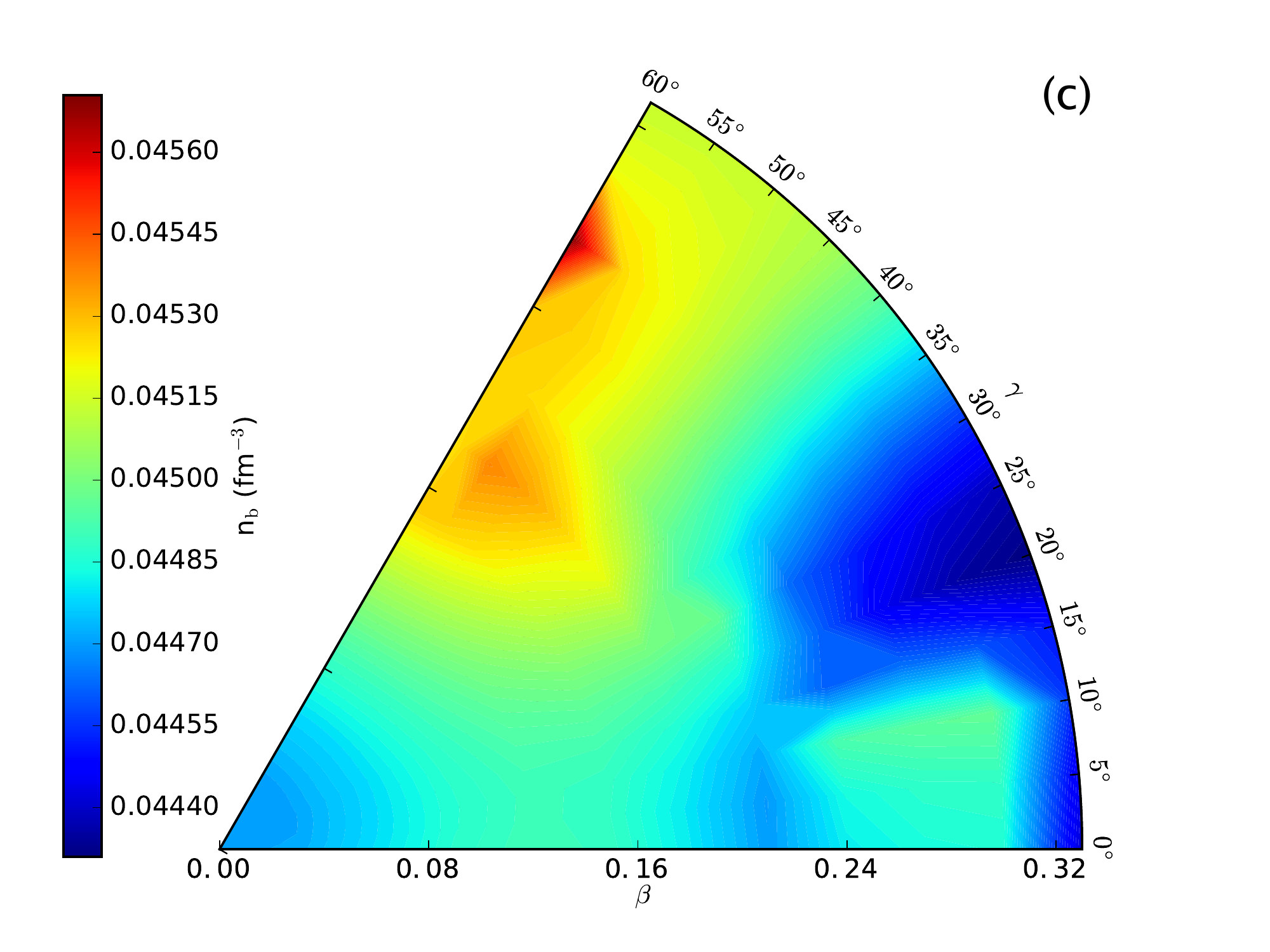}}
\caption{Average baryon density surfaces (bottom) at a constant pressure of $0.15$ MeV fm$^{-3}$, corresponding to a baryon density of $\approx 0.045$fm$^{-3}$, for cells containing $A_{\rm cell}$=454 nucleons. Results are shown for proton fraction of 0.018 (a), 0.022 (b) and 0.026 (c). The local minima differ in baron density by of order 5\%.} \label{fig:10}
\end{figure*}

\begin{figure*}[!t]
\centerline{\includegraphics[scale=0.3]{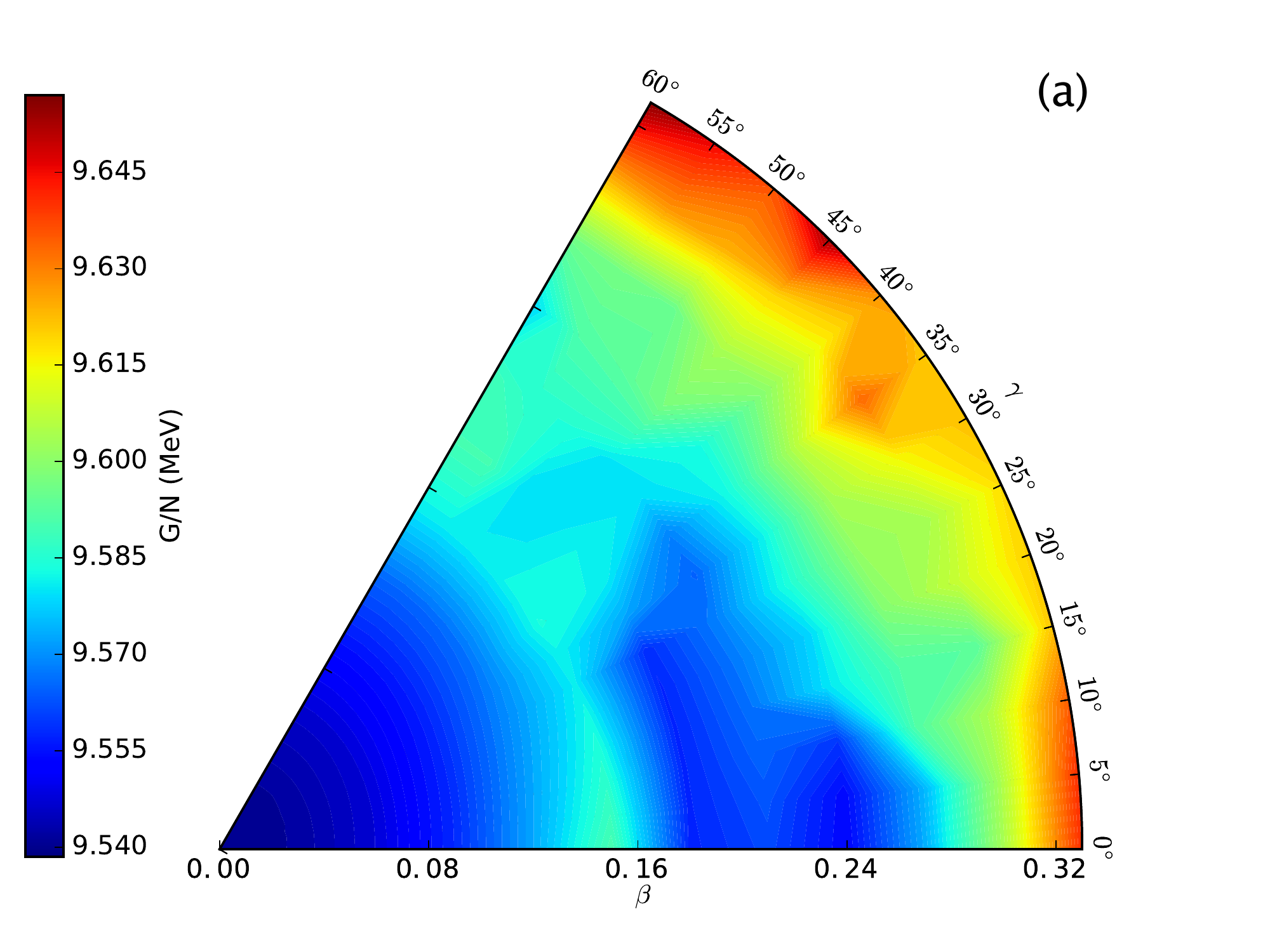}\includegraphics[scale=0.3]{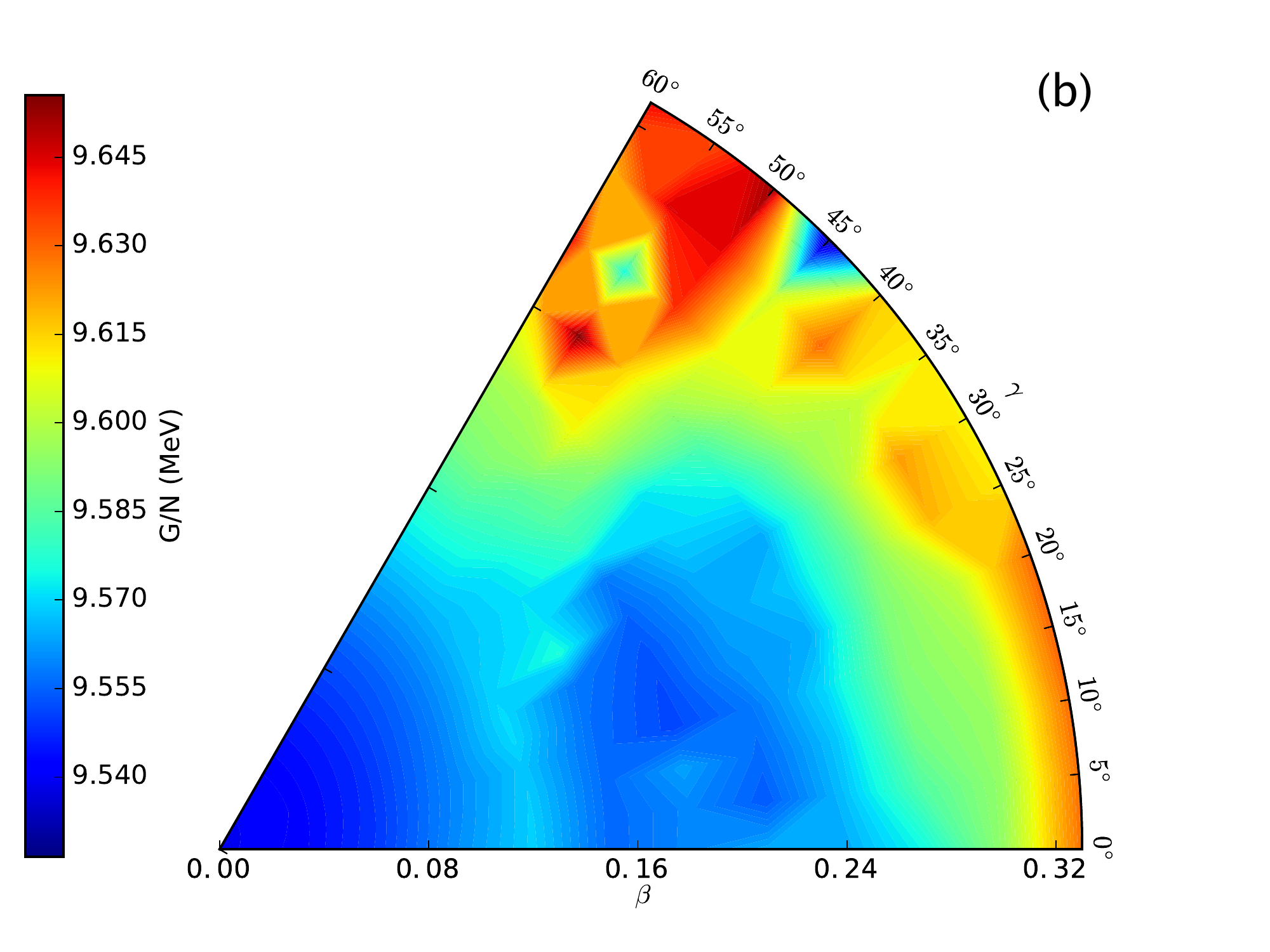}\includegraphics[scale=0.3]{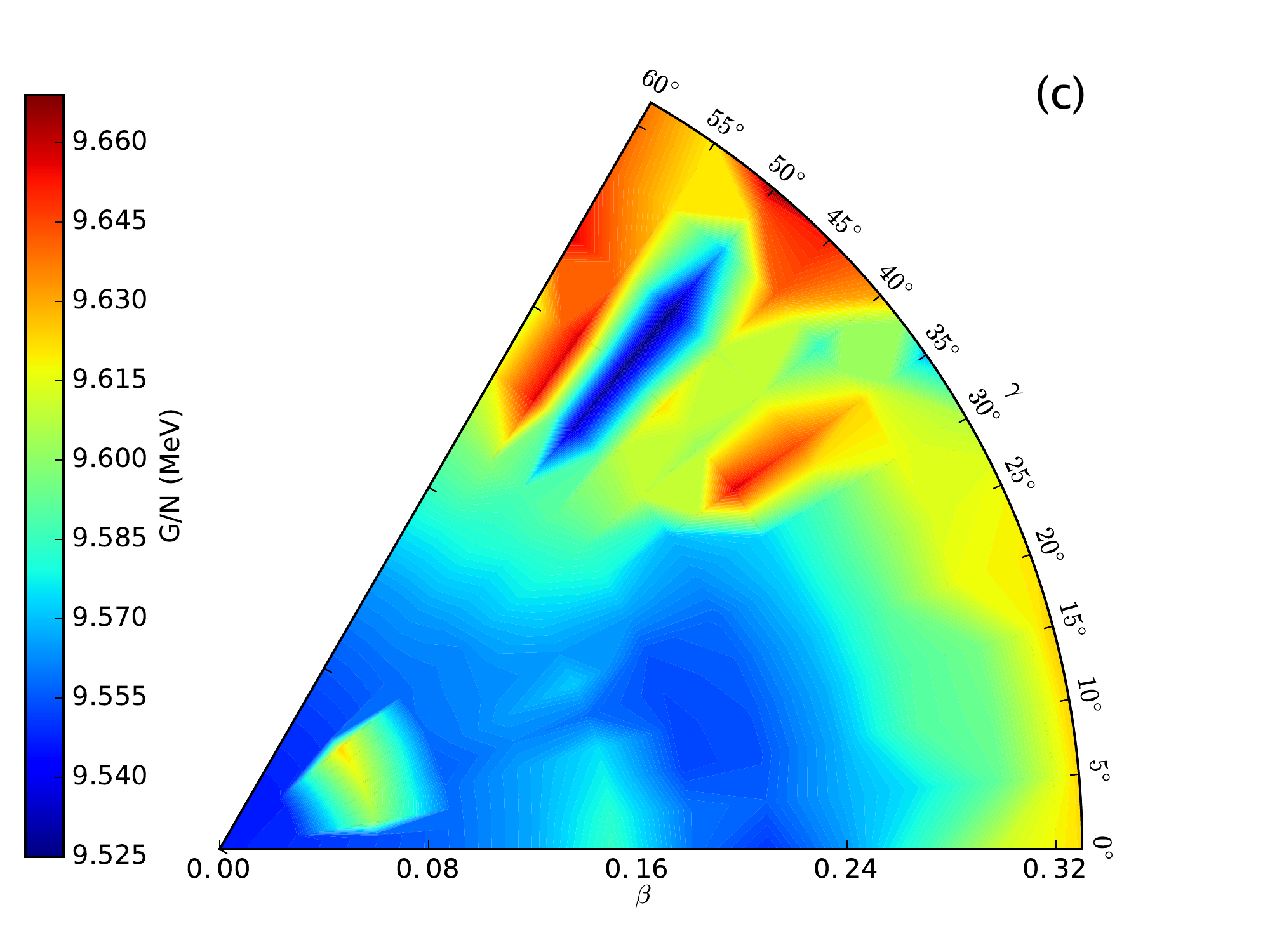}}
\centerline{\includegraphics[scale=0.47]{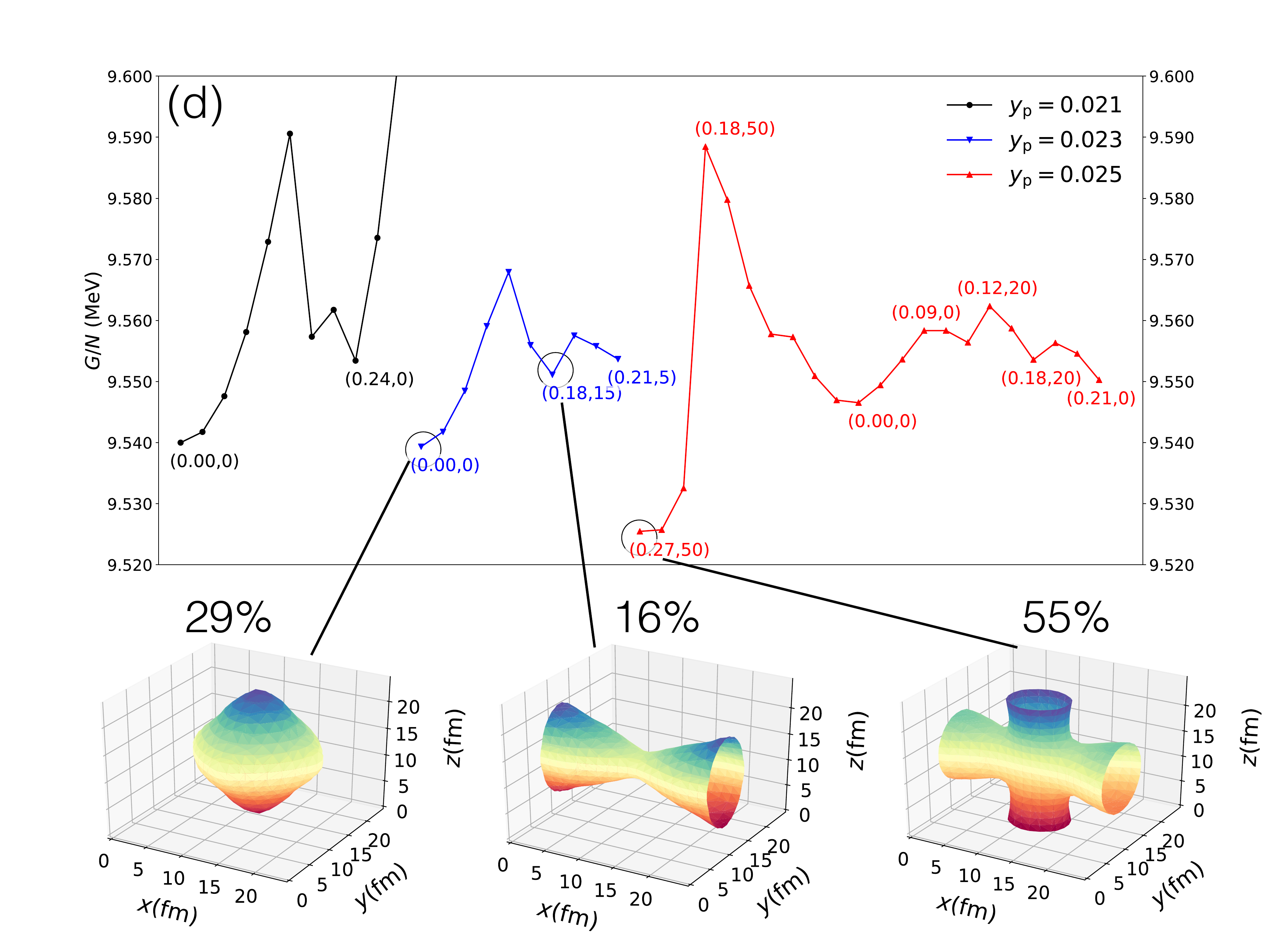}}
\caption{Top row: Gibbs free energy surfaces at a pressure of 0.184 MeV fm$^{-3}$, corresponding to a baryon density of $\approx 0.05$fm$^{-3}$, at proton fractions of 0.021 (a), 0.023 (b) and 0.025 (c), for cells containing $A_{\rm cell}$=956 nucleons. Below is the Gibbs free energy variation along one dimensional paths passing through the energy minimum (d). Selected $(\beta,\gamma)$ coordinates are shown along the one-dimensional plots. Visualizations of the minimum energy nuclear shapes are shown, obtained by plotting a surface of constant neutron density corresponding to the average neutron density in the cell. The minimum energy nuclear shape is a nuclear `waffle', with an isolated nuclear phase and spaghetti phase as local minima at a different proton fraction. The relative abundances of the phases at a temperature equal to the fictive temperature is shown below the visualizations of the phases.} \label{fig:11}
\end{figure*}

\begin{figure*}[!t]
\centerline{\includegraphics[scale=0.3]{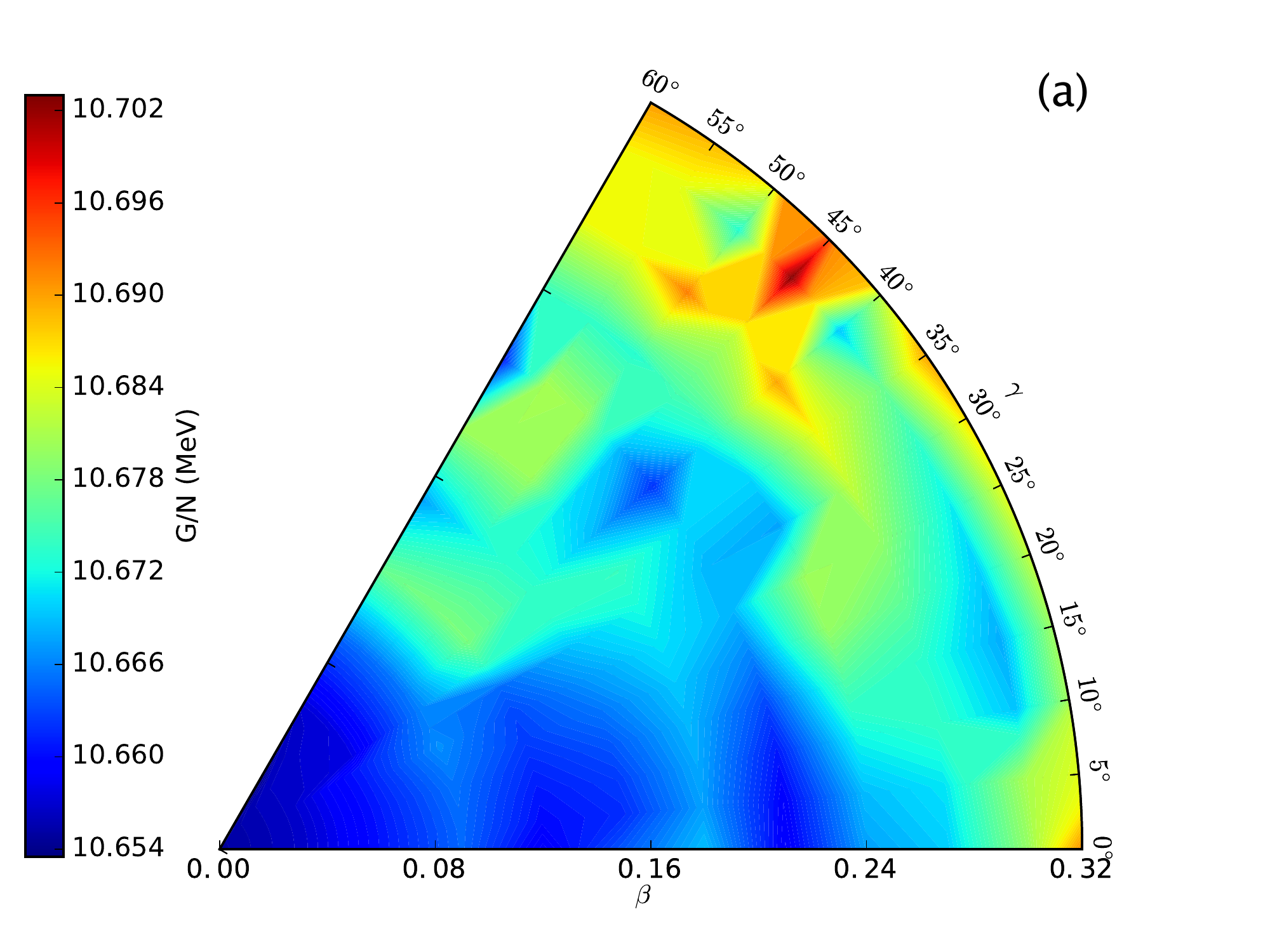}\includegraphics[scale=0.3]{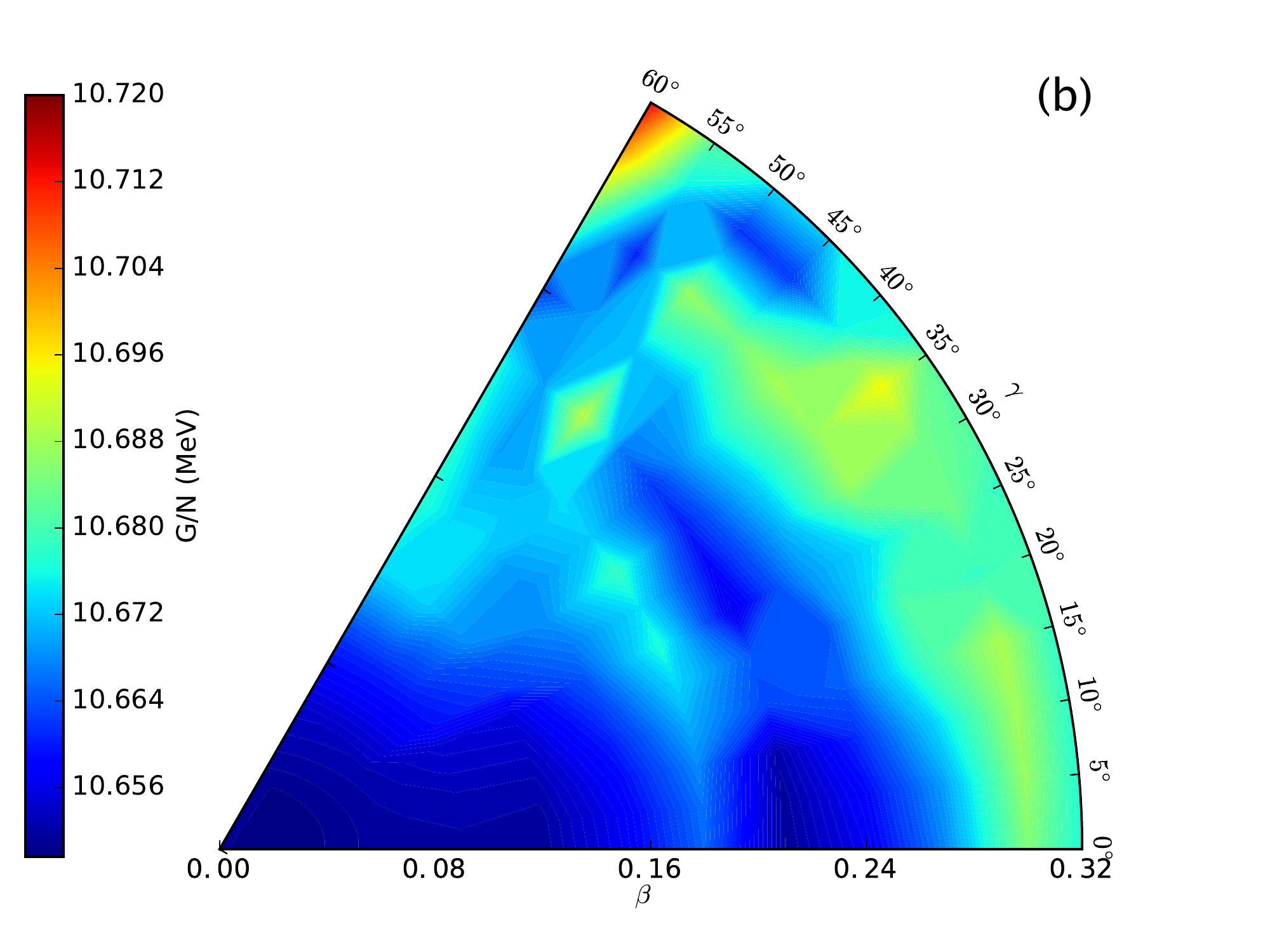}\includegraphics[scale=0.3]{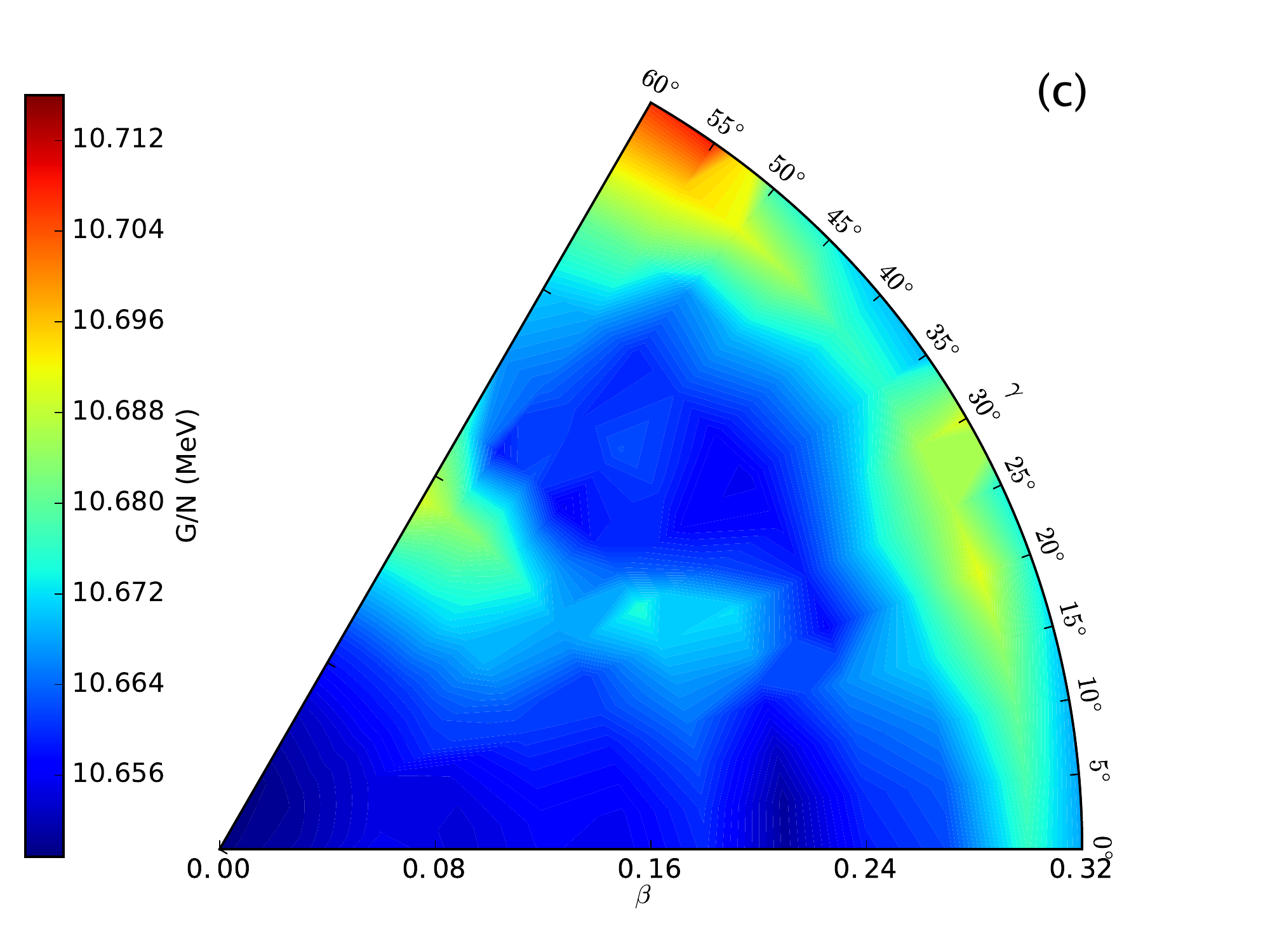}}
\centerline{\includegraphics[scale=0.47]{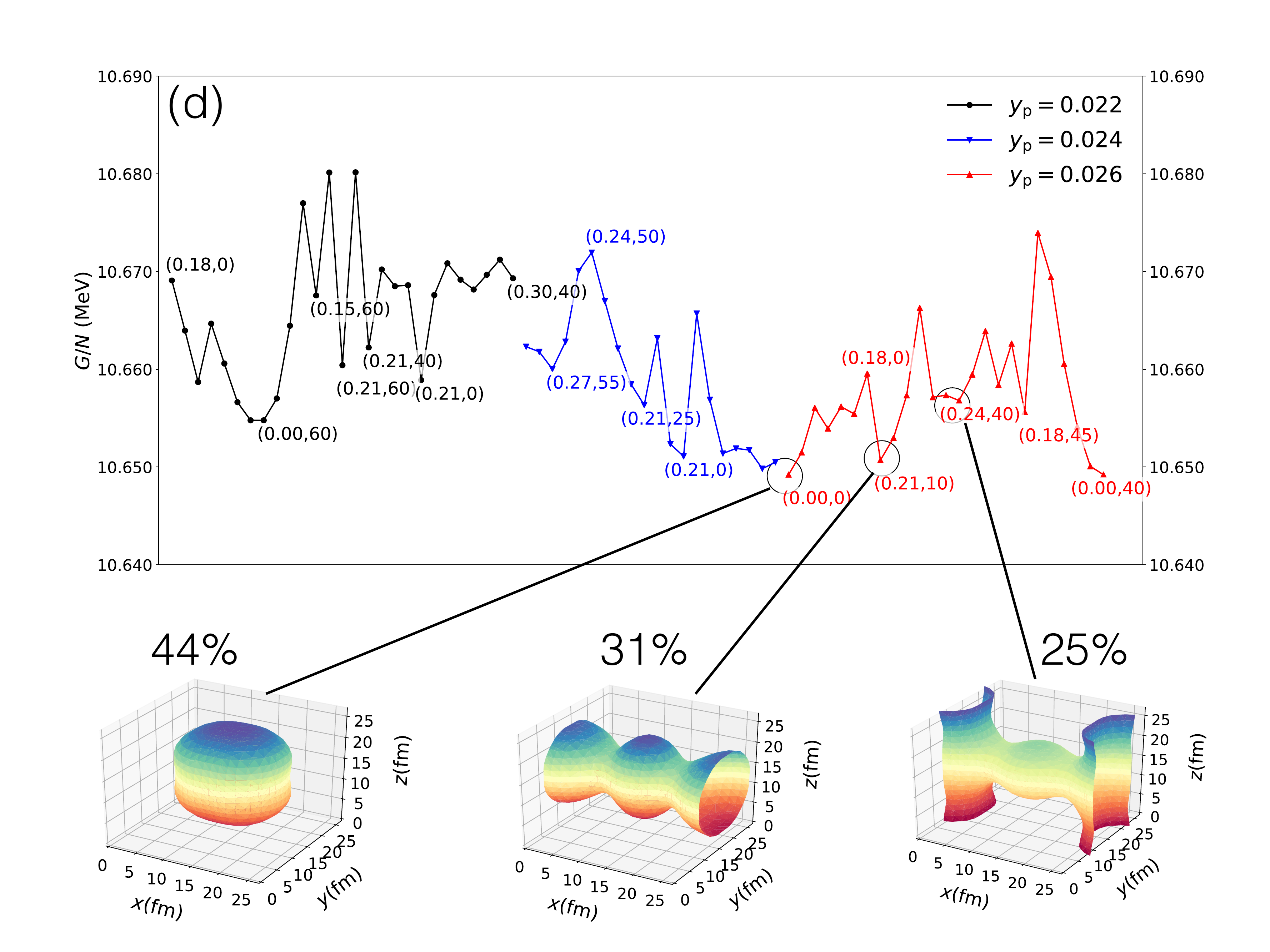}}
\caption{Top row: Gibbs free energy surfaces at a pressure of 0.234 MeV fm$^{-3}$, corresponding to a baryon density of $\approx 0.054$fm$^{-3}$, at proton fractions of 0.022 (a), 0.024 (b) and 0.026 (c), for cells containing $A_{\rm cell}$=1166 nucleons. Below is the Gibbs free energy variation along one dimensional paths passing through the energy minimum (d). Selected $(\beta,\gamma)$ coordinates are shown along the one-dimensional plots. Visualizations of the minimum energy nuclear shapes are shown, obtained by plotting a surface of constant neutron density corresponding to the average neutron density in the cell. The minimum energy nuclear shape is roughly spherical, followed by cylindrical and waffle shapes. The relative abundances of the phases at a temperature equal to the fictive temperature is shown below the visualizations of the phases.} \label{fig:12}
\end{figure*}

\begin{figure*}[!t]
\centerline{\includegraphics[scale=0.35]{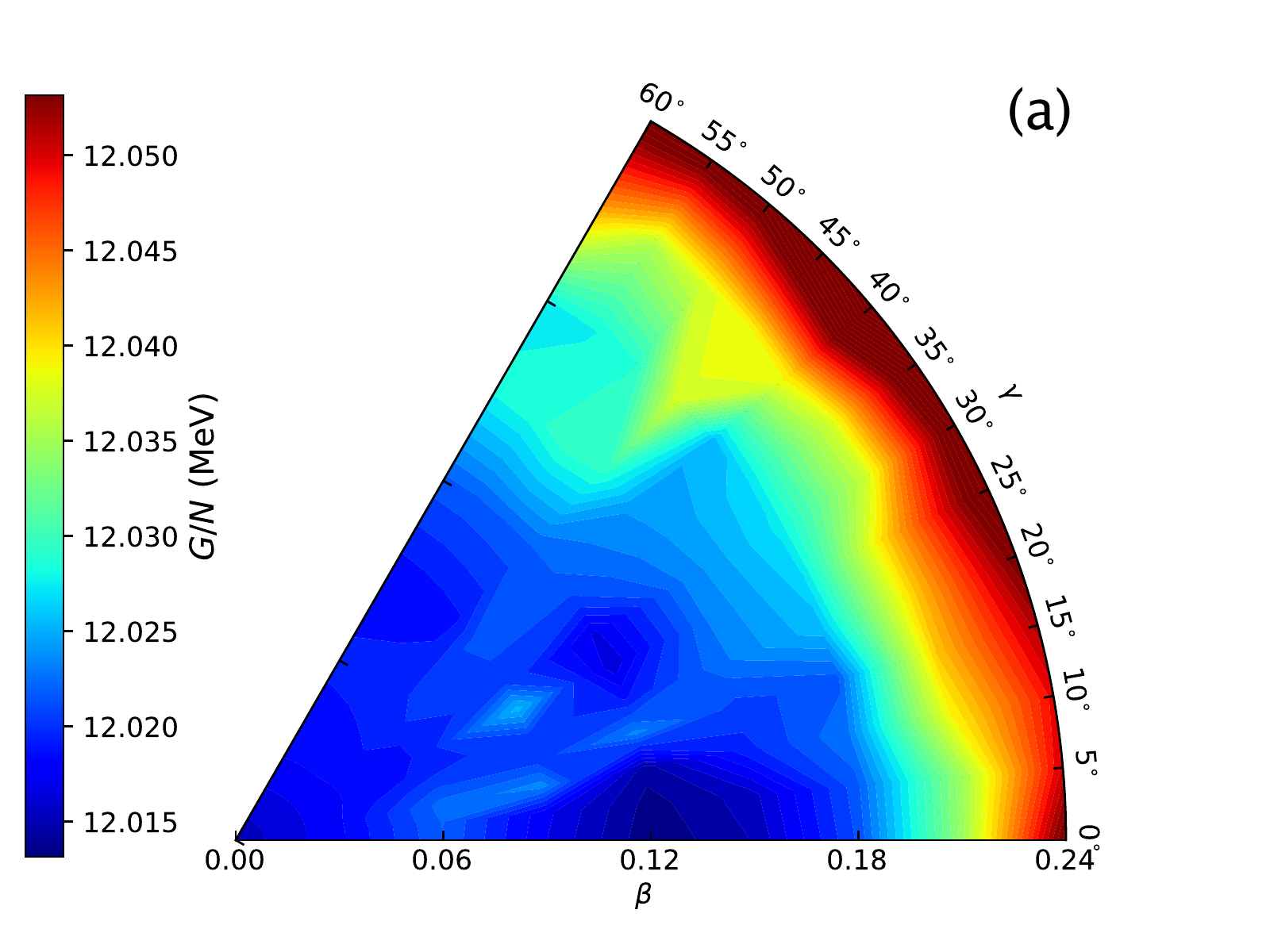}\includegraphics[scale=0.35]{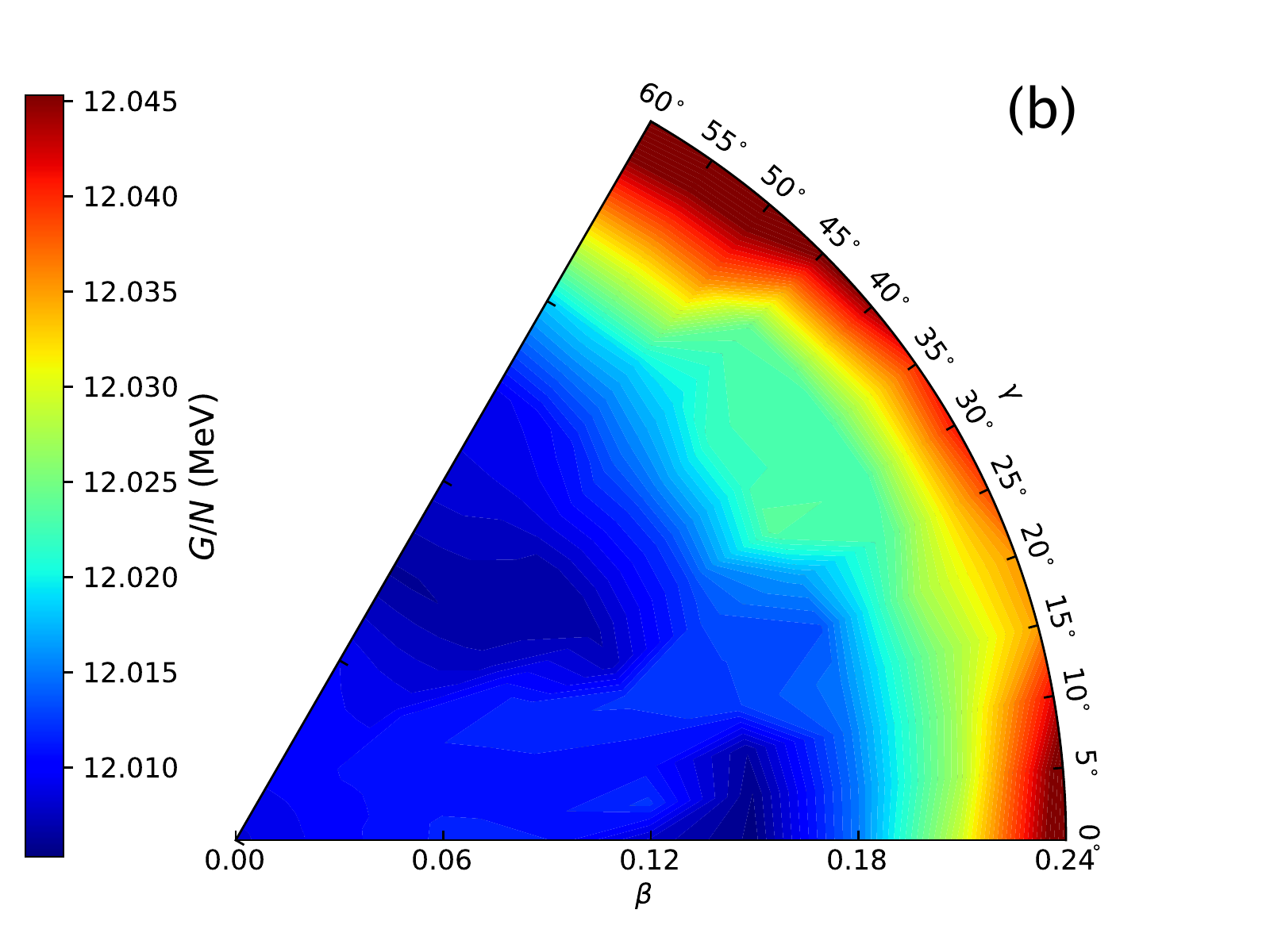}\includegraphics[scale=0.35]{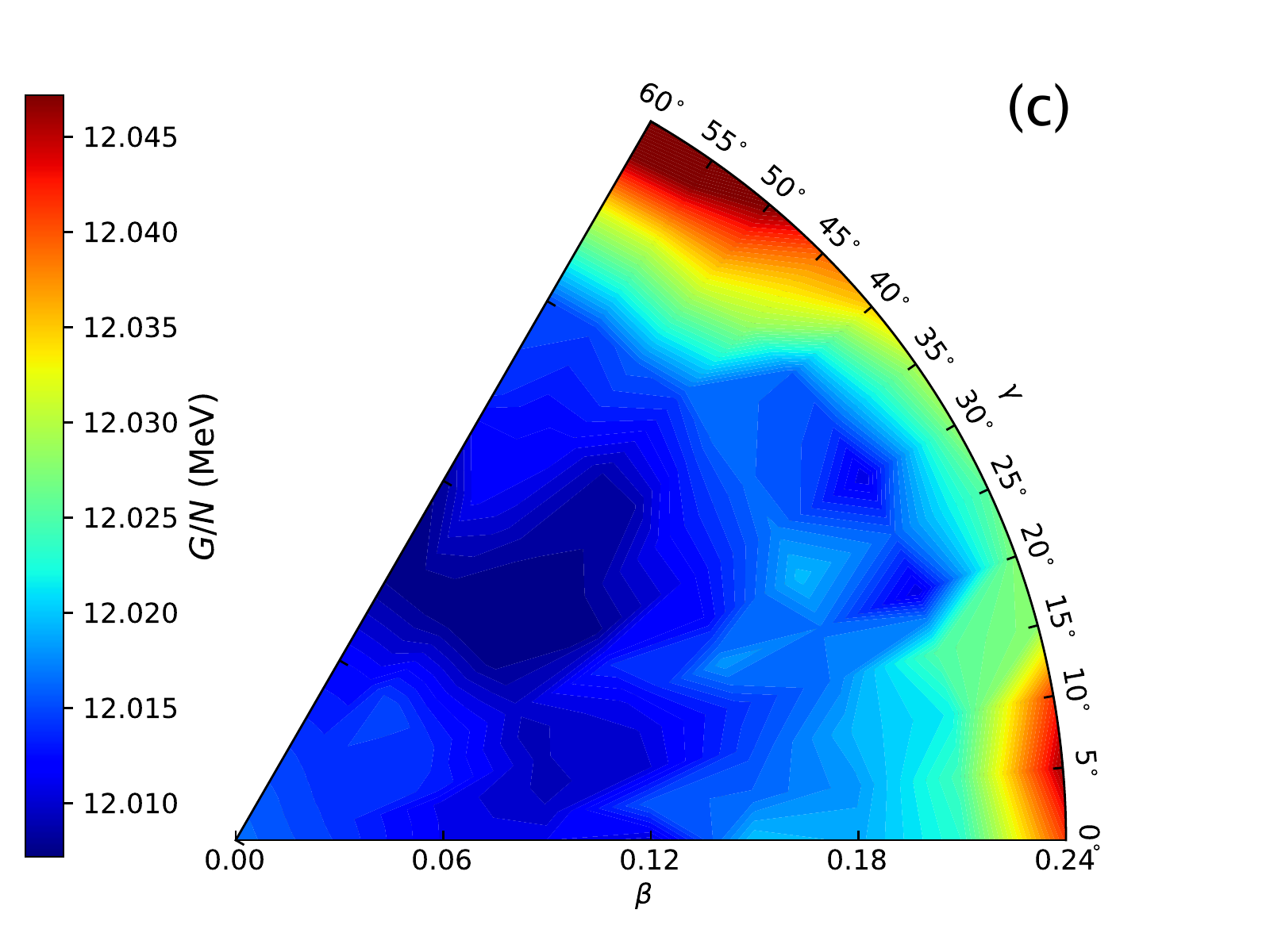}}
\centerline{\includegraphics[scale=0.47]{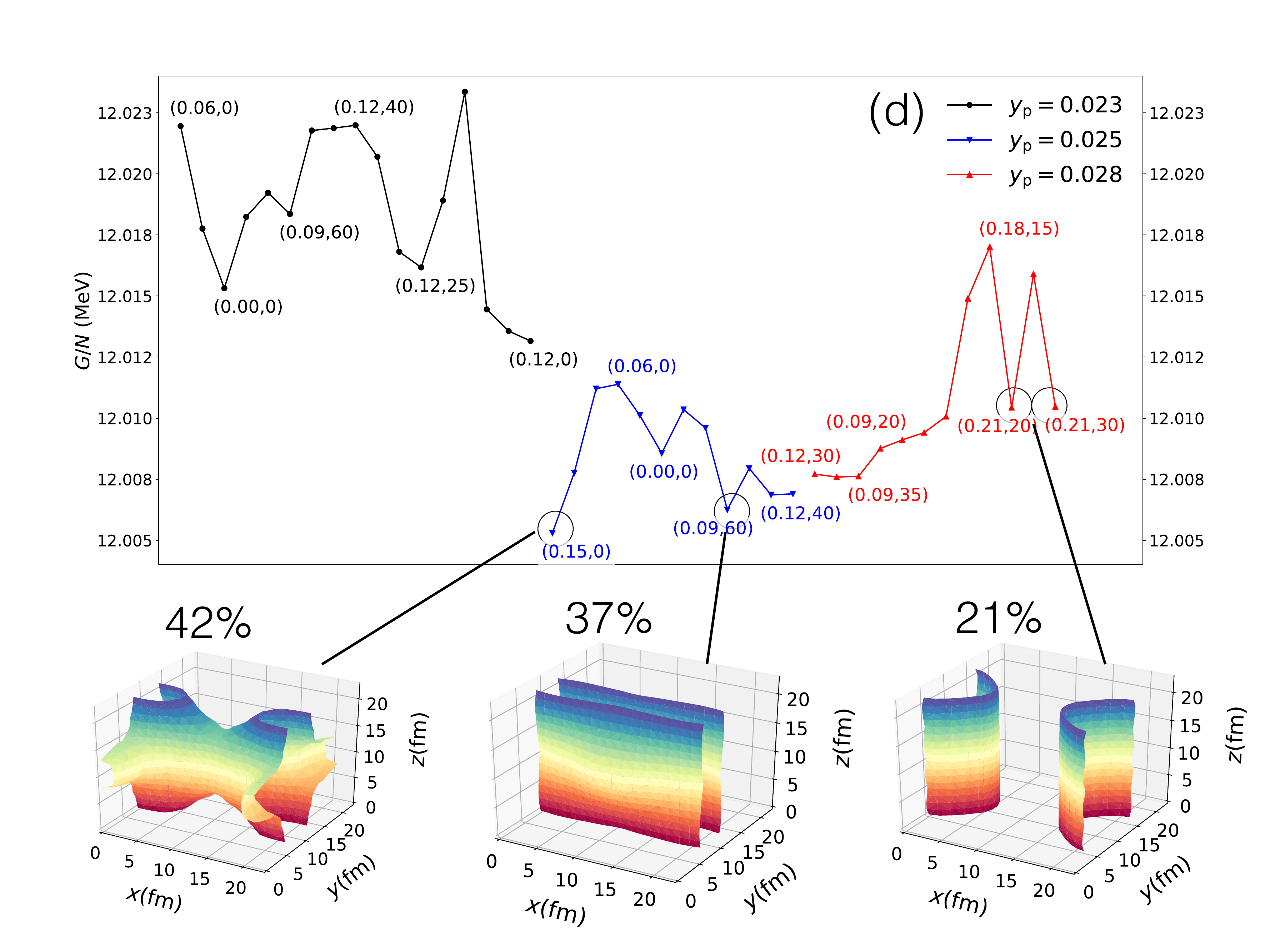}}
\caption{Top row: Gibbs free energy surfaces at a pressure of 0.30 MeV fm$^{-3}$, corresponding to a baryon density of $\approx 0.066$fm$^{-3}$, at proton fractions of 0.022 (a), 0.025 (b) and 0.028 (c), for cells containing $A_{\rm cell}$=784 nucleons. Below is the Gibbs free energy variation along one dimensional paths passing through the energy minimum (d). Selected $(\beta,\gamma)$ coordinates are shown along the one-dimensional plots. Visualizations of the minimum energy nuclear shapes are shown, obtained by plotting a surface of constant neutron density corresponding to the average neutron density in the cell. The minimum energy nuclear configuration is the bi-continuous cubic-P phase, followed by planar and cylindrical geometries. The relative abundances of the phases at a temperature equal to the fictive temperature is shown below the visualizations of the phases.} \label{fig:13}
\end{figure*}


\begin{figure*}[!t]
\centerline{\includegraphics[scale=0.35]{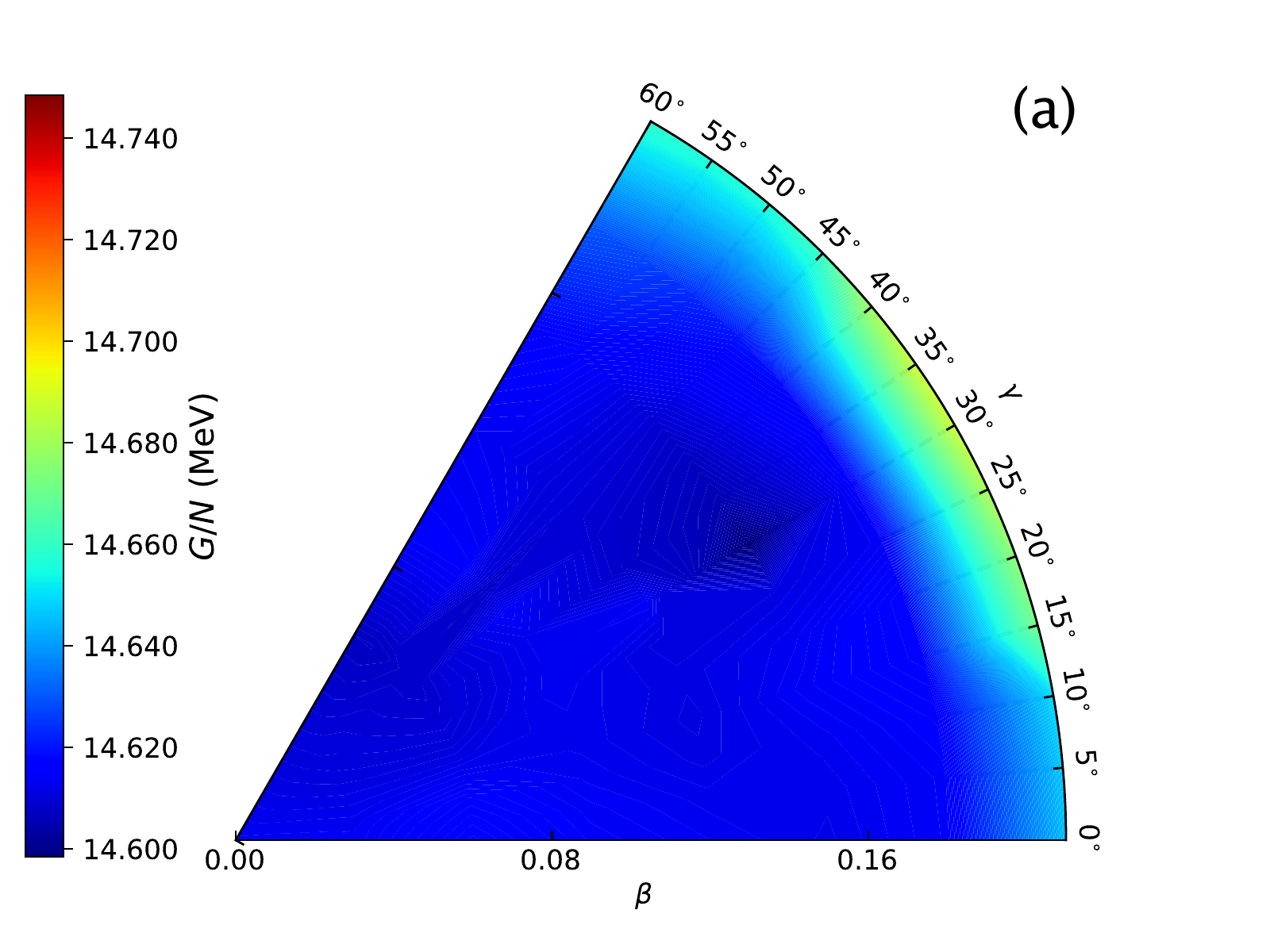}\includegraphics[scale=0.35]{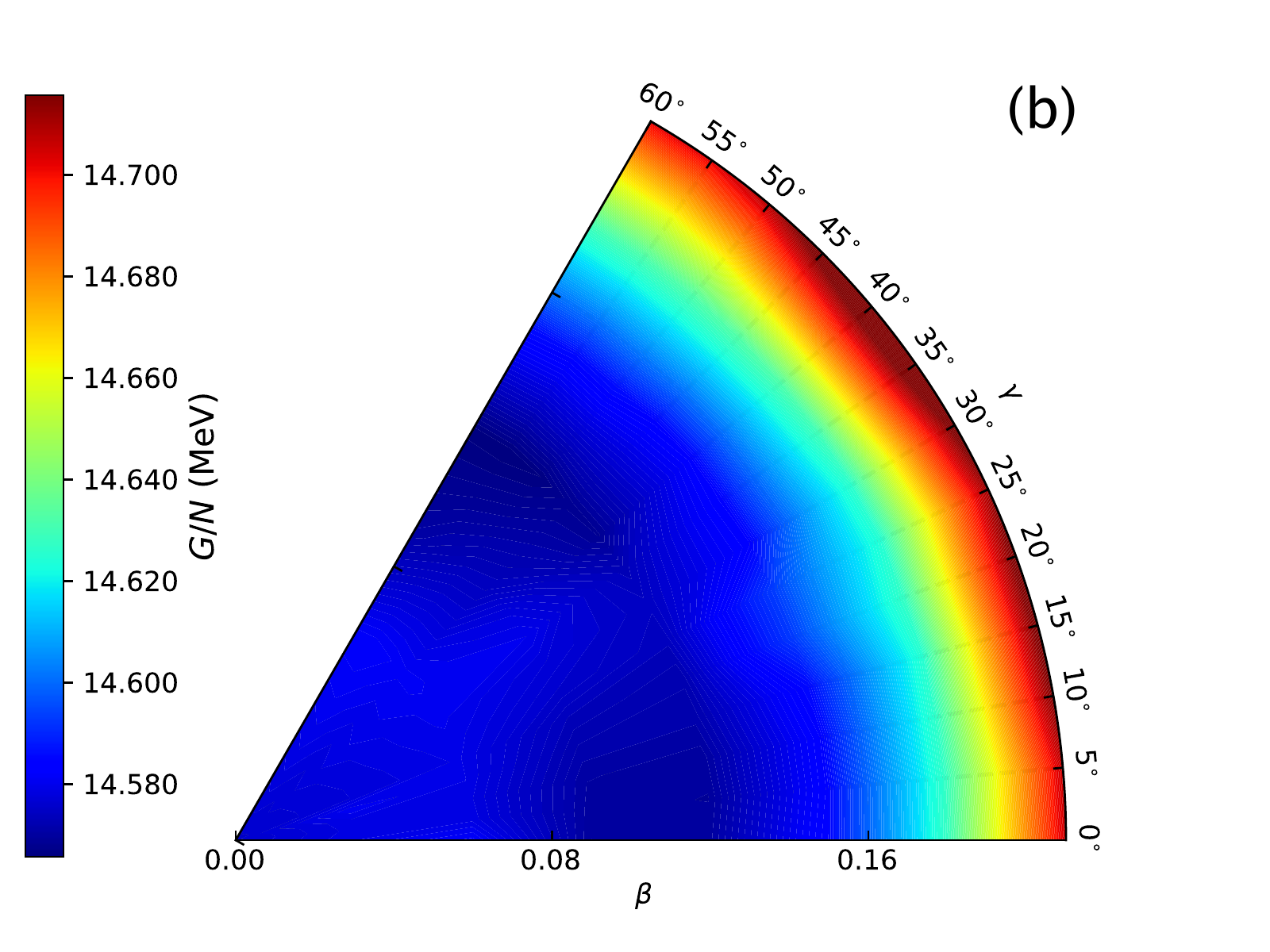}\includegraphics[scale=0.35]{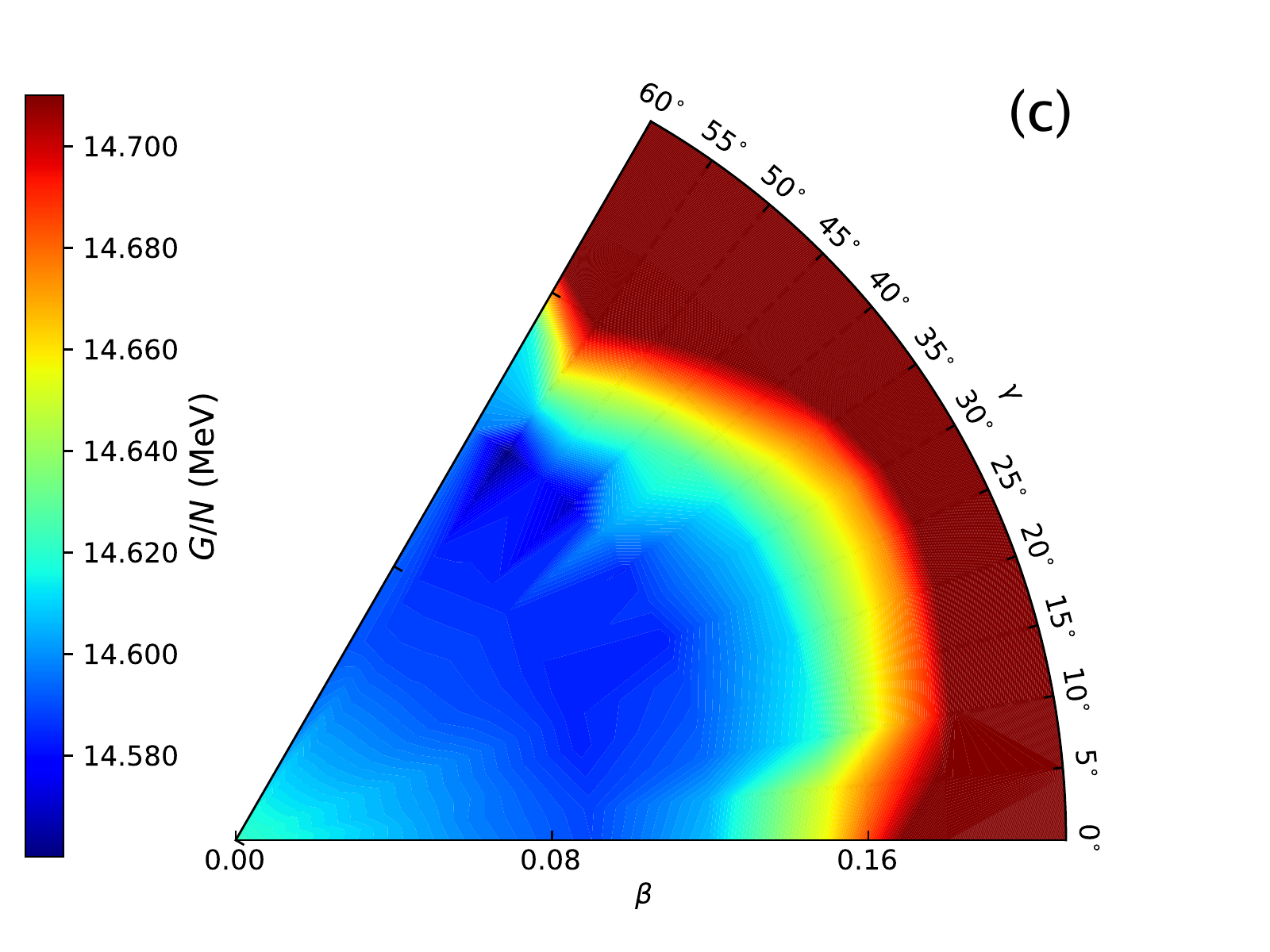}}
\centerline{\includegraphics[scale=0.47]{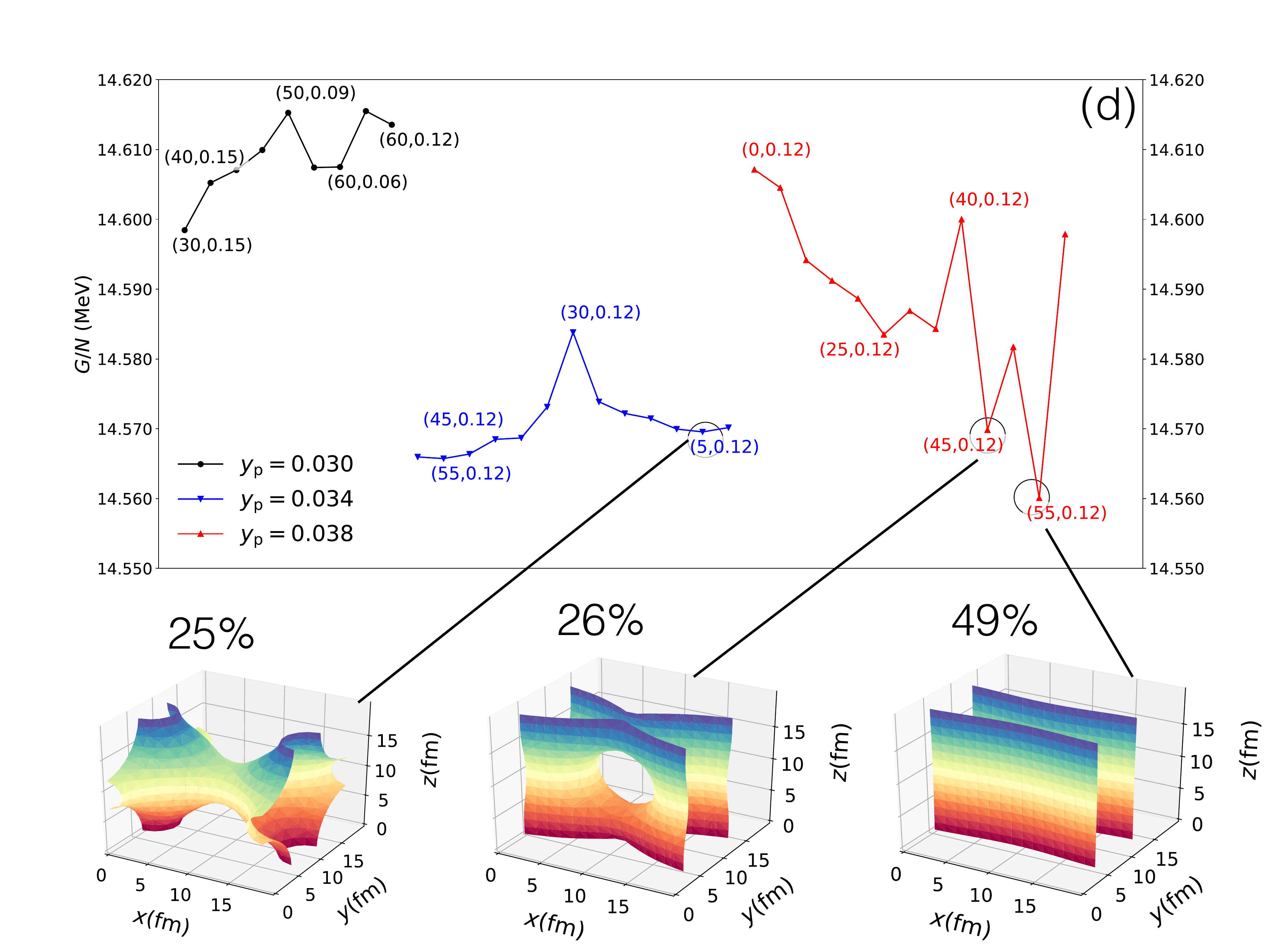}}
\caption{Top row: Gibbs free energy surfaces at a pressure of 0.32 MeV fm$^{-3}$, corresponding to a baryon density of $\approx 0.07$fm$^{-3}$, at proton fractions of 0.03 (a), 0.034 (b) and 0.038 (c), for cells containing $A_{\rm cell}$=532 nucleons. Below is the Gibbs free energy variation along one dimensional paths passing through the energy minimum. Selected $(\beta,\gamma)$ coordinates are shown along the one-dimensional plots. Visualizations of the minimum energy nuclear shapes are shown, obtained by plotting a surface of constant neutron density corresponding to the average neutron density in the cell. The minimum energy nuclear configuration is planar, followed by waffle and bi-continuous cubic-P phases. The relative abundances of the phases at a temperature equal to the fictive temperature is shown below the visualizations of the phases.} \label{fig:14}
\end{figure*}


\begin{figure*}[!t]
\centerline{\includegraphics[scale=0.35]{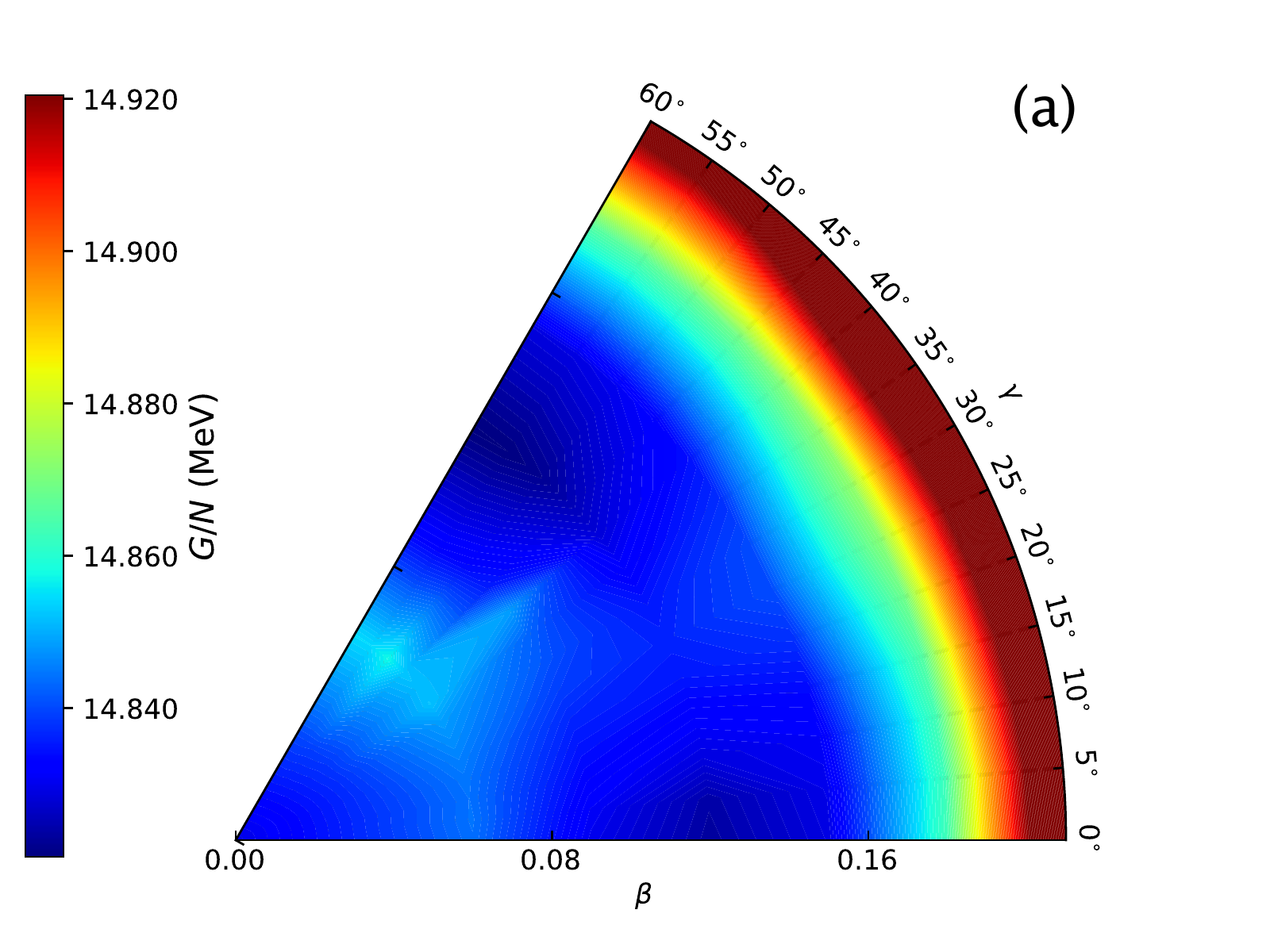}\includegraphics[scale=0.35]{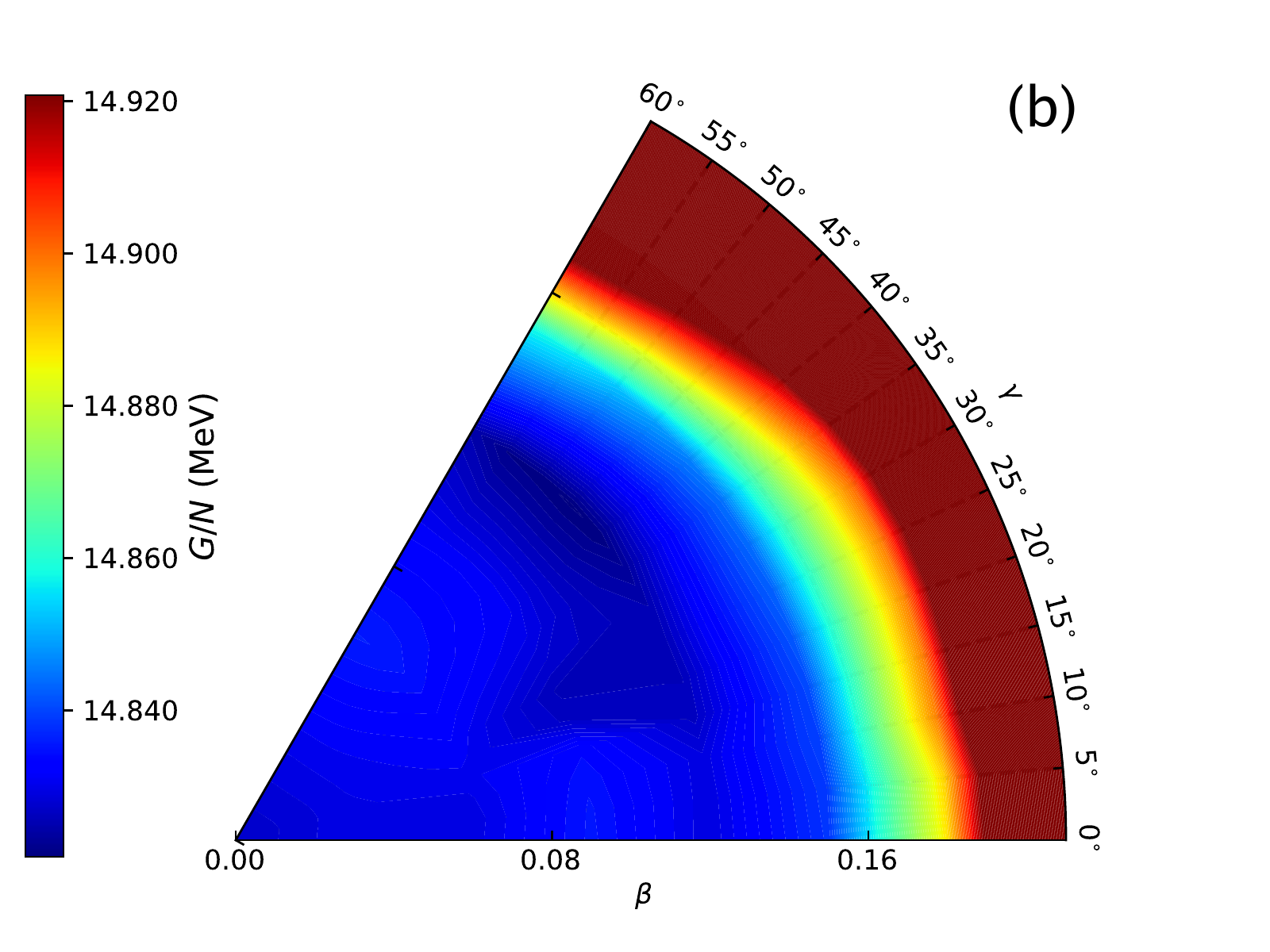}\includegraphics[scale=0.35]{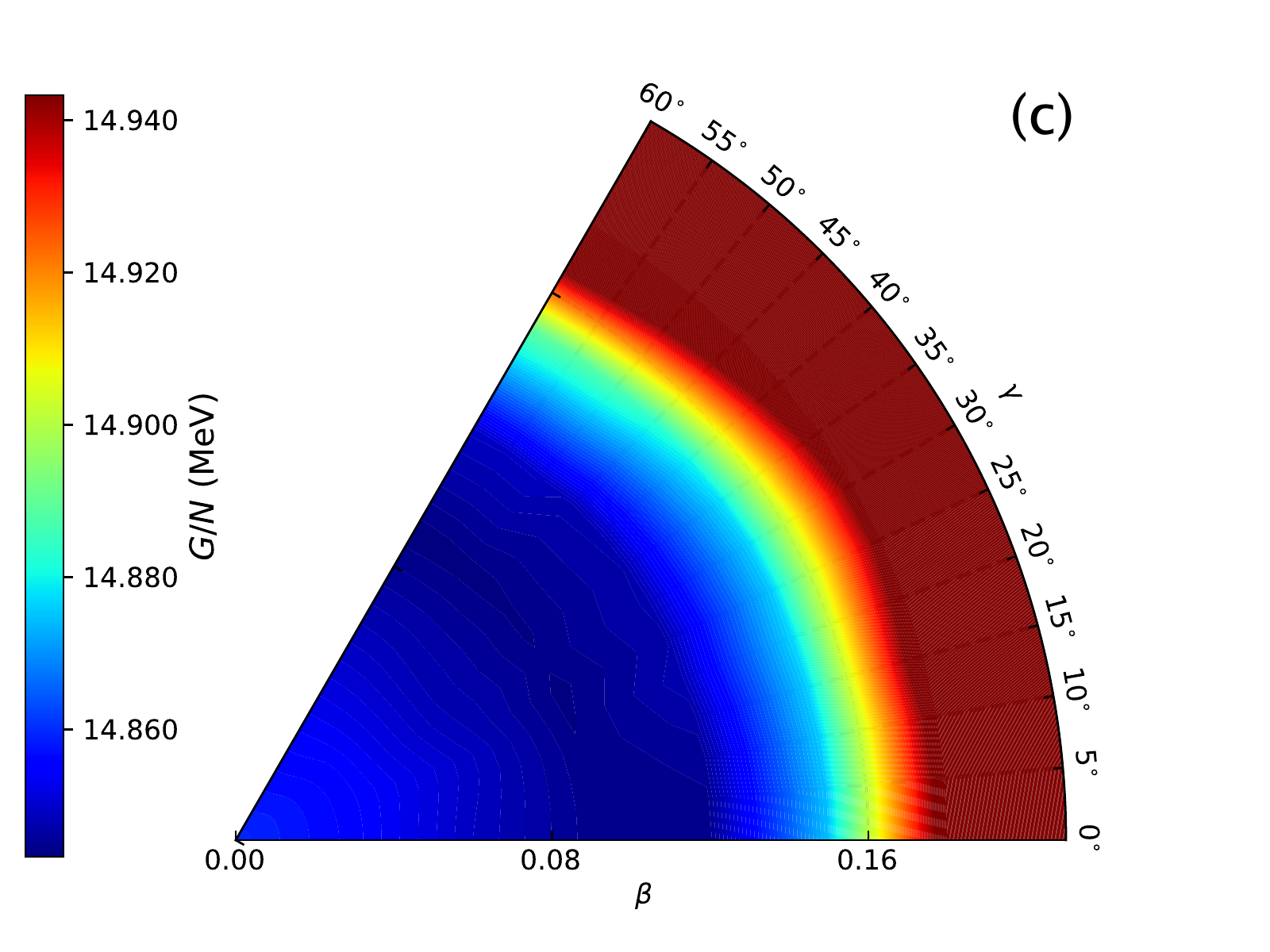}}
\centerline{\includegraphics[scale=0.47]{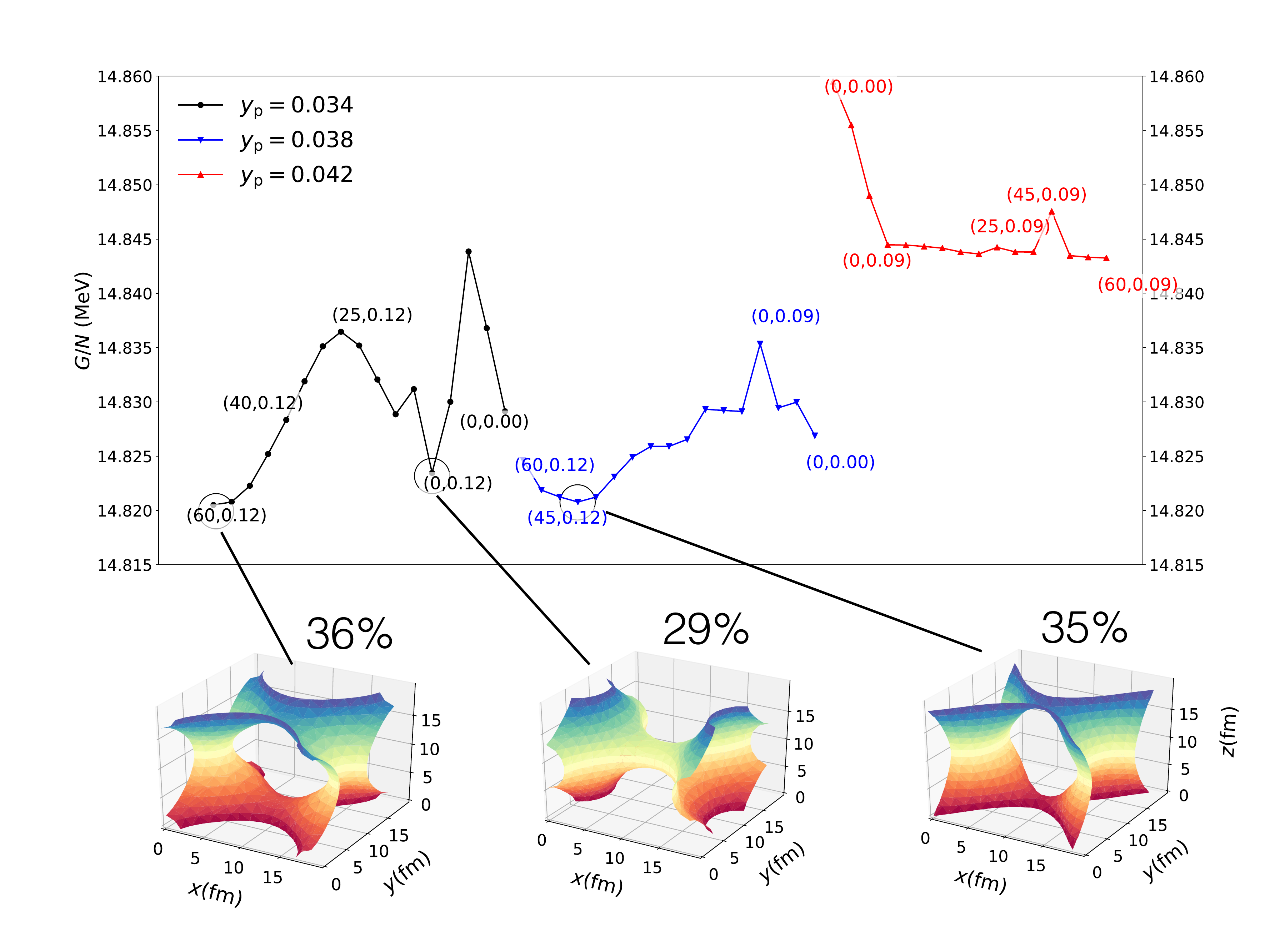}}
\caption{Top row: Gibbs free energy surfaces at a pressure of 0.34 MeV fm$^{-3}$, corresponding to a baryon density of $\approx 0.076$fm$^{-3}$, at proton fractions of 0.034 (a), 0.038 (b) and 0.042 (c), for cells containing $A_{\rm cell}$=532 nucleons. Below is the Gibbs free energy variation along one dimensional paths passing through the energy minimum (d). Selected $(\beta,\gamma)$ coordinates are shown along the one-dimensional plots. Visualizations of the minimum energy nuclear shapes are shown, obtained by plotting a surface of constant neutron density corresponding to the average neutron density in the cell. All three minima are variations of the bi-continuous cubic-P phase, albeit at different proton fractions. The relative abundances of the phases at a temperature equal to the fictive temperature is shown below the visualizations of the phases.} \label{fig:15}
\end{figure*}

\begin{figure*}[!t]
\centerline{\includegraphics[scale=0.35]{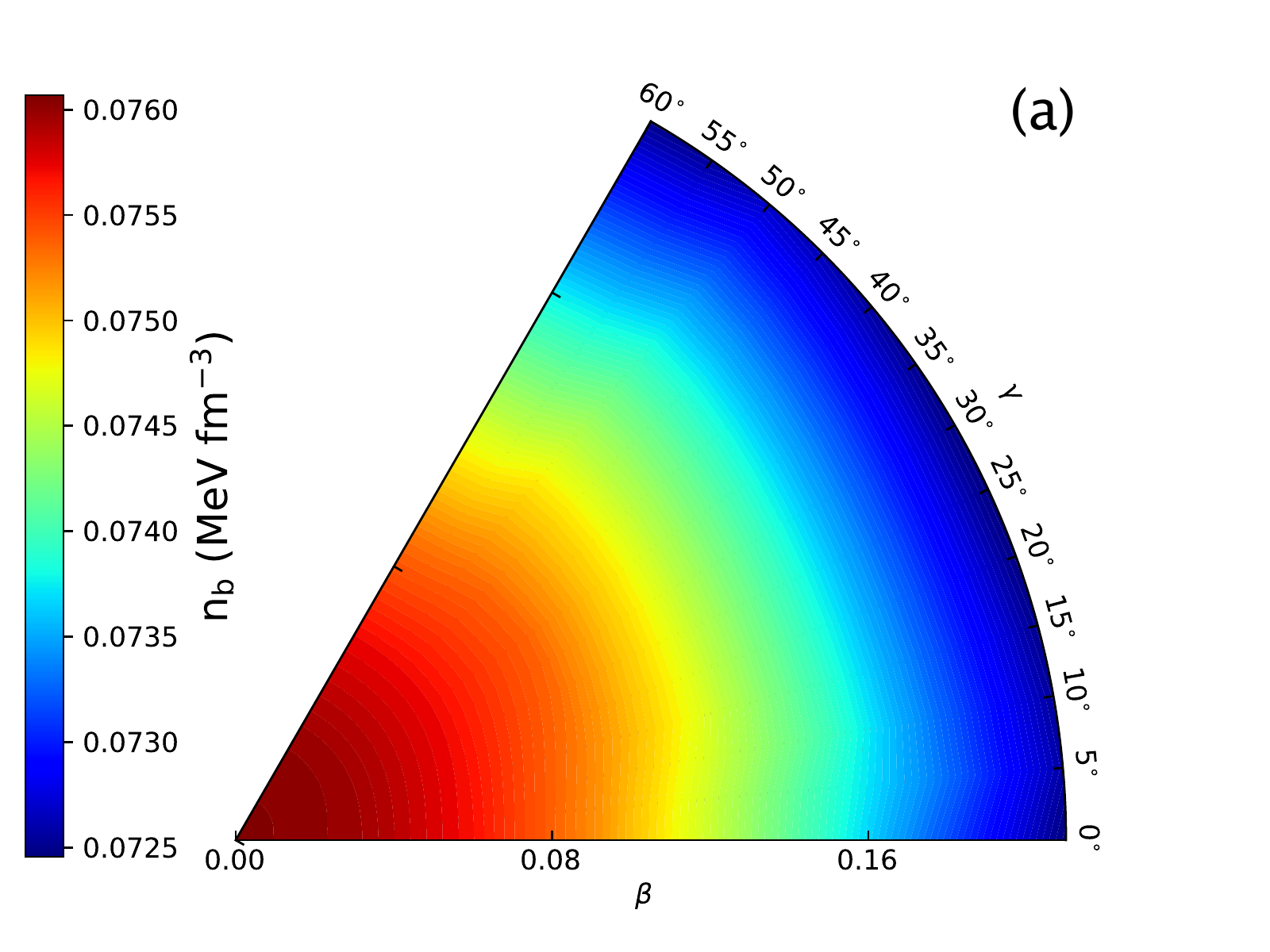}\includegraphics[scale=0.35]{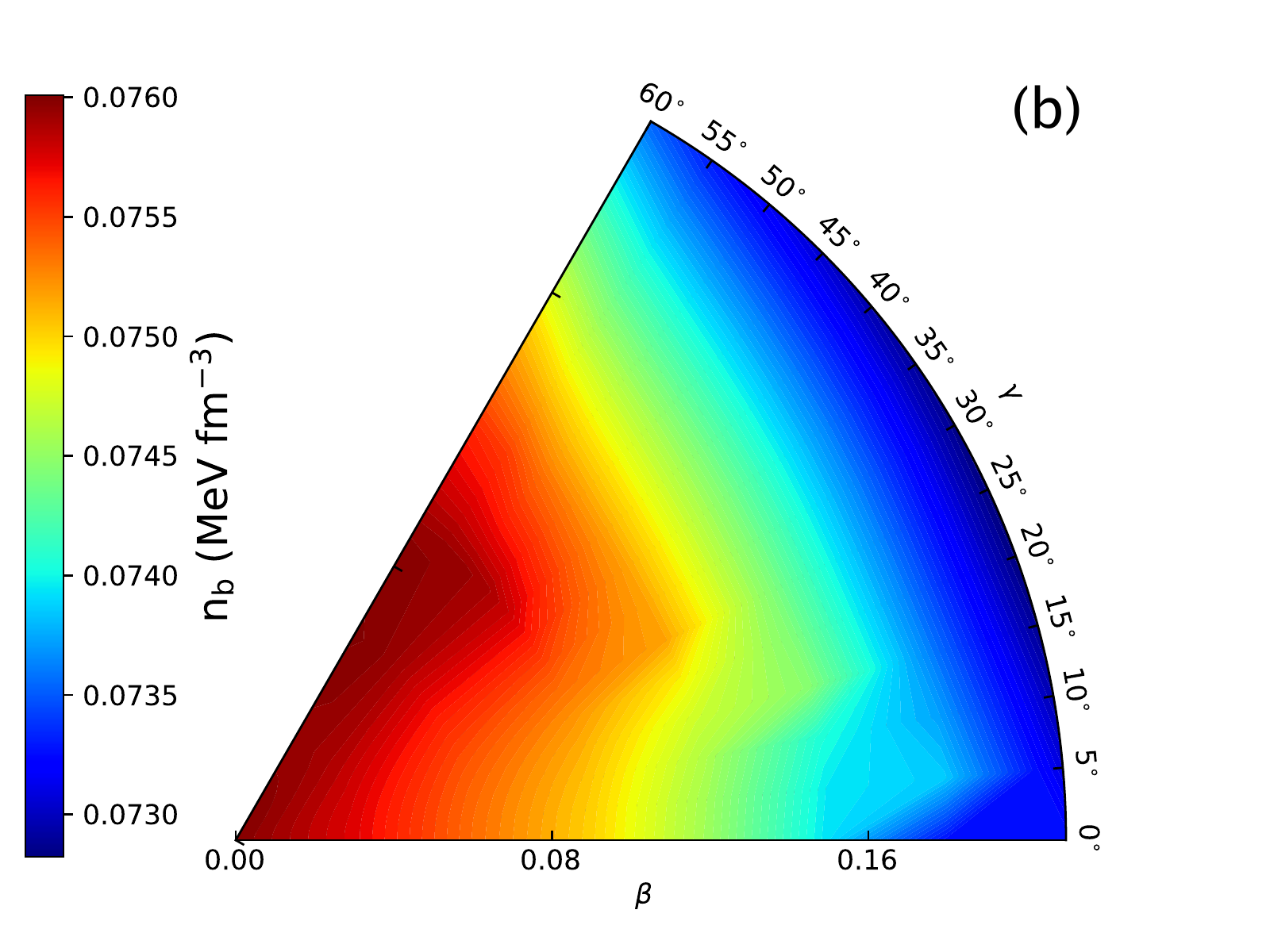}\includegraphics[scale=0.35]{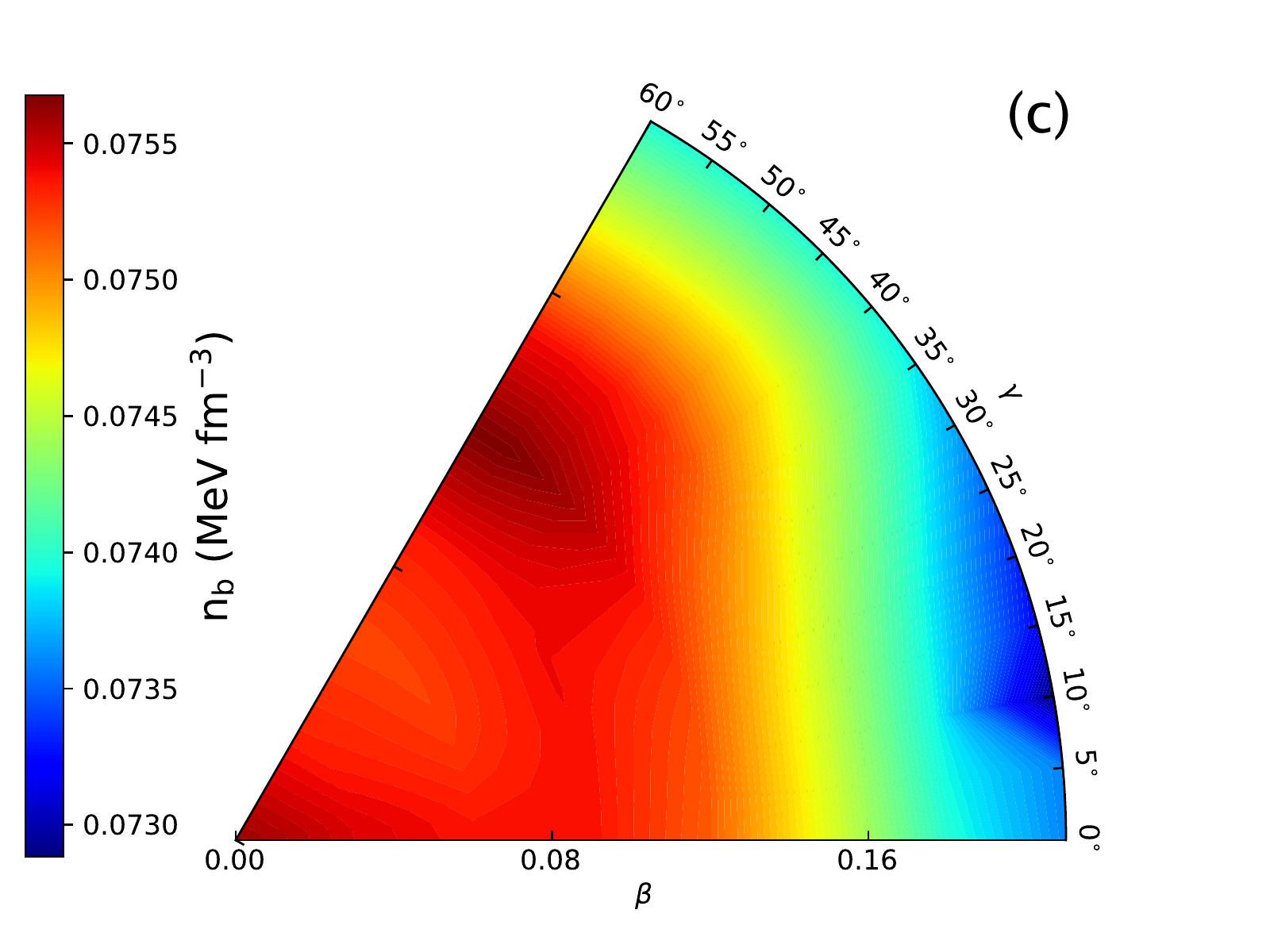}}
\caption{Average baryon density surfaces (bottom) at a constant pressure of $0.524$ MeV fm$^{-3}$, corresponding to a baryon density of $\approx 0.076$fm$^{-3}$, for cells containing $A_{\rm cell}$=532 nucleons. Results are shown for proton fraction of 0.034 (a), 0.038 (b) and 0.042 (c). The local minima differ in baron density by of order 5\%.} \label{fig:16}
\end{figure*}


\begin{figure*}[!t]
\centerline{\includegraphics[scale=0.35]{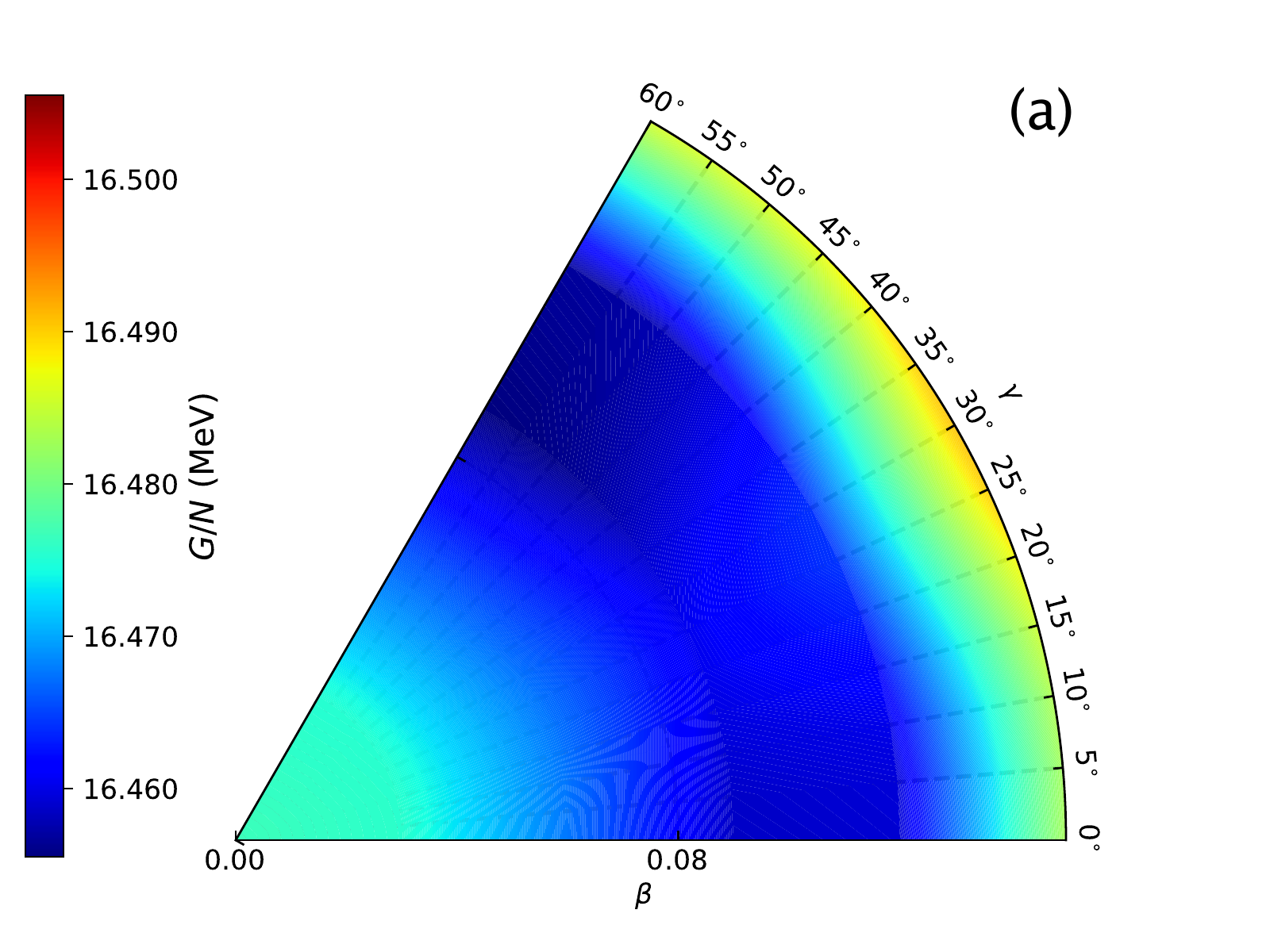}\includegraphics[scale=0.35]{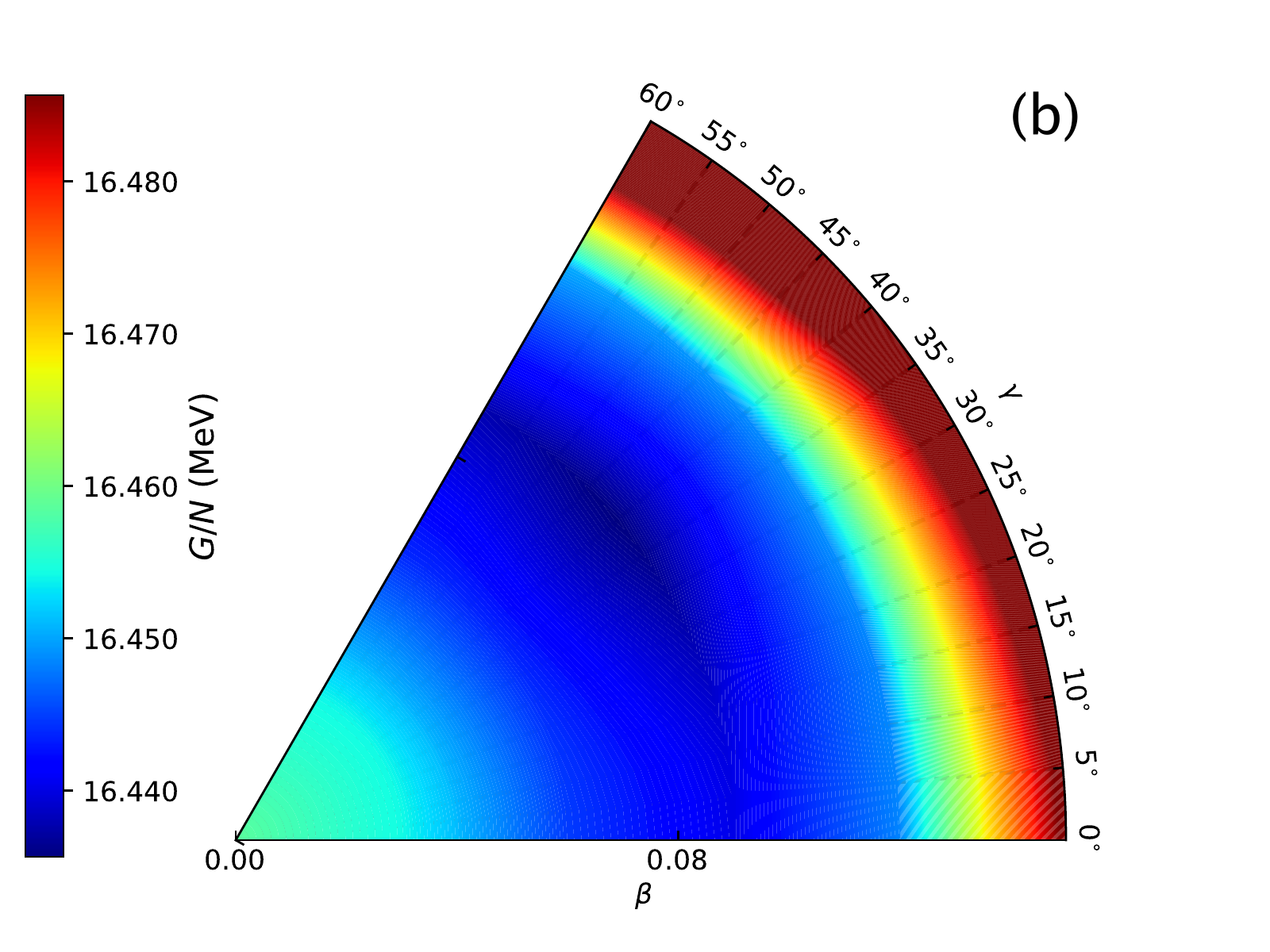}\includegraphics[scale=0.35]{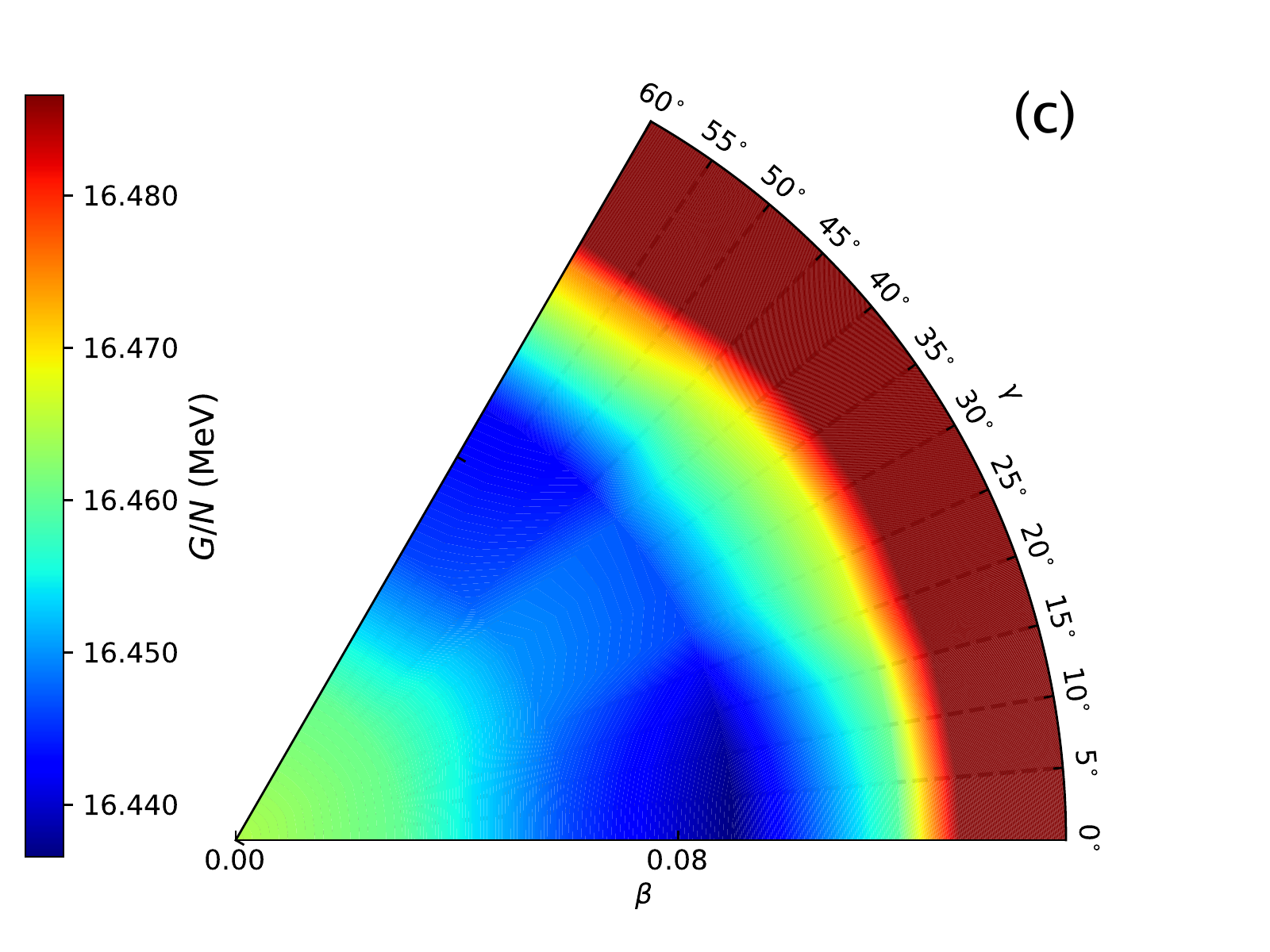}}
\centerline{\includegraphics[scale=0.47]{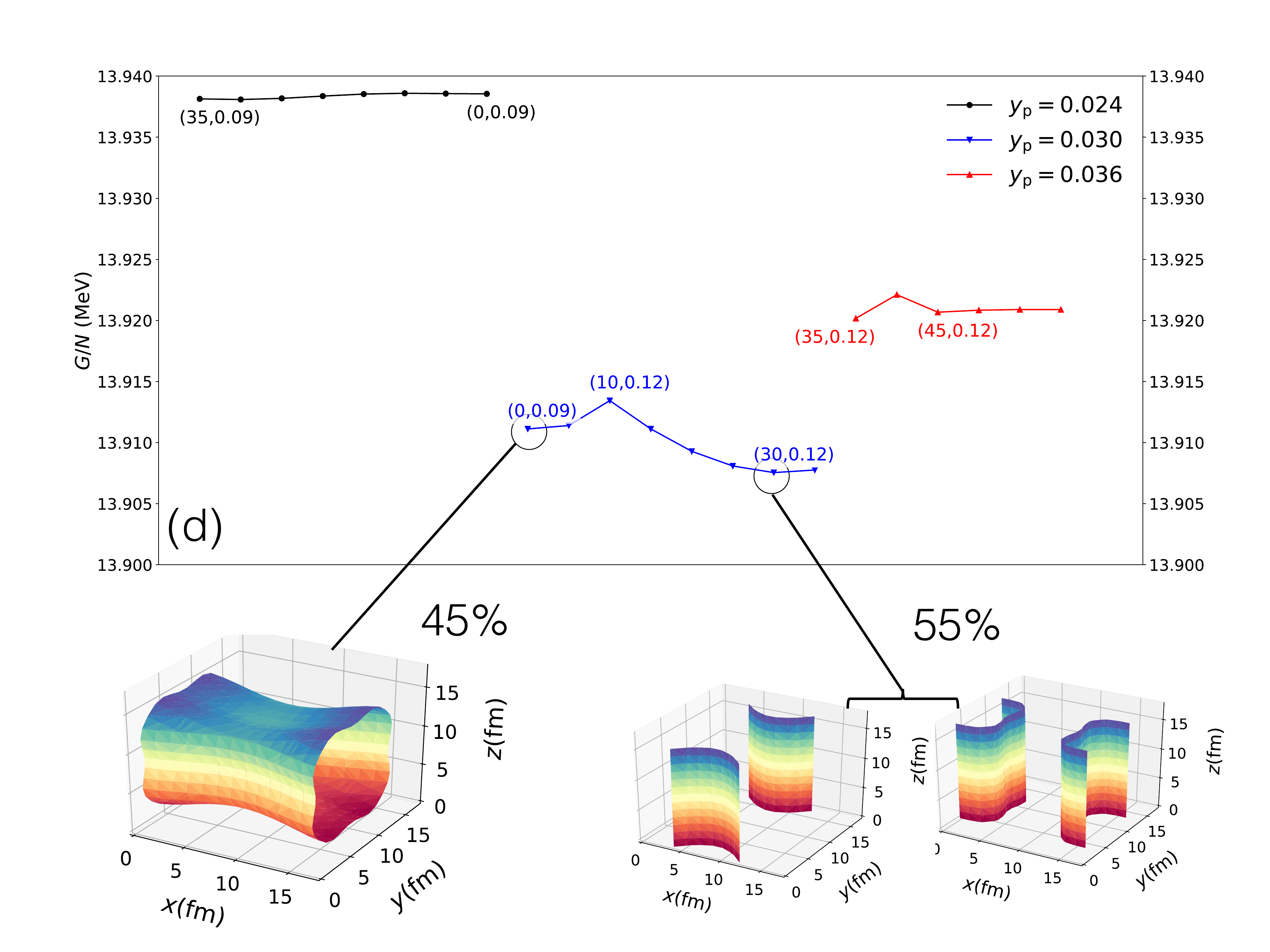}}
\caption{Top row: Gibbs free energy surfaces at a pressure of 0.45 MeV fm$^{-3}$, corresponding to a baryon density of $\approx 0.082$fm$^{-3}$, at proton fractions of 0.024 (a), 0.030 (b) and 0.036 (c), for cells containing $A_{\rm cell}$=332 nucleons. Below is the Gibbs free energy variation along one dimensional paths passing through the energy minimum (d). Selected $(\beta,\gamma)$ coordinates are shown along the one-dimensional plots. Visualizations of the minimum energy nuclear shapes are shown, obtained by plotting a surface of constant neutron density corresponding to the average neutron density in the cell. The first minimum is a combination of both a cylindrical high density and low density region (spaghetti and anti-spaghetti), whereas the second minimum is cylindrical. The relative abundances of the phases at a temperature equal to the fictive temperature is shown below the visualizations of the phases. In the simplified model we take a single barrier height to be the fictive temperature. These abundances will be frozen in as the temperature drops further unless quantum tunneling processes anneal the matter.} \label{fig:17}
\end{figure*}

\begin{figure*}[!t]
\centerline{\includegraphics[scale=0.3]{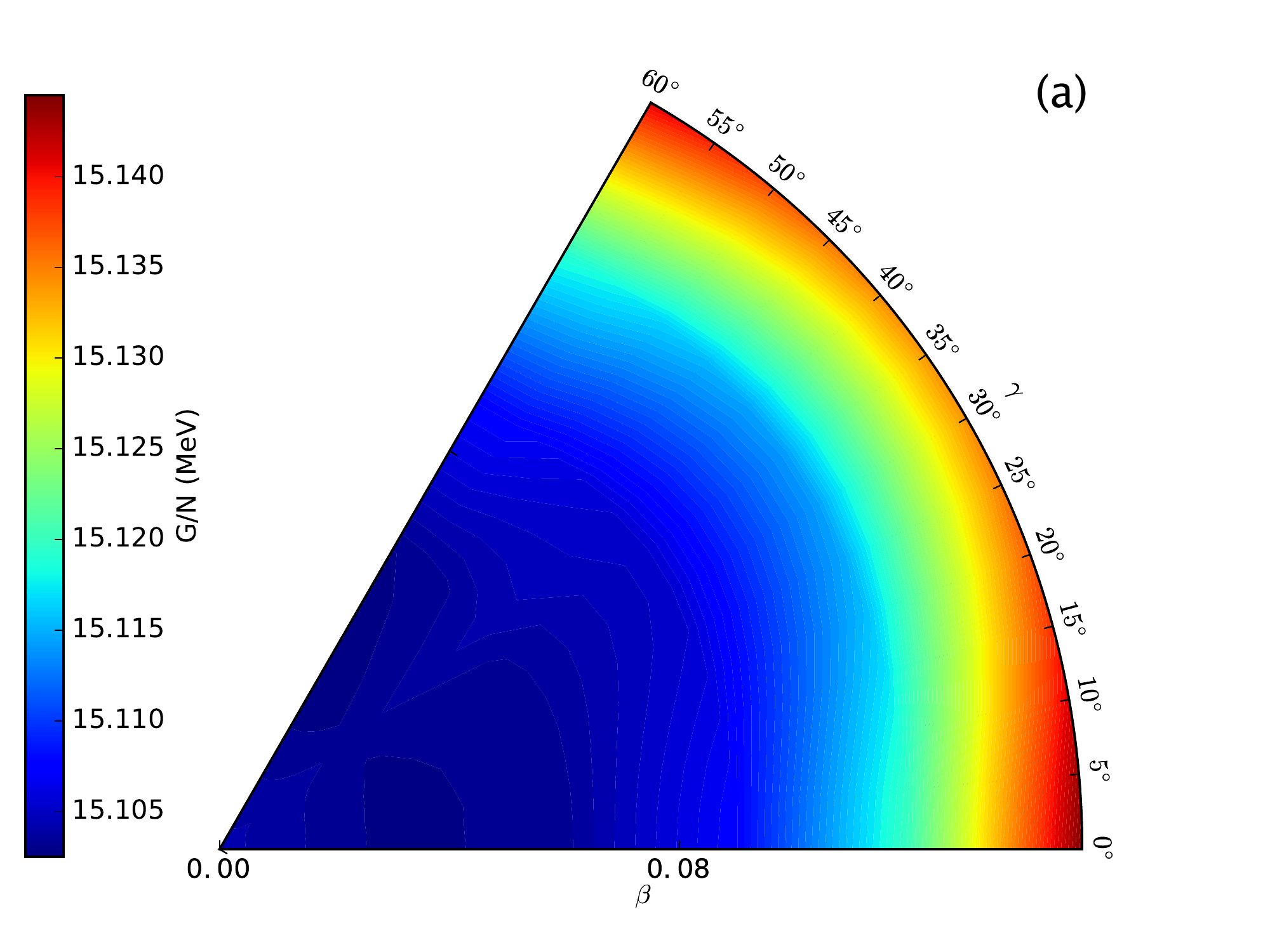}\includegraphics[scale=0.3]{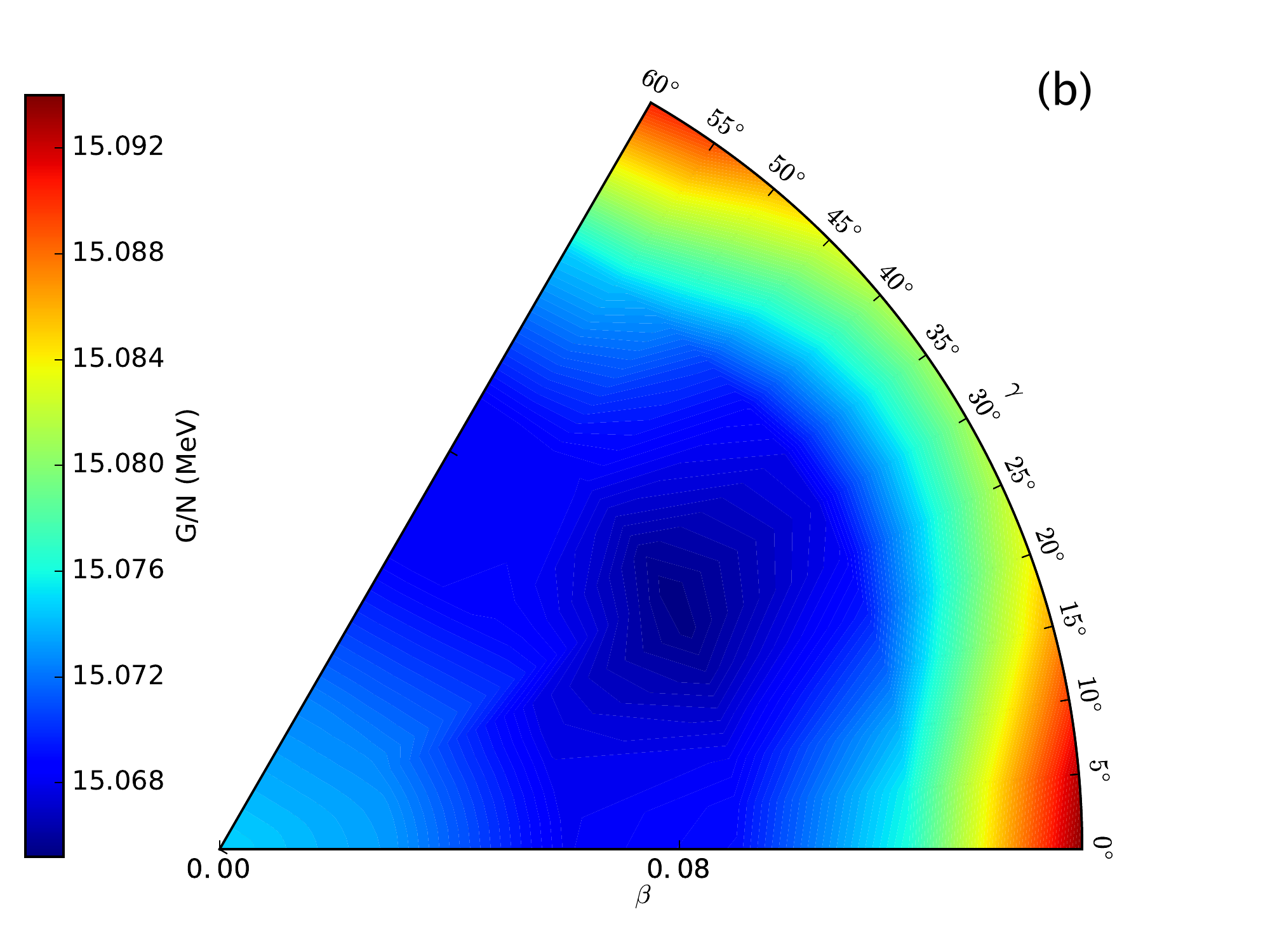}\includegraphics[scale=0.3]{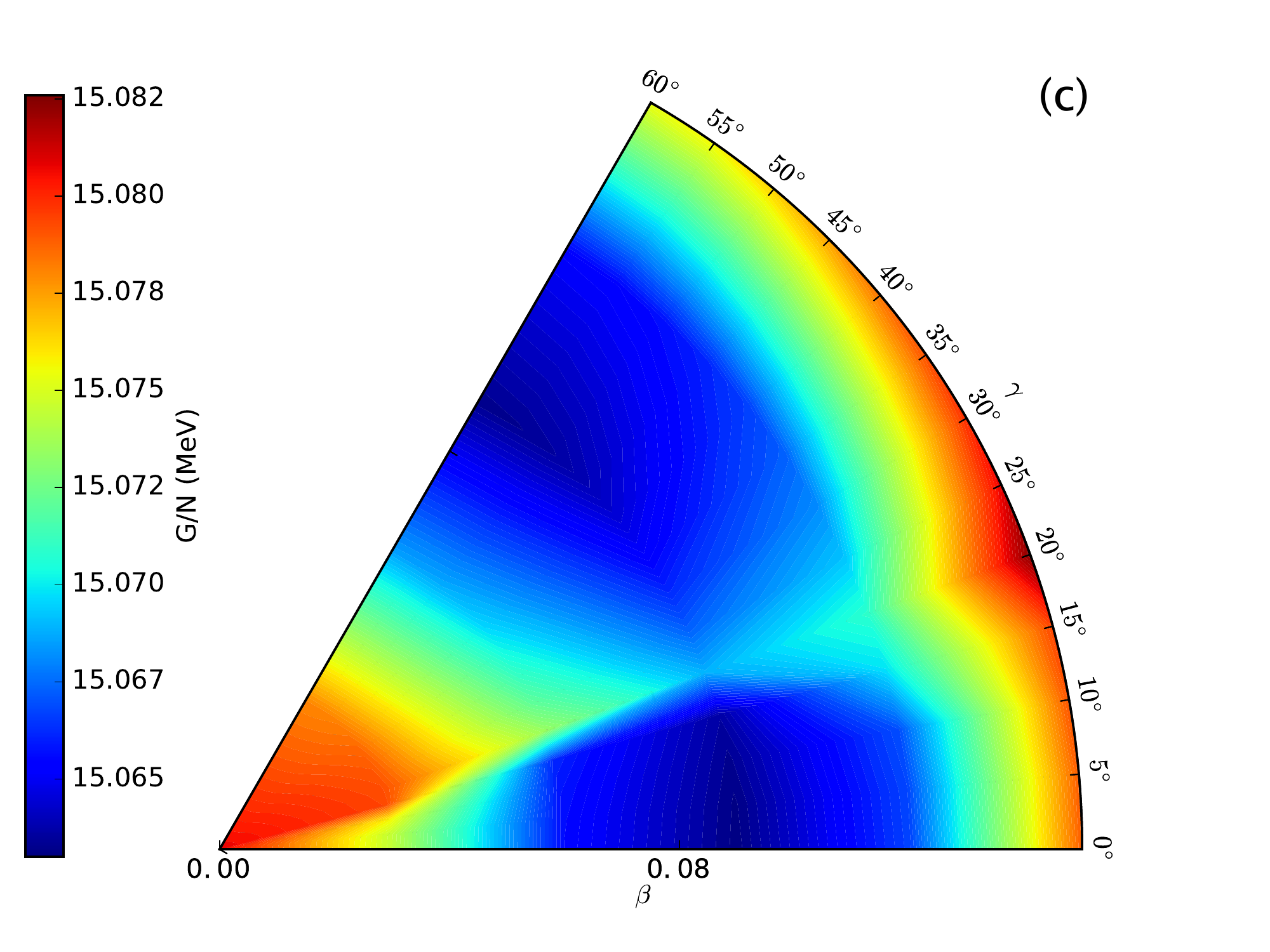}}
\centerline{\includegraphics[scale=0.47]{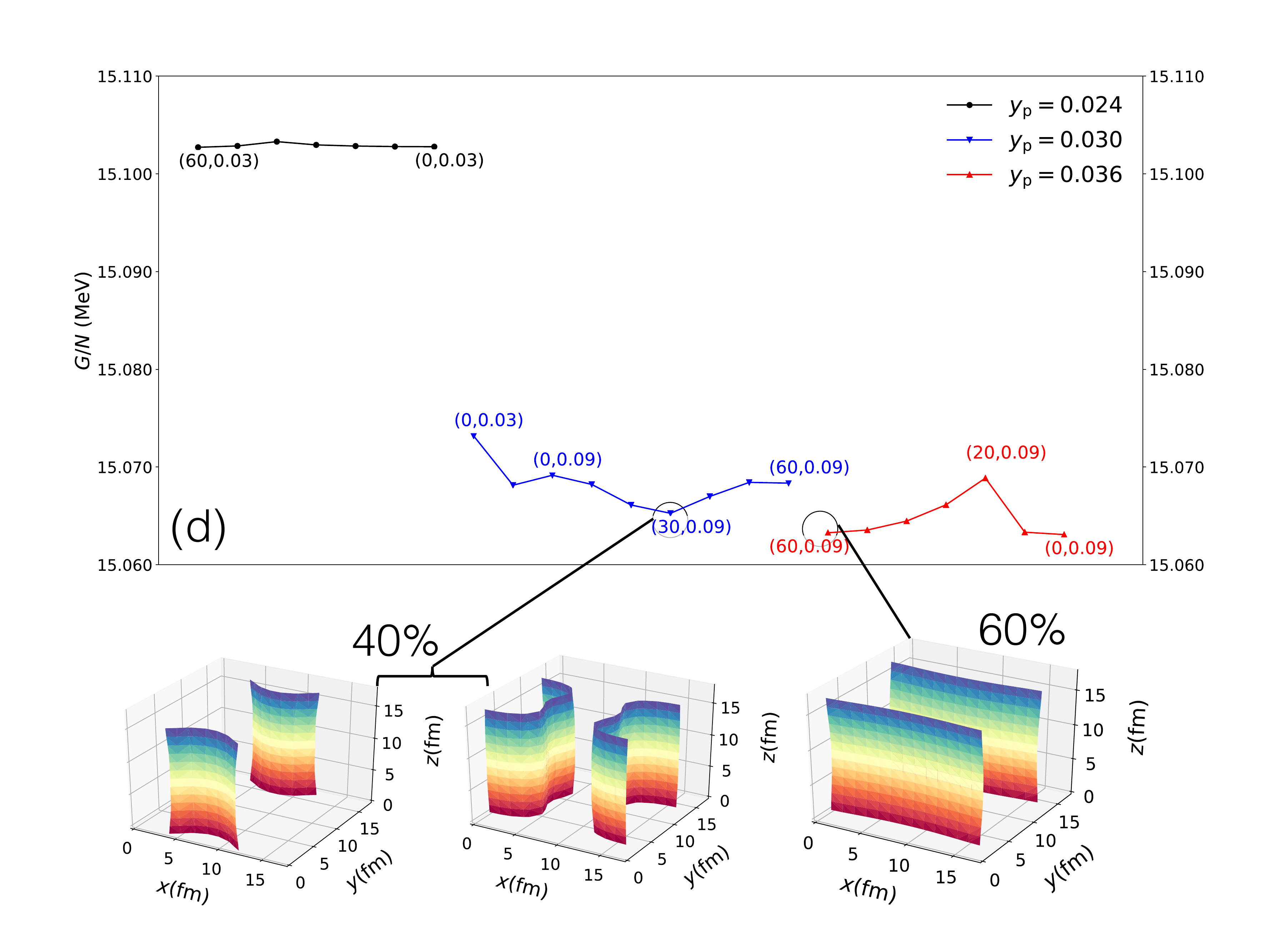}}
\caption{Top row: Gibbs free energy surfaces at a pressure of 0.53 MeV fm$^{-3}$, corresponding to a baryon density of $\approx 0.088$fm$^{-3}$, at proton fractions of 0.024 (a), 0.030 (b) and 0.036 (c), for cells containing $A_{\rm cell}$=332 nucleons. Below is the Gibbs free energy variation along one dimensional paths passing through the energy minimum (d). Selected $(\beta,\gamma)$ coordinates are shown along the one-dimensional plots. Visualizations of the minimum energy nuclear shapes are shown, obtained by plotting a surface of constant neutron density corresponding to the average neutron density in the cell. The first minimum is planar, and the second minimum is a combination of both a cylindrical high density and low density region (spaghetti and anti-spaghetti). The relative abundances of the phases at a temperature equal to the fictive temperature is shown below the visualizations of the phases.} \label{fig:18}
\end{figure*}

\begin{figure*}[!t]
\centerline{\includegraphics[scale=0.5]{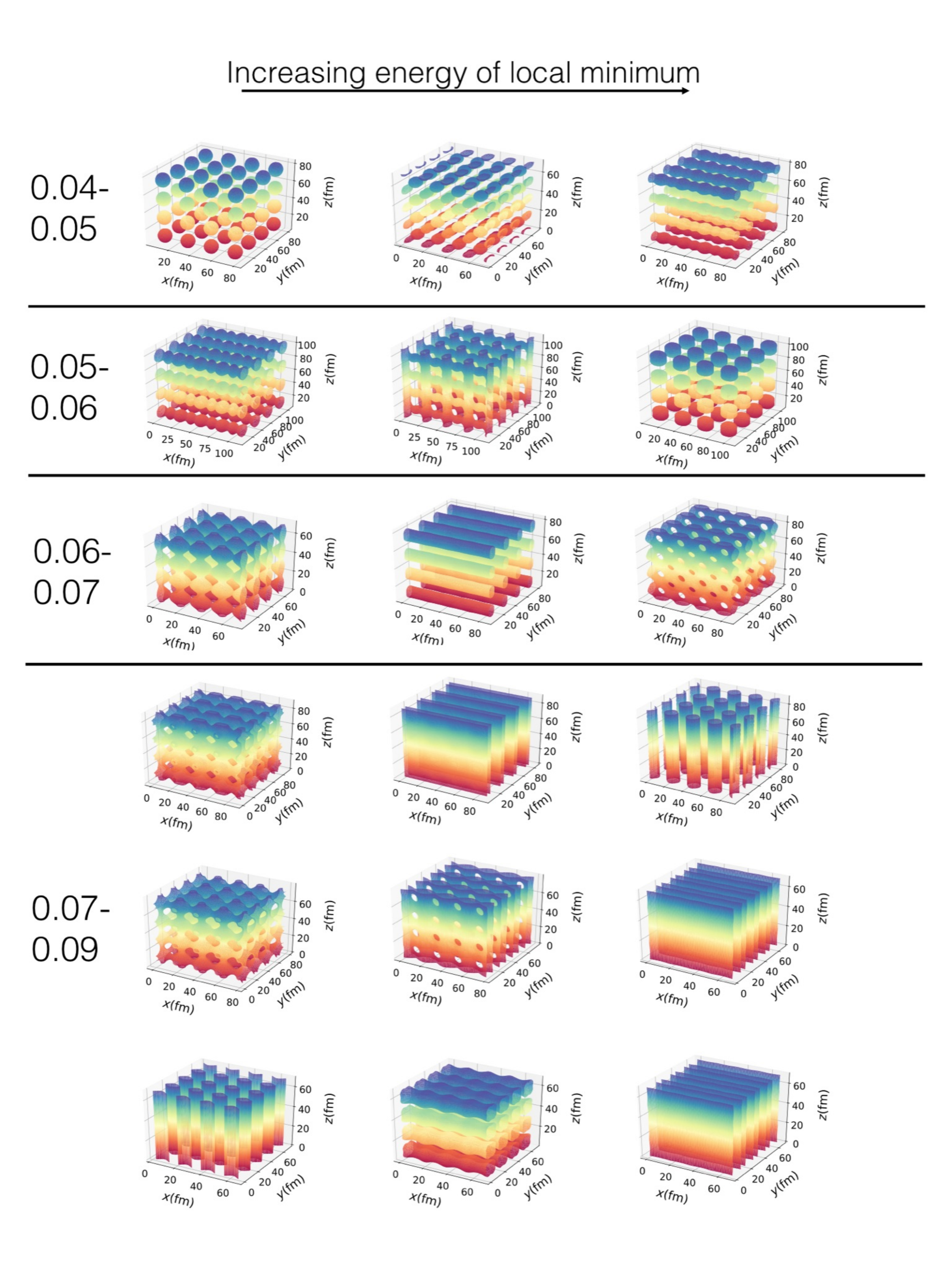}}
\caption{Overview of all nuclear geometries that emerge from our calculations. In order to visualize the phases better we have plotted a region of 4 unit cells squared, but it is important to remember we only calculate one single unit cell. The colors are just for clarity of visualization and have no physical meaning. The left column gives the approximate baryon densities the configurations to the right cover. The minimum energy configurations are the leftmost pictures in each row, with the energy of the minimum increasing rightwards. We display only a representative selection of configurations. Horizontal lines divide the graphic into the four distinct regions of pasta: from low to high density, the region where pasta first appears as a local minimum at higher energy than the isolated nuclear phase; the region where pasta shapes become the absolute minimum, but isolated nuclei remain as higher energy local minima; the region where only pasta shapes exist, and the region where only pasta shapes exists, and protons have become delocalized in all dimensions.} \label{fig:19}
\end{figure*}


\begin{figure*}[!t]
\centerline{\includegraphics[scale=0.45]{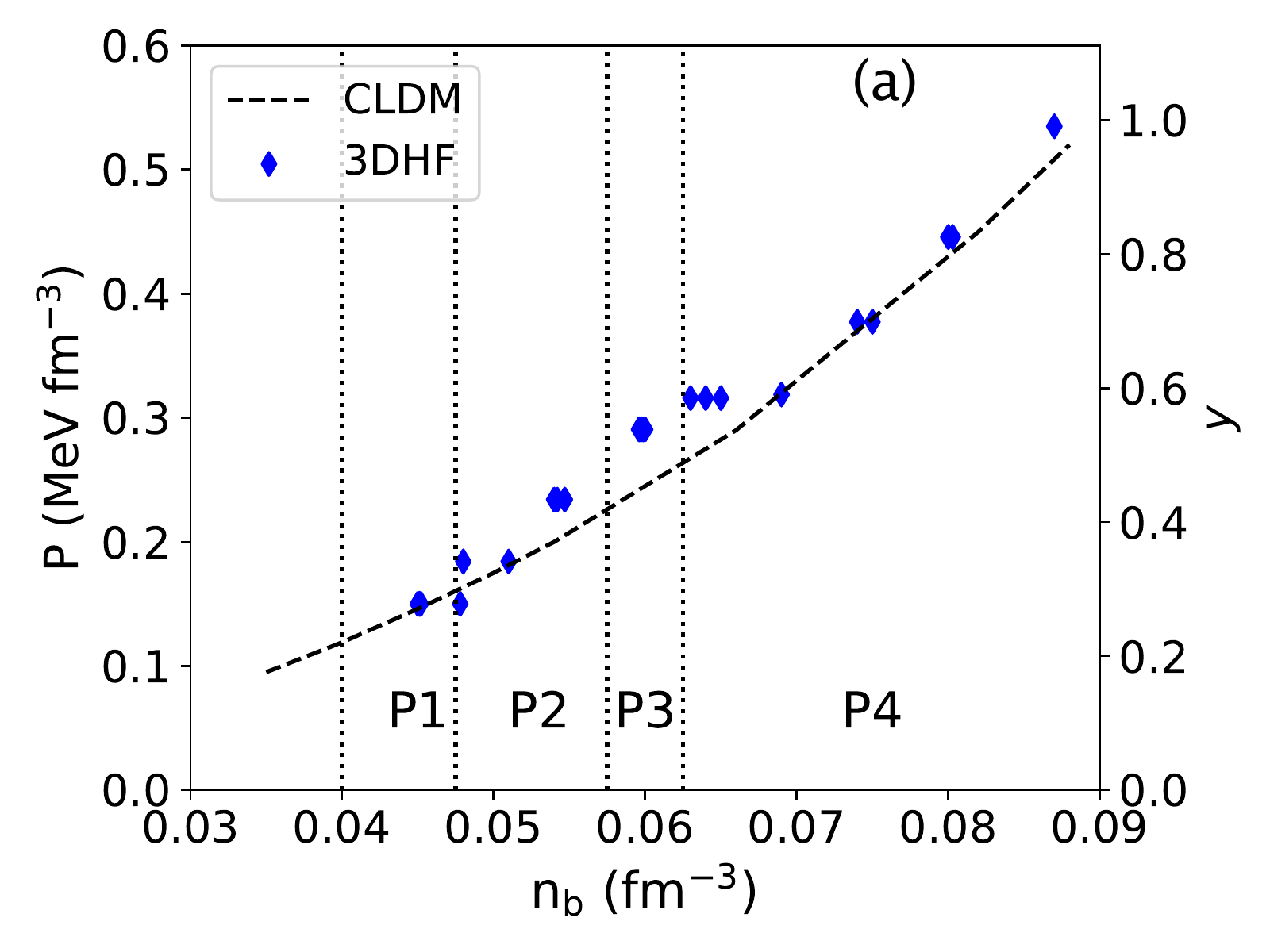}\includegraphics[scale=0.45]{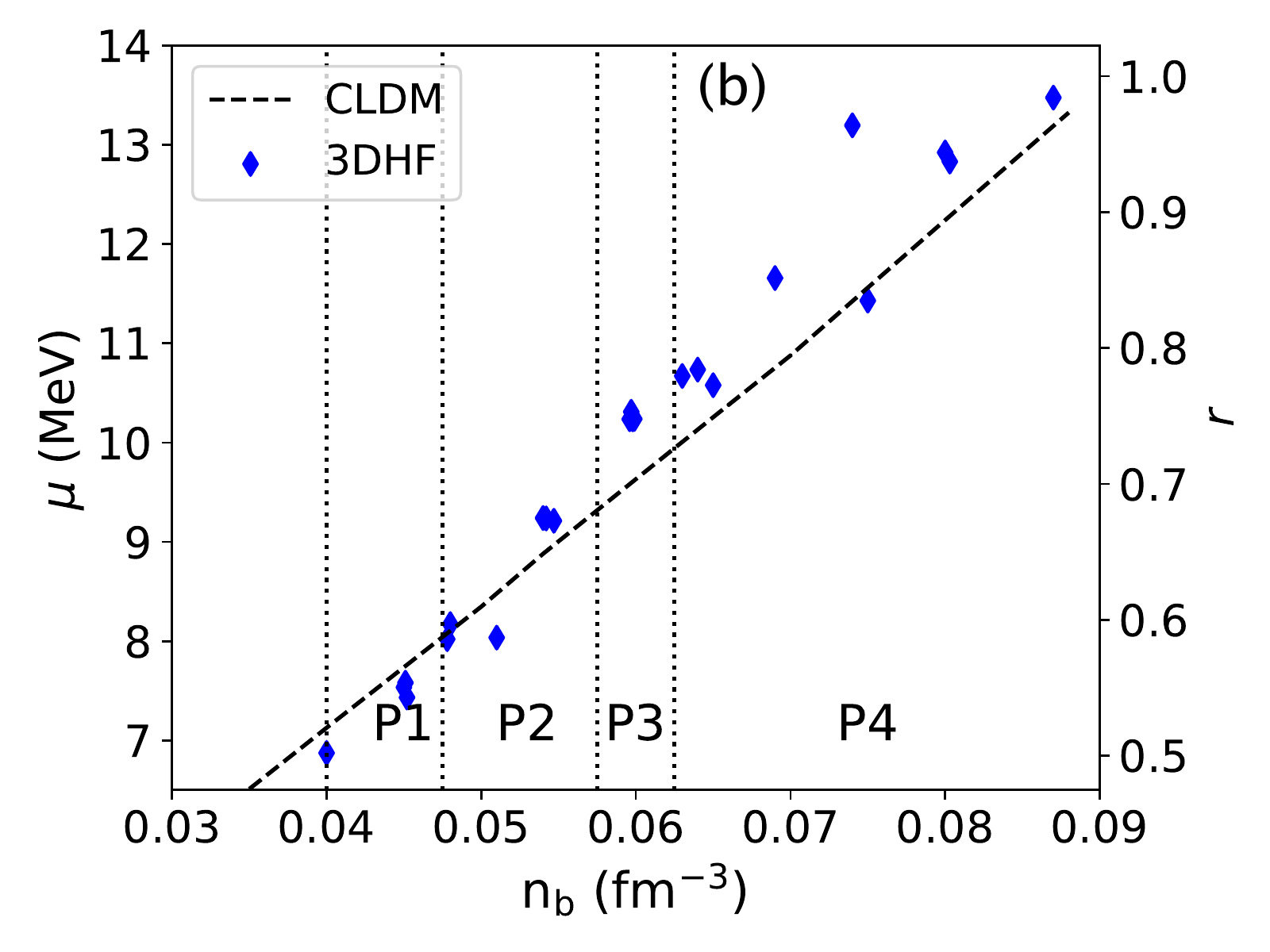}}
\caption{(a) The pressure of matter in the crust as calculated by the compressible liquid drop model (dashed line), and the pressure of each of the configurations examined in our 3DHF calculations (blue diamonds). (b) The chemical potential from the CLDM and our 3DHF calculations. Vertical dotted lines divide the four regions of pasta P1-P4 as described in the text. The mass coordinate $y$ calculated from the CLDM pressure is given in the right vertical scale in (a), and the radial coordinate $r$ calculated from the CLDM chemical potential is given in the right vertical scale in (b).} \label{fig:20}
\end{figure*}

\begin{figure}
\centerline{\includegraphics[scale=0.45]{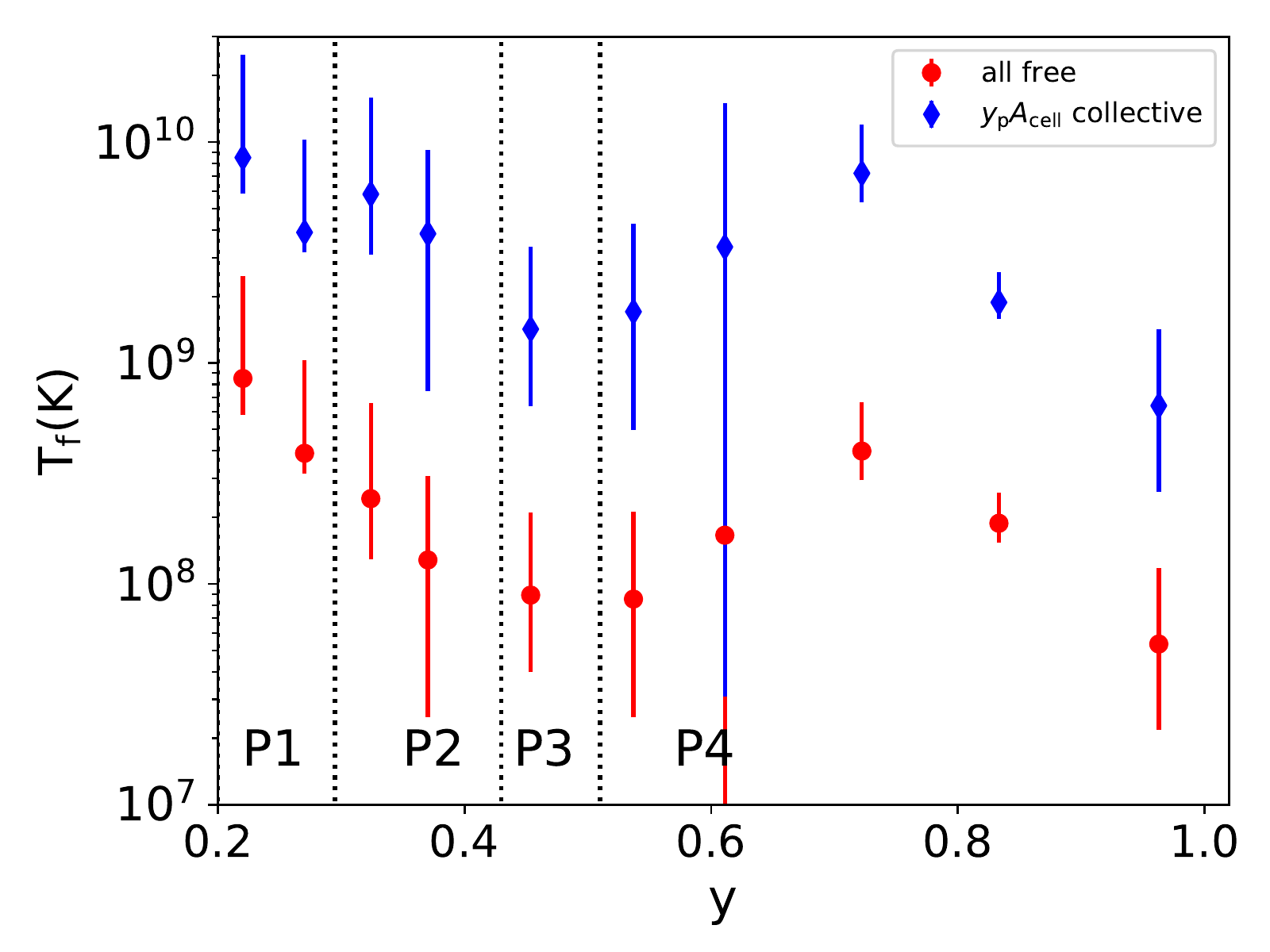}\hspace{1cm}}
\caption{The fictive temperature $T_{\rm f}$ as a function of mass coordinate in the crust $y$ assuming that the relevant degrees of freedom are individual nucleons (red points), or a collection of nucleons of order the proton number $y_{\rm p} A_{\rm cell}$ (blue diamonds). The bars span the range of of barrier heights from our calculations. Vertical dotted lines divide the four regions of pasta P1-P4 as described in the text.} \label{fig:21}
\end{figure}

\begin{figure*}[!t]
\centerline{\includegraphics[scale=0.45]{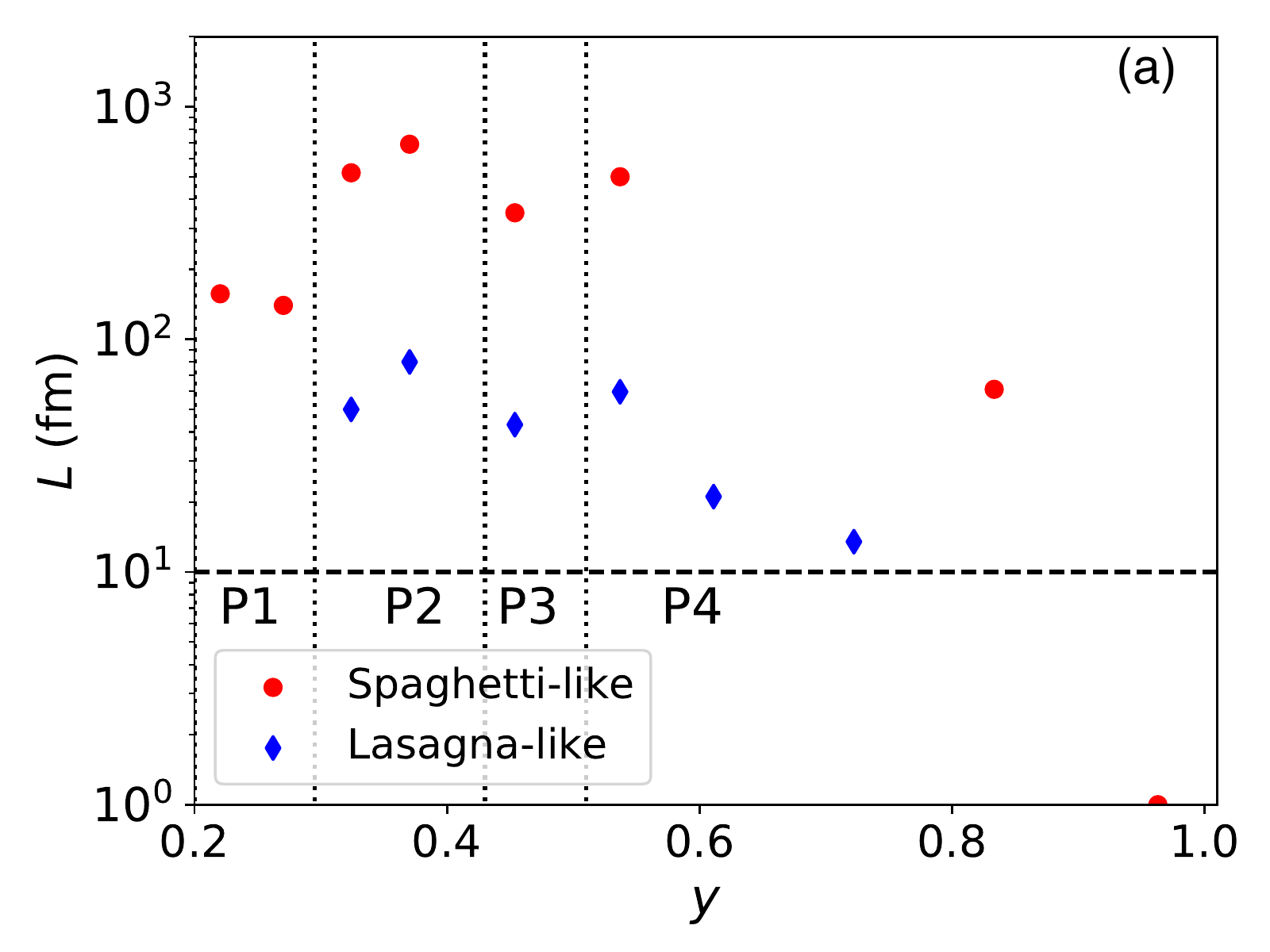}\includegraphics[scale=0.45]{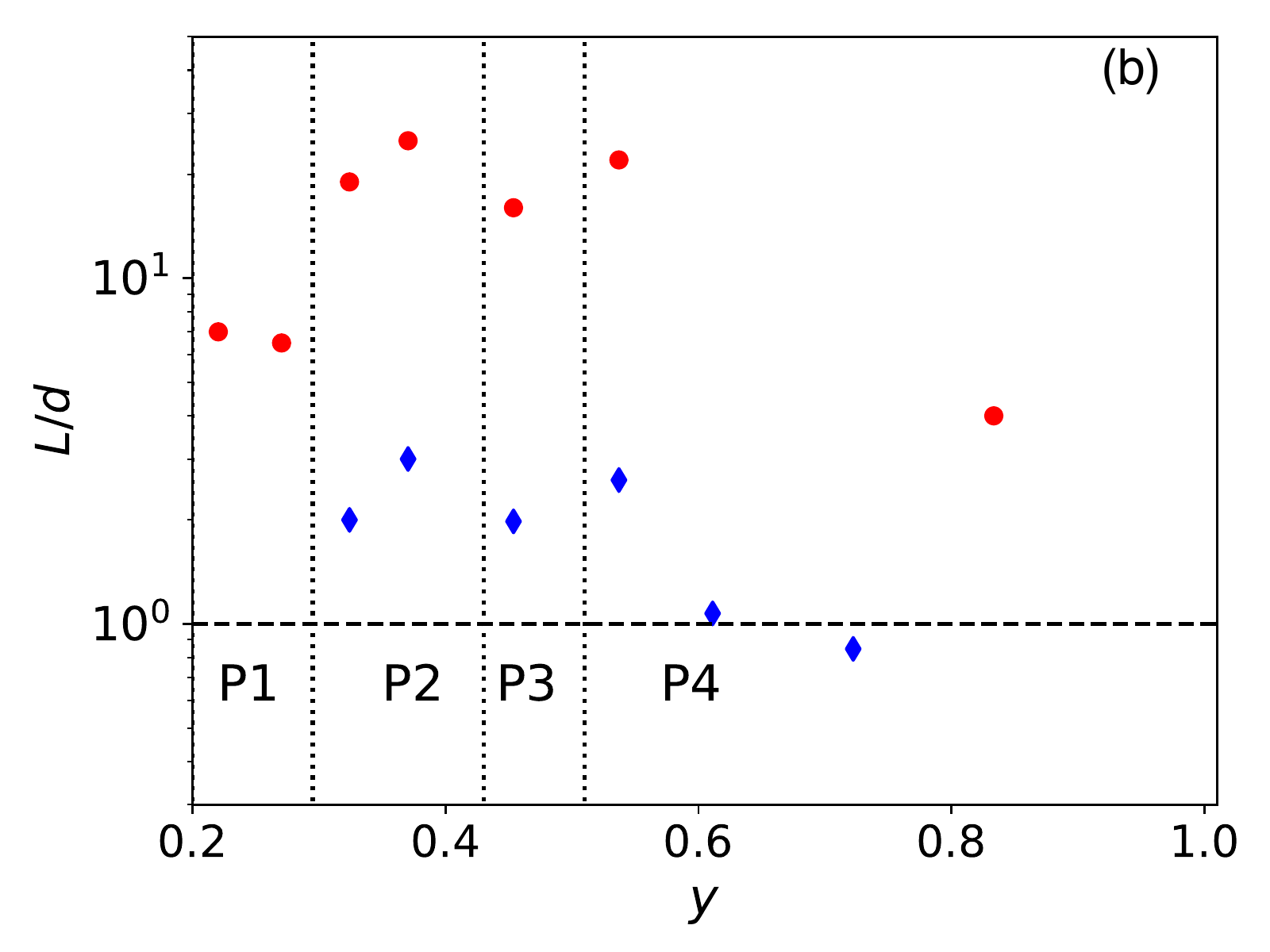}}
\caption{The long range order of the pasta phases $\Delta L$ calculated at the fictive temperature, assuming individual nucleons are the relevant degrees of freedom, as a function of mass parameter. We give the length scale in absolute terms (a) and in units of cell size (b). The horizontal dashed lines indicate the point where the order drops below 10fm - approximately the characteristic width of nuclear clusters (a) - and below one cell size (b). Spaghetti-like configurations are shown as red pints, and lasagna-like configurations are shown by blue diamonds. Below the dashed lines, matter is completely disordered. These are \emph{upper} limits on the long range order: if we assume that clusters of nucleons are the relevant degrees of freedom, the fictive temperature increases (see Figure~\ref{fig:21}) and the long range order drops. Vertical dotted lines divide the four regions of pasta P1-P4 as described in the text.} \label{fig:22}
\end{figure*}

\begin{figure}
\centerline{\includegraphics[scale=0.45]{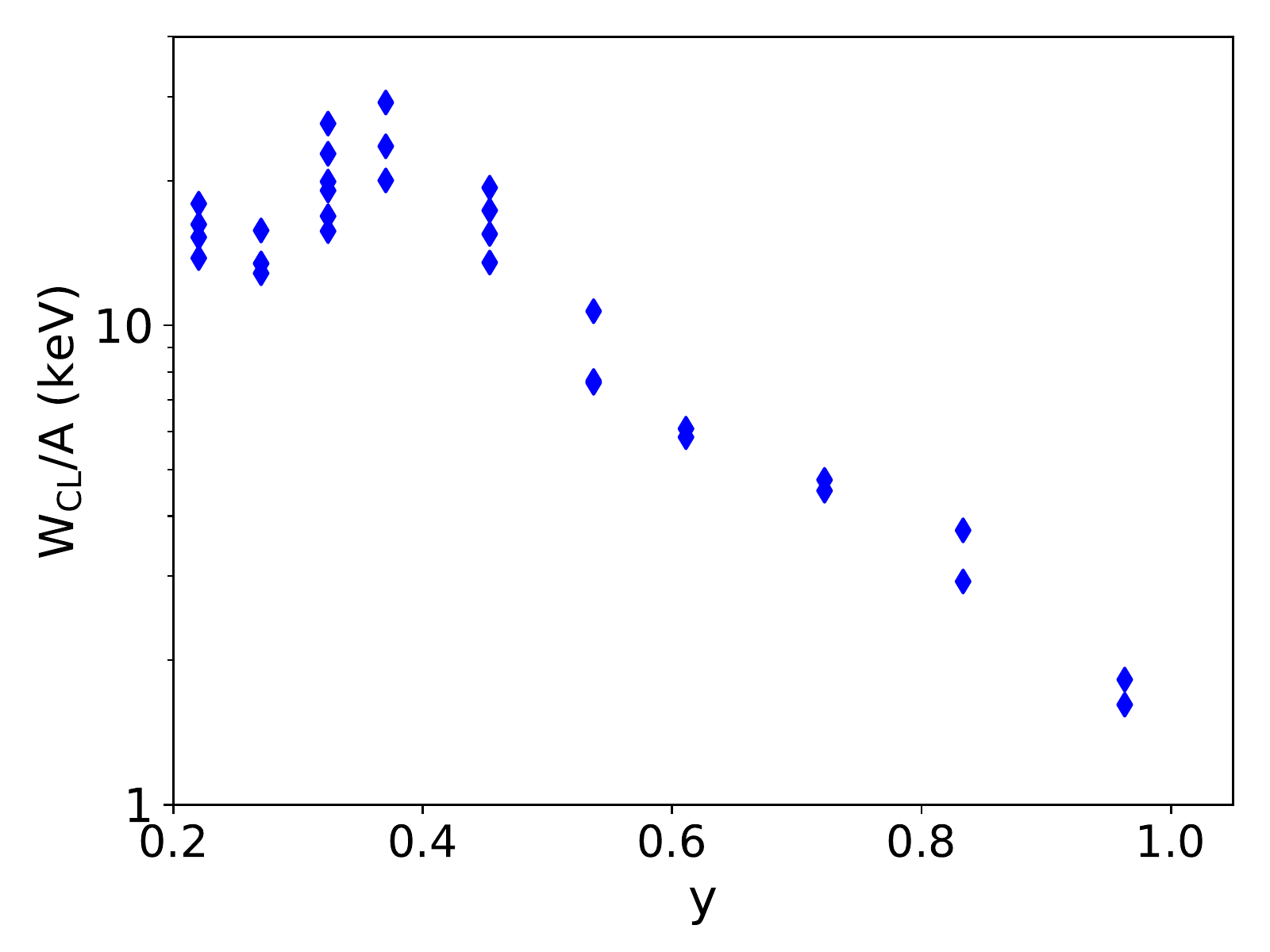}}
\caption{The lattice energy per nucleon in the unit cell for minimum energy configurations as a function of mass coordinate $y$. The lattice energy drives the stability of the phases; it peaks at around $y=0.4$ (a baryon density of $\approx 0.06$fm$^{-3}$), where isolated nuclei disappear from the minima in the energy landscape, and then decreases roughly exponentially with density thereafter.} \label{fig:23}
\end{figure}

\subsubsection{The size of domains}

The scale on which the pasta phases are ordered can be found by estimating the length $L$ over which thermal fluctuations disrupt the long-range order of spaghetti-like and lasagna-like configurations. Since the waffle phases are planar, they can be approximated here as lasagna phases. We follow the formalism laid out in \cite{Watanabe:2000aa}. Relative to the cell spacing $d=2r_{\rm c}$ where $r_{\rm c}$ is the cell radius, it is given as follows:

For the spaghetti-like structures,

\be
\bigg({L \over d}\bigg)_{\rm spaghetti} \approx \bigg[ { (B_{\rm 2d} +2C_{\rm 2d}) (\pi \lambda a)^{1/2} \over k_{\rm B} T} \bigg]^{1/2}
\ee

\noindent where

\be
B_{\rm 2d} = 1.5 w_{\rm C+L}\;\;\;\; C_{\rm 2d} \approx 10^{2.1(u-0.3)} w_{\rm C+L}
\ee 

\be
K_{\rm 3} \approx 0.0655 w_{\rm C+L} r_{\rm c}^2 \;\;\;\; a \approx 2 r_{\rm c}
\ee

For 1D lasagna-like structures,

\be
\bigg({L \over d}\bigg)_{\rm lasagna} \approx \bigg[ { 4 \pi (BK_1)^{1/2} \over k_{\rm B} T \ln({R \over 2r_{\rm c}})} \bigg]^{1/2}
\ee

\noindent where

\be
B_{\rm 1d} = 6 w_{\rm C+L} \;\;\;\;\;\;\; K_1 \approx {2 \over 15} w_{\rm C+L} (1 + 2u - 2u^2) r_{\rm c}^2
\ee

\noindent and $R$ is the typical length of the structure, $u$ the volume fraction of the pasta. $w_{\rm C+L}$ is the total electrostatic energy density, including the all important lattice contribution which, in addition to quantum shell effects, drives the stability of the phases. The electrostatic energy density $w_{\rm C+L}$ is extracted from the simulation. The factors involving $u$ change only by a factor of at most 2 over a reasonable range of $u$ (as they enter as powers of 1/4), so for simplicity and since we are interested only in a rough estimate of length scales, we take the average of that range and fold it into the numerical prefactor.

At low temperatures $<10^7$K, this overestimates the order, because quantum fluctuations become important, but our lower limits on the fictive temperature are all significantly above $10^7$K.

The electrostatic energy per nucleon is $W_{\rm C+L}$ = $ Vw_{\rm C+L}/A$ where $V$ is the volume of the cell and $A_{\rm cell}$ the number of nucleons in the cell. $V = r_{\rm c}^3$, and using $k T_{\rm f}$ = $A_{\rm collective} G_{\rm barrier}$ and evaluating the numerical factors, 

\begin{equation}
\bigg({L \over d}\bigg)_{\rm spaghetti} \approx \bigg( {r_{\rm c} - r_{\rm N} \over r_{\rm c} } \bigg) \bigg( {A_{\rm cell} \over A_{\rm collective}} \bigg)^{1/2} \bigg( {W_{\rm C+L} \over G_{\rm barrier} } \bigg)^{1/2}
\end{equation}

and

\be
\bigg({L \over d}\bigg)_{\rm lasagna} \approx \bigg({L \over d}\bigg)_{\rm spaghetti} {1 \over [ 3.5 \ln{(R/r_{\rm c}})]^{1/2} }
\ee

Varying $R$ between scales of 1 m to $10^{-12}$m varies the logarithmic factor between around 2 to 7, and 

\be
{({L / d})_{\rm spaghetti} \over ({L / d})_{\rm lasagna}} \approx 7 - 25
\ee

\noindent i.e. the order of the spaghetti-type phases will be an order of magnitude larger than the order of the lasagna-type phases.

The configurations corresponding to the local minima in Figure~6 have $W_{\rm C+L}$ = 13-19 keV, $(r_{\rm c} - r_{\rm N})/ r_{\rm c}$ = 0.21 for spaghetti, and 0.5 for lasagna (taking the RMS radius of the cluster in our simulations). Assuming no collectivity, $A_{\rm cell}/A_{\rm collective}$=614 and 

\be
({L / d})_{\rm spaghetti} \approx 40
\ee

Assuming complete collectivity, $A_{\rm cell}/A_{\rm collective}$ = 1, 

\be
({L / d})_{\rm spaghetti} \lesssim 2
\ee

\noindent and matter is completely disordered.

From this, the domains containing the spaghetti phase have a length scale of forty lattice spacings, and the two waffle phases which are similar to lasagna phases (with holes in), are ordered on the length scale ${L / d}_{\rm lasagna} \sim$ two lattice spacings (and are therefore very disordered) at the fictive temperature.

The more nucleons behave collectively, the temperature at which matter becomes frozen into local minima will increase, and the scale on which the matter is ordered decreases. Therefore our estimates above give \emph{upper} limits on the distance scales over which pasta is ordered. 

We can also conclude that fluctuations in the average density and electron fraction, of order 5-10\% occur on microscopic scales of no more than around 10 lattice spacings at this depth.


In the next section, we now go through the crust from the lowest to highest densities we performed calculations at, spanning the pasta phases. We will present the Gibbs free energy surfaces calculated in the same way as presented in this section, the resulting equilibrium fractions of the different phases, and the upper limit to the length scale of their domains derived from the fictive temperature.

\subsection{The onset of pasta: $n_{\rm b}\approx$ 0.035 fm$^{-3}$ - 0.045 fm$^{-3}$.}

In Figures \ref{fig:7}-\ref{fig:9}, we show the Gibbs free energy surfaces at proton fractions of $y_{\rm p}$=0.018, 0.022 and 0.026 in a cell with 454 nucleons in total for calculations performed around $n_{\rm b}$=0.035 fm$^{-3}$, $n_{\rm b}$=0.04 fm$^{-3}$ and $n_{\rm b}$=0.045 fm$^{-3}$ respectively. The Gibbs energy surfaces are calculated at constant pressures of $P$=0.094, 0.12 and 0.15 MeV fm$^{-3}$ respectively (corresponding to the pressure of the spherical configuration at $y_{\rm p}$=0.022 and a density of $n_{\rm b}$=0.035 fm$^{-3}$, $n_{\rm b}$=0.04 fm$^{-3}$ and $n_{\rm b}$=0.045 fm$^{-3}$).

At $P$=0.094 MeV fm$^{-3}$, $n_{\rm b}\approx$ 0.035 fm$^{-3}$, there is a single global minimum at each proton fraction at $(\beta,\gamma)=(0,0\degree)$ corresponding to a spherical nucleus. No other local minima are present. We have not yet entered the pasta phases.

At both $P$=0.12 MeV fm$^{-3}$, $n_{\rm b}\approx$ 0.04 fm$^{-3}$ and $P$=0.15 MeV fm$^{-3}$, $n_{\rm b}\approx$ 0.045 fm$^{-3}$, more structure starts to appear in the energy surface. Some of these structures are local minima corresponding to the first appearance of the nuclear pasta configuration. The variation in Gibbs free energy along trajectories in the energy surface at $P$=0.12 MeV fm$^{-3}$ and 0.15 MeV fm$^{-3}$ are shown in Figures~\ref{fig:8} and \ref{fig:9}. 

At $P$=0.12 MeV fm$^{-3}$ the fictive temperature is $kT_{\rm f}$=73 keV$\to T_8$=8.5 (where $T_8=T/10^8$K), and we see the minimum is still a spherical nuclear phase. A local minima 41 keV above the global minimum corresponding to a spaghetti phase appears at $y_{\rm p}$=0.022. Two local minima corresponding to elongated isolated nuclei in a 2D lattice, analogous to the smectic-B phases of liquid crystals, appear at $y_{\rm p}$=0.022 and $y_{\rm p}$=0.026, at heights 50 and 64 keV above the global minimum. Using these numbers, the relative abundances of the spherical:spaghetti:deformed nucleus($y_{\rm p}$=0.022): deformed nucleus($y_{\rm p}$=0.026) phases are 0.40:0.23:0.20:0.17.

At $P$=0.15 MeV fm$^{-3}$ the fictive temperature is $kT_{\rm f}$ = 34 keV $\to T_8$=4.0, we see the same three phases, with the elongated nuclear phase appearing 21 keV and the spaghetti phase 25 keV above the global minimum of spherical nuclei. Note the fictive temperature and energy separation of minima is decreasing with density. The ratios of spherical nuclei:elongated nuclei:spaghetti are 0.50:0.27:0.24.

In Figure~10 we show a representative set of baryon density surfaces, for calculations performed around $n_{\rm b}$=0.045 fm$^{-3}$. The fluctuations in baryon density between local minima are small at these depths, of order 1\% of the average density.

The lattice energy of the spaghetti phases from the quantum calculations is 18keV per nucleon and 13keV per nucleon for pressures of $P$=0.12 MeV fm$^{-3}$ and 0.15 MeV fm$^{-3}$ respectively, and the corresponding length scale of the spaghetti domains is $\approx$ 7 times the lattice spacing in both cases ($\approx 150$fm). In the case of $P$=0.12 MeV fm$^{-3}$ 10\% of the composition is at a proton fraction of 0.026, so fluctuations in average proton fraction occur of order $20\%$ on length scales of $\sim$ 150fm.

This region in which pasta first appears as local minima above a ground state is the first of four distinct pasta regimes we will encounter, and ranges from $n_{\rm b} \approx$ 0.04 fm$^{-3}$ up to $n_{\rm b} \approx$ 0.05 fm$^{-3}$. It is characterized by spherical nuclei being the absolute ground state, but spaghetti phases and highly deformed nuclei appearing as closely separated local minima, accounting for about over half of the material in the crust at the fictive temperature. Spaghetti phases constitute $\approx$25\% of the material at these densities at the fictive temperatures of $4-8 \times10^8$K. If annealing takes place over sufficiently long timescales, in this regime all matter will eventually be converted to spherical nuclei.

\subsection{Pasta is established, but spherical nuclei still exist: $n_{\rm b}\approx$  0.05 fm$^{-3}$ and 0.054fm$^{-3}$.}

In Figure~\ref{fig:11} we show the Gibbs free energy surfaces at proton fractions of $y_{\rm p}$=0.021, 0.023 and 0.025 at a constant pressure $P$=0.184 MeV fm$^{-3}$ (densities around $n_{\rm b}$=0.05 fm$^{-3}$) in a cell containing 956 nucleons. In Figure~\ref{fig:12} we show the Gibbs free energy surfaces at proton fractions of $y_{\rm p}$=0.022, 0.024 and 0.026 at a constant pressure $P$=0.234 MeV fm$^{-3}$ (densities around $n_{\rm b}$=0.054 fm$^{-3}$) in a cell containing 1166 nucleons. In both cases, one can see that the energy surfaces are becoming much richer in structure, with numerous local minima. In each case we identify three minima accessible to all others by continuous deformation and adjustment of proton fraction without increasing energy.

We see that at both densities isolated nuclei are still present, but they are no longer always the absolute minimum. At $P$=0.184 MeV fm$^{-3}$ the absolute minimum is the nuclear `waffle' phase, with a corrugated spaghetti phase also appearing as a local minimum. The minima occur at two different proton fractions, pointing to possible fluctuations in average proton fraction of order 10\%. At $P$=0.234 MeV fm$^{-3}$ the isolated nuclear phase is - just - the absolute minimum, almost equal in energy to the spaghetti phase. Again, a waffle phase appears as a local minimum.

At $P$=0.184 MeV fm$^{-3}$ the fictive temperature $kT_{\rm f}$ is $21$keV $\to T_8$=2.4, and $\Delta G_{21}=$14keV and $\Delta G_{31}=$26keV. The relative abundances at $T_{\rm f}$ are waffle:spherical:spaghetti are 0.55:0.29:0.16.

At $P$=0.234 MeV fm$^{-3}$ the fictive temperature $kT_{\rm f}$ is $11$keV $\to T_8$=1.3 and $\Delta G_{21}=$4keV and $\Delta G_{31}=$6keV. The ratios of minima 1(spherical):2 (spaghetti):3(waffle) are 0.44:0.31:0.25. 

The lattice energy of the spaghetti and waffle phases from the quantum calculations are 23keV per nucleon and 17keV per nucleon respectively for $P$=0.184 MeV fm$^{-3}$ and 24keV per nucleon and 20keV per nucleon respectively for $P$=0.234 MeV fm$^{-3}$. 

Waffles are really a form of lasagna with a density modulation along its surface, and we can estimate its long range order as that of lasagna.
It is of order 2-3 lattice spacings for $P$=0.184 MeV fm$^{-3}$ and $P$=0.234 MeV fm$^{-3}$ - indicating these phases are almost completely disordered. The corresponding length scale of the spaghetti domains from is 19 and 25 times the lattice spacing (500-600fm) for $P$=0.184 MeV fm$^{-3}$ and $P$=0.234 MeV fm$^{-3}$ respectively. Fluctuations of proton fraction of order 10\% occur on these length scales.

In this regime of pasta, the pasta shapes have for the most part become the ground state, but spherical nuclei still appear as local minima (and can occasionally appear as the global minimum.) If annealing takes place, on sufficiently long timescales, all matter in this regime will be converted to pasta. The fictive temperature has dropped to a 1-3$\times10^8$K. At each density in this regime, isolated nuclei, spaghetti and waffles coexist at the fictive temperature.

\subsection{Exclusively pasta, but protons are still localized to the pasta structures:}

In the density region between around 0.055 fm$^{-3}$ to 0.065 fm$^{-3}$, isolated nuclei cease to be local minima at all, so we have entered the regime of pure pasta. We have already detailed a representative set of configurations at 0.06 fm$^{-3}$. Waffles and spaghetti are the coexisting phases in this density window. Protons are still localized in at least one dimension.

\subsection{Protons are delocalized}

In Figures~\ref{fig:13}-\ref{fig:15}, we show the Gibbs free energy surfaces at pressures of $P$=0.30 MeV fm$^{-3}$,0.32 MeV fm$^{-3}$ and 0.34 MeV fm$^{-3}$ corresponding to densities around $n_{\rm b}$=0.066 fm$^{-3}$, 0.07 fm$^{-3}$ and $n_{\rm b}$=0.076 fm$^{-3}$. The cells contain 784, 532 and 532 nucleons respectively. 

At $P$=0.30 MeV fm$^{-3}$, $n_{\rm b}$=0.066 fm$^{-3}$, we perform calculations at proton fractions of $y_{\rm p}$=0.022, 0.025 and 0.028. The minimum energy configuration is the bicontinuous cubic-P (BCP) phase - essentially a nucleus joined in all three directions to adjacent nuclei, allowing the protons to be entirely delocalized. Both cluster and external neutron fluid form two interlaced, continuous domains through space. Lasagna and spaghetti phases make up the two remaining unique local minima. At $P$=0.32 MeV fm$^{-3}$ and $P$=0.34 MeV fm$^{-3}$ fm$^{-3}$, ($n_{\rm b}$=0.07 fm$^{-3}$ and $n_{\rm b}$=0.076 fm$^{-3}$) we perform calculations at proton fractions of $y_{\rm p}$=0.030, 0.034, 0.038 and 0.42. At the lower of those two densities, the absolute minimum corresponds to the lasagna phase, with waffle and BCP phases making up the remaining unique local minima. At the higher density, all minima correspond to variations of the BCP phases at two different proton fractions $y_{\rm p}$=0.034, 0.038.
Therefore the three main phases that coexist in this range of densities are spaghetti, waffles, lasagne and the BCP phase.

At $P$=0.30 MeV fm$^{-3}$, $n_{\rm b}$=0.066 fm$^{-3}$, the fictive temperature $kT_{\rm f}$ is 7.4keV $\to T_8$=0.85, and $\Delta G_{21}=$1keV and $\Delta G_{31}=$5keV. The ratios of BCP:lasagna:spaghetti are 0.42:0.37:0.21. The lattice energy of the lasagna phase from the quantum calculations is 7keV per nucleon and for the spaghetti phase it is 11keV per nucleon. For the spaghetti phases, our estimated domain sizes at the fictive temperature is 22 lattice spacings, and for the lasagna is about 3 lattice spacings.

At $P$=0.32 MeV fm$^{-3}$, $n_{\rm b}$=0.07 fm$^{-3}$, the fictive temperature $kT_{\rm f}$ is 14keV $\to T_8$=1.7, and $\Delta G_{21}=$10keV and $\Delta G_{31}=$10keV. The ratios of lasagna:waffle:BCP are 0.49:0.26:0.25. The lattice energy of both lasagna and waffle phases is 6keV per nucleon, and their estimated domain size is approximately one lattice spacing.

At $P$=0.34 MeV fm$^{-3}$, $n_{\rm b}$=0.076 fm$^{-3}$ the fictive temperature $kT_{\rm f}$ is 16keV $\to T_8$=2.0, and $\Delta G_{21}=$0.3keV and $\Delta G_{31}=$3keV. The ratios of the abundances of all three BCP phases are 0.36:0.35:0.29. The BCP phase is stabilized with respect to thermodynamic fluctuations of the type explored in this paper, so in this density region it is possible that the order of matter is much longer than density regions where the BCP phase is not the only nuclear geometry.

In \ref{fig:16} we show the baryon density surfaces at a pressure of $P$=0.34 MeV fm$^{-3}$. Variations from local minimum to local minimum are around 5\% that of the average baryon density.

Moving up in density, in Figures \ref{fig:17}-\ref{fig:18}, we show the Gibbs free energy surfaces at pressures of $P$=0.45 MeV fm$^{-3}$ and 0.53 MeV fm$^{-3}$ corresponding to densities around $n_{\rm b}$=0.082 fm$^{-3}$ and 0.088 fm$^{-3}$. In each case the cells contain 332 nucleons, and we calculate energy surfaces at proton fractions of $y_{\rm p}$=0.24, 0.3 and 0.36. We see that as we get close to the crust-core transition, the energy surfaces are becoming simpler in their structure. In both cases, two minima appear.

At $P$=0.45 MeV fm$^{-3}$, the two minima are at a proton fraction of 0.03. The higher-lying one is a spaghetti like configuration; the absolute minimum is a more complex configuration, consisting of cylindrical holes (anti-spaghetti) and spaghetti. To reveal the structure requires plotting the two isosurfaces at low and high density, as shown in Figure \ref{fig:17}; the low density isosurface (about half the average neutron density) is shown on the left and reveals the cylindrical hole; on the right, at high density (about three-quarters the average neutron density in the cell) the spaghetti structure is revealed. In the bulk, protons as well as neutrons are delocalized.

At $P$=0.53 MeV fm$^{-3}$, the absolute minimum is at a proton fraction of 0.036 and corresponds to a lasagna structure. A local minimum appears at a proton fraction of 0.3 and corresponds to the same spaghetti-anti-spaghetti configuration seen at the previous density. 

At $P$=0.45 MeV fm$^{-3}$, the fictive temperature $kT_{\rm f}$ is 16keV $\to T_8$=1.9, and $\Delta G_{21}=$3.5 keV. The ratios of spaghetti hole:spaghetti are 0.55:0.45. The lattice energy of the spaghetti phase is 3.7keV per nucleon, and their estimated domain size is approximately 4 lattice spacings. The lattice energy of the spaghetti hole is 2.9 keV per nucleon, with a domain size similar order.

At $P$=53 MeV fm$^{-3}$, the fictive temperature $kT_{\rm f}$ is 4.6keV $\to T_8$=0.53, and $\Delta G_{21}=$2 keV. The ratios of lasagna:spaghetti hole are 0.6:0.4. The lattice energy of the lasagna phase is 1.8 keV per nucleon, and the lattice energy of the spaghetti hole is 1.6keV per nucleon. The estimated domain size is approximately 2 lattice spacings for the lasagna and significantly less than 1 lattice spacing for the spaghetti hole. 

The spaghetti phases, the first pasta to appear as one descends through the crust, can persist to relatively deep layers, making them the most robust of the pasta phases.

\section{Summary and Discussion}

In Figure~\ref{fig:19} we show a visual summary of the pasta phases found to coexist at increasing depth in the crust. Laterally, they are arranged with the minimum energy configuration on the left. In order to give a visualization of how the phases appear over longer length scales than a single cell, we have stacked 4 cells in each direction, but it's important to remember we only every calculate one unit cell. We have not included every single density and configuration detailed in the previous section, but instead selected those that best represent that region of density. Horizontal lines demarcate the four distinct regions we have identified, which we label P1-P4 in what follows. With increasing density, (P1) the region where pasta first appears as a local minimum, but isolated nuclei occupy the absolute minimum (P2) pasta phases have become consistently the absolute minimum but isolated nuclei still occupy local minima, (P3) all local minima correspond to pasta phases, while the protons are localized to the clusters in at least one dimension, (P4) all local minima correspond to pasta phases and protons are delocalized in all three dimensions.

One can trace the appearance and fate of one's favorite pasta shape with density. Isolated nuclei persist up to around 0.06 fm$^{-3}$, but are demoted to local minimum at around 0.05 fm$^{-3}$. Isolated nuclei themselves may have exotic properties at high densities, with highly deformed nuclei aligning like liquid crystals. Spaghetti appears at low densities $\approx$0.04 fm$^{-3}$ as a local minimum, and is the most robust pasta shape, making appearances right up to densities of $\approx$0.088 fm$^{-3}$. It is most likely to be the absolute minimum at densities 0.05-0.06 fm$^{-3}$, appearing at local minima at most densities thereafter. Next, nuclear waffles appear as a local minimum 0.05-0.06 fm$^{-3}$, and persist over a wide density region up to close to 0.08 fm$^{-3}$. In our calculations, they appear as the global minimum at densities around 0.06 fm$^{-3}$. Nuclear waffles are a lasagna like phase - essentially lasagna modulated by density fluctuations along the plane of the shape. Lasagna appear at around 0.066 fm$^{-3}$ and persist to the highest densities as local minima. Interestingly, lasagna never appear as the minimum energy configuration in our calculations; at the same density region they appear, so does the bicontinuous cubic-P phase, which appears to be energetically more stable and is the absolute minimum in the region 0.07-0.08 fm$^{-3}$, until being usurped by the phase that contains cylindrical holes. As discussed, the cylindrical hole phase we discover is actually a phases consisting of alternating cylindrical under - and over-densities, with neutrons and protons delocalized in the bulk. This is the absolute minimum up to the highest density we consider.

The exact phases that appear at each density are driven by the proton fraction and modulated by shell effects. In order to evolve through the pasta phases, one can either increase the density or the proton fraction, since protons promote clustering.

In Figure~\ref{fig:20} we plot the pressure (Figure~\ref{fig:20}a) and chemical potential (Figure~\ref{fig:20}b) at which we perform each of our calculations as a function of the densities of each local minimum as blue diamonds. We compare this with the pressure and neutron chemical potential as calculated with the compressible liquid drop model (CLDM) using the surface parameters that give closest agreement to 3DHF results (as described in \cite{Balliet:2020aa}). First, we note that spurious shell effects caused by the discretization of the neutron energy spectrum and corresponding change in the density of states means we don't expect the pressure and neutron chemical potential to match exactly the CLDM results. Also, due to real shell effects, the equilibrium proton fraction differs from the CLDM result, and the pressure and neutron chemical potential depends on the proton fraction. Having said that, for most densities the pressures and neutron chemical potentials obtained in the 3DHF calculations well match the CLDM result, with the major differences being at intermediate densities. 

The pressure can be translated approximately into a mass coordinate for the crust $y$, which is also approximately the column depth in the crust - the fraction of the mass of the crust above a given layer at a given pressure. It is given by

\begin{equation}
y[P(\rho)] \equiv \frac{M_{\star}-M(P)}{M_{\star}} \approx 1 - \bigg( \frac{P_{\rm cc}-P(\rho)}{P_{\rm cc}} \bigg)
\end{equation}

Similarly the neutron chemical potential at a particular density approximately determines the depth of the layer, and we define the radial coordinate

\begin{equation}
r[\mu(\rho)] \equiv \frac{R_{\star}-R(\mu)}{R_{\star}} \approx 1 - \bigg( \frac{\mu_{\rm cc}-\mu(\rho)}{\mu_{\rm cc}} \bigg)
\end{equation}

\noindent where the subscript `cc' denotes the quantity at the crust-core transition, $M_{\star}$, $R_{\star}$ is the stellar radius and mass and $M(P)$ and $R(\mu)$ the radius and mass out to the layer of the star characterized by the pressure $P$ and chemical potential $\mu$ respectively.

Our 3DHF calculations suggest that the crust-core transition density is $\approx$0.09fm$^{-3}$. The CLDM results presented in Figure~20 also give a transition density of 0.09fm$^{-3}$. The CLDM crust-core transition pressure $P_{\rm cc}$ and chemical potential $\mu_{\rm cc}$ are 0.54MeV fm$^{-3}$ and 13.6 MeV respectively. The mass and radial coordinates $y$ and $m$ we display from now on are obtained from the CLDM results. The mass and radius coordinates are a better measure of location in the crust, since density isn't directly related to the depth in the crust (different EOSs will give different densities at a given depth or mass coordinate in the crust). In what follows, we plot various quantities versus $y$.

The plot of pressure versus density illustrates the size of the local average density fluctuations (the variation over the scale of the domain sizes). At $P$=0.33 MeV fm$^{-3}$, for example, the density of the coexisting phases have a spread of 0.003 fm$^{-3}$, about 5\% of the average density. Typically the density fluctuations are between 2\% and 5\%.

In Figure~20, the dotted vertical lines delineate the 4 distinct regions of pasta we identify: (P1) (pasta is a local minimum, spherical nuclei the ground state) accounts for about 10\% of the crustal mass and thickness, and spans a region between $\approx$ 20\% and 30\% of the way into the crust by mass and $\approx$ 50\% and 60\% by depth; (P2) (pasta is the ground state, spherical nuclei are local minima) accounts for about 15\% of the crustal mass and 10\% the crustal thickness, and spans a region between $\approx$ 30\% and 45\% of the way into the crust by mass and $\approx$ 60\% and 70\% by depth; (P3) (all local minima are pasta, protons are localized) accounts for about 5\% of the crustal mass and 5\% the crustal thickness, and spans a region between $\approx$ 45\% and 50\% of the way into the crust by mass and $\approx$ 70\% and 75\% by depth; and (P4) (all local minima are pasta, protons are delocalized) accounts for about 50\% of the crustal mass and 25\% the crustal thickness, and spans a region between $\approx$ 55\% and 100\% of the way into the crust by mass and $\approx$75\% and 100\% by depth).

In Figure~\ref{fig:21}, we plot the fictive temperature as a function of column depth $y$. The lower red points show the temperatures assuming the nucleons behave independently, and the upper blue diamonds assume that of order the proton number in the unit cell behave collectively when matter is rearranged from one pasta shape to another. There are a two main contributors to the energy barriers: the lattice energy caused by the Coulomb repulsion of the protons in adjacent pasta structures, and the shell energy. Rearranging the pasta configuration requires a rearrangement of nucleons in the vicinity of the Fermi surface of order the proton number.

The fictive temperature is calculated based on the average barrier heights between local minima in our calculations; the bars on each points indicate the range of barrier heights at each depth. 

The lower limit to the barrier heights given by the red points start around $10^9$ K and drop to $\sim 10^8$K throughout most of the pasta region, dropping down below $10^8$K at in the final 20\% of the crust by mass. The estimates assuming a realistic number of nucleons behaving collectively are an order of magnitude higher.

When the crust temperature drops below this temperature, the coexisting pasta domains are frozen in with respect to thermal fluctuations, at their equilibrium abundances and characteristic lengths at the fictive temperature. Quantum tunneling could then anneal the crust on a timescale that is yet to be determined, the end result being a single phase of pasta at each depth. Future heating could then repopulate the local minima.

The length scales of the domains at the fictive temperature are plotted in Figure~\ref{fig:22}. Here we take the low end of the bound on the fictive temperature from the red points in Figure~21: this gives an \emph{upper} limit to the domain sizes at the time the domains freeze. We plot the size of the domains as a function of mass coordinate in absolute terms (Fig. ~\ref{fig:22}a) and in units of lattice spacings $d$ (Fig. ~\ref{fig:22}b). Dashed horizontal lines indicate where the order falls below 10fm (Fig. ~\ref{fig:22}a) - a typical width of pasta structures - and below one lattice spacing (Fig. ~\ref{fig:22}b). These lines indicate the scale below which the phases become completely disordered - essentially liquid. The temperature at which the order corresponds to these lines is the melting temperature of the glassy pasta. Spaghetti-like configurations are shown as red points and lasagna-like configurations as blue diamonds. Spaghetti phases generally have long range order of at most between 100 and 1000fm (10-50 lattice spacings), with the longest range order occurring in regions P2 and P3. Lasagna-like phases are an order of magnitude less ordered, peaking at 80fm ($\approx 3$ lattice spacings) and dipping down to below the melting point in the deepest layers.  

Given these are \emph{upper} limits, we can conclude that the pasta phases are highly disordered at the fictive temperature, which supports the hypothesis that they have very high thermal and electrical resistivity. Additionally, given the domains are only ordered over a short range, electron scattering from domain boundaries could contribute significantly to the thermal and electrical resistivity of the deep layers of the crust.

As the temperature drops below the fictive temperature, the domains are frozen and cannot immediately grow in size, being bounded by adjacent domains. If annealing occurs, the domains containing lower energy phases will gradually grow as they convert surrounding higher energy phases. 

If the crust gets heated above the fictive temperature, thermal fluctuations can once again repopulate the local minima. The base of the crust is expected to be heated up to temperatures of order $10^8$K during accretion, enough to enter the regime where different phases could be repopulated and coexist. If a stage is reached where an entire layer has been completely annealed, increases in temperature even below the fictive temperature will decrease the long-range order. In either case, the thermal conductivity could be temperature dependent, something that might be explored in crust cooling simulations.

One of the crucial quantities in determining the stability of phases is the lattice Coulomb energy per particle $W_{\rm CL}/A$, which we plot in Figure~\ref{fig:23} as a function of column density $y$ for all local minima. The lattice energy rises to a peak at a mass fraction of around 0.4 (a density of around 0.06fm$^{-3}$), corresponding to the region where isolated nuclei have vanished from the energy landscape, and the energy surfaces have their most structure. From then on, the lattice energy decreases roughly exponentially up to the crust-core transition. It should vanish completely at the crust-core transition, but density fluctuations due to the spurious shell effects mean that it remains finite in our calculations. 

\section{Conclusions}

We have conducted a large set of quantum calculations of nuclear pasta using the three-dimensional Hartree-Fock method with the NRAPR Skyrme interaction and BCS pairing. A quadrupole constraining potential allowed us to control the shape of the nuclear deformation, and therefore systematically probe the energy landscape of nuclear pasta. Although the constraint determines the shape we get, by ranging over the whole parameter space of the quadrupole deformation, we were able to reproduce and compare all pasta shapes in an unbiased way.

We selected 11 different density regimes to perform calculations at, ranging from 0.035fm$^{-3}$ to 0.088fm$^{-3}$. At each density we selected a single cell size, chosen based on the range of cell sizes predicted by calculations using the compressible liquid drop model (CLDM). In each density regime we performed calculations at five different, closely spaced densities and used the results to interpolate the energy at a given reference pressure, chosen to be the pressure for a zero-deformation configuration at the central density in the range. We thus calculated the Gibbs free energy, and were able to compare configurations in equilibrium at constant pressure (and therefore at the same depth in the crust). In each density regime we also performed calculations over a range of proton fractions to locate the $\beta$-equilibrium proton fraction. In total, the results we show are obtained from approximately 30,000 calculations and 300,000 CPU hours.

We show that the energy landscape consists of multiple local minima corresponding to different nuclear geometries with very similar energies, separated by barriers of 1-100keV which generally decrease with depth. These findings are consistent with earlier, more limited explorations of the energy landscape of nuclear pasta \cite{Magierski:2001yq}. The various minima at a given depth can be at different proton fractions and densities. Therefore, at a given depth in the star, multiple nuclear pasta shapes may coexist, and there may be local variations of the average electron fraction and baryon density of order 10\% and 1-5\% respectively.

The following nuclear geometries were found: spherical and deformed nuclei appear up to $n_{\rm b}$=0.06fm$^{-3}$ as either the minimum energy configuration (up to $n_{\rm b}\approx$0.05fm$^{-3}$) or a local minimum. Spaghetti appears over the range 0.04-0.088fm$^{-3}$, the waffle phase (perforated planes) appear over the range 0.05-0.08fm$^{-3}$, lasagna over 0.065-0.088fm$^{-3}$, the bi-continuous cubic-P (BCP) phase over the range 0.066-0.08fm$^{-3}$ and cylindrical holes appear at 0.08-0.088 fm$^{-3}$. The BCP phase consists of both continuous neutron matter and nuclear matter interlaced; the protons are therefore delocalized in all directions. The cylindrical holes coexist with spaghetti phases, with protons delocalized in the bulk. At each depth in the crust we identify the phases that coexist and estimate their relative abundances. Spaghetti appears over the widest range of densities. 

Four distinct regions can be identified, which we denote P1-4: (P1) roughly spherical nuclei are the minimum energy configuration, but pasta appears as local minima; (P2) pasta phases become the minimum energy configuration, but spherical nuclei still occupy some local minima; (P3) all local minima correspond to pasta configurations, and protons are localized in at least one dimension, and (P4) all local minima correspond to pasta configurations, and the appearance of the BCP phase indicates protons are delocalized in all dimensions. We find the BCP phase is particularly stable and the delocalized proton region accounts for at least half of the pasta layers by mass and depth. The regions P2-P4 - where pasta is the ground state - occupy about 70\% of the whole crust by mass and 40\% by depth. In total, P1-P4 accounts for almost 80\% of the mass of the crust and 50\% of its thickness.

The nature of the energy landscape suggests the pasta phases are glassy: they undergo a transition to an amorphous solid at a certain temperature below their melting temperature. We take the barrier heights between local minima to set the temperature scale - the so-called fictive temperature, to borrow a term from condensed matter physics, above which matter is an amorphous solid. As matter cools below the fictive temperature, we posit that it becomes frozen into domains with a length scale set by stability against thermal fluctuations at the fictive temperature. These length scales are of order 10 times the cell size at lower depths, falling to below the cell size at higher depths, depending on whether the nuclear geometry is spaghetti-like or lasagna-like. If the length scale is below the cell size, the matter remains a liquid (no long-range order at all) until the temperature cools significantly below the fictive temperature. 

Once matter is frozen into domains, it is possible annealing will begin to homogenize the matter at a given depth, with domains corresponding to the marginally energetically preferred configuration expanding their volume, converting the other phases. The conversion process could be similar to pycnonuclear fusion \cite{Yakovlev:2006aa,Yakovlev:2010aa}, whose timescale uncertainty is many orders of magnitude (see for example the table 2 in \citep{Yakovlev:2006ab}). Additionally, domains many times the size of a single unit cell will need to be converted. An analysis of the uncertainty this leads to on the annealing timescale would be an important future study to undertake. Annealing would release heat in the deep layers of the crust of order 1-50keV/nucleon - the typical energy differences between local minima and the ground state.

Although we perform a very large number of calculations, we still do not sample widely the unit cell size at any given depth, and although it is reasonable to expect that our results do not qualitatively depend on cell size, it is likely that some phases we don't see (notably bubble) are missed due to this restriction. Phases that never appear as the minimum energy configuration in our calculations - like spaghetti - likely would in a larger set of calculations. In a recent set of calculations using the same method at fixed proton fractions, and generally larger cell sizes, a similar set of pasta geometries were found \cite{Fattoyev:2017aa}. Our results should be taken as an illuminating snapshot of the pasta landscape. 

The phases of pasta present depend on the proton fraction, which in turn depends on the symmetry energy at pasta densities. An examination of the dependence of our results on the EOS - particularly, the extent of the four distinct regions of pasta - will be presented in an upcoming work.

As we explore the nuclear pasta phases in more detail, their structure at both micro- and meso-scales becomes richer and more complex, presenting a serious challenge to modeling the material and transport properties near the crust-core boundary. However, the fact they likely occupy a large mass fraction and thickness of the crust, and that they mediate the transition from solid crust to liquid core, means they remain an essential ingredient in the modeling of many macroscopic, observable neutron star phenomena.

\acknowledgements
WGN and AS acknowledge support from NASA grant 80NSSC18K1019. This work benefited from discussions at the 2018 Frontiers in Nuclear Astrophysics Conference supported by the National Science Foundation under Grant No. PHY-1430152 (JINA Center for the Evolution of the Elements). In particular, the authors would like to thank Jerome Margueron and Sanjay Reddy for useful discussions.

\bibliographystyle{unsrtnat}

\end{document}